\documentclass[superscriptaddress,aps,preprintnumbers,amsmath,showpacs,amssymb,prd,nofootinbib,preprint]{revtex4-1}
\pdfoutput=1
\usepackage{booktabs} 
\usepackage{array} 
\usepackage[table,xcdraw]{xcolor} 
\usepackage{amsmath,amssymb}
\usepackage{graphicx}
\usepackage{fancyhdr}
\usepackage{bm, color}
\usepackage{slashed,amsthm,amsfonts,empheq}
\usepackage[caption=false]{subfig}
\usepackage{hyperref}
\usepackage{booktabs}
\usepackage{multirow}
\usepackage[left=2cm,right=2cm,top=2cm,bottom=2cm,includefoot,a4paper]{geometry}
\usepackage{tikz}
\usetikzlibrary{arrows.meta}

\newcommand{\Slash}[1]{{\ooalign{\hfil#1\hfil\crcr\raise.167ex\hbox{/}}}}

\newcommand{\beq}{\begin{equation}}  \newcommand{\eeq}{\end{equation}}
\newcommand{\bef}{\begin{figure}}  \newcommand{\eef}{\end{figure}}
\newcommand{\bec}{\begin{center}}  \newcommand{\eec}{\end{center}}

\newcommand{\laq}[1]{\label{eq:#1}}  
\newcommand{\Eq}[1]{Eq.(\ref{eq:#1})}
\newcommand{\Eqs}[1]{Eqs.(\ref{eq:#1})}
\newcommand{\eq}[1]{(\ref{eq:#1})}

\newcommand{\vev}[1]{\left\langle {#1} \right\rangle}

\newcommand{\lac}[1]{\label{chap:#1}}
\newcommand{\SU}[1]{{\rm SU{#1} } }

\def\({\left(}
\def\){\right)}

\def\O{\mathcal{O}}
\def\U{\mathop{\rm U}}

\newcommand{\OR}{~{\rm or}~}
\newcommand{\AND}{~{\rm and}~}

\newcommand{\GEV}{\,{\rm GeV}}

\def\d{\delta}

\def\f{\phi}

\def\l{\lambda}

\def\x{\xi}

\def\F{\Phi}

\def\*{\dagger}

\begin{document}

\title{
PBHs and GWs from Scaling Monopoles
}
\author{Daiki Aburatani}
\affiliation{Department of Physics, Tokyo Metropolitan University, Minami-Osawa, Hachioji-shi, Tokyo 192-0397, Japan}

\author{Wakutaka Nakano}
\affiliation{Department of Physics, Tokyo Metropolitan University, Minami-Osawa, Hachioji-shi, Tokyo 192-0397, Japan}

\author{Wen Yin}
\affiliation{Department of Physics, Tokyo Metropolitan University, Minami-Osawa, Hachioji-shi, Tokyo 192-0397, Japan}

\begin{abstract}
Monopoles with sufficiently weak gauge couplings, or from global symmetries, can form scaling networks in the early Universe whose average energy density tracks the cosmological background.
In this work, we find, by performing
classical lattice simulations to estimate the overdensities, that primordial black holes (PBHs) with a broad
mass spectrum can be produced during this evolution if the Higgs expectation
value $v$ satisfies
$
v\gtrsim 0.1 M_{\rm pl}$.  The formation is driven by the
stochastic realization of the monopole number in Hubble patches causing the overdensities.
We also show that gravitational waves (GWs) generated by the scaling dynamics are
produced at the same epoch, with spectra correlated with the PBH spectra and
with amplitudes testable in future observations.  Interestingly, if the scaling
regime is terminated by the gauge boson mass for the gauged monopole, a non-negligible fraction of the PBHs can carry magnetic charge, and the resulting magnetic Coulomb force between such
charged PBHs is predicted to be comparable to the gravitational force.  Together with the PBH and GW signals, this provides a smoking-gun signature of the scenario.  We also point out simple cosmological scenarios, which may also
apply to PBH formation from scaling cosmic strings, that allow PBHs to constitute dominant dark
matter.
\end{abstract}
\maketitle

 \section{Introduction}

Primordial black holes (PBHs) are among the simplest candidates for dark matter and have been extensively studied in a wide range of cosmological scenarios. Their formation typically requires large density fluctuations in the early Universe, such as those generated during inflation or phase transitions \cite{Carr:1974nx,Carr:1975qj,Sasaki:2018dmp,Carr:2020gox}.  (See Refs.\,\cite{Villanueva-Domingo:2021spv,Green:2020jor,Carr:2026hot} for recent review on PBH.)

Another well-known possibility is PBH formation from topological defects,
especially cosmic strings and domain walls
\cite{Hawking:1987bn,Polnarev:1988dh,Caldwell:1995fu,MacGibbon:1997pu,Helfer:2018qgv,Garriga:1993gj,James-Turner:2019ssu,Ferrer:2018uiu,Liu:2019lul,Gouttenoire:2023naa,Kitajima:2023cek,Gouttenoire:2023ftk,Ferreira:2024eru,Masubuchi:2026eau,Sugeno:2025kwx,Miyazaki:2025tvq}. 
In the conventional loop-collapse picture for cosmic strings, a closed string
loop forms a black hole only if its shape becomes sufficiently compact during
the oscillation.  The resulting PBH abundance is therefore governed by the
collapse fraction of such compact loop configurations, which has been estimated
numerically from realistic loop samples.  Since defect networks can operate
over an extended period, they may lead to an extended PBH mass spectrum rather
than a monochromatic one.  In this work, we propose a distinct mechanism in
which PBHs are produced by a scaling monopole network.\footnote{PBH formation has also been discussed
in hybrid monopole-string systems
\cite{Matsuda:2005ez}, where the monopoles can enhance PBH
formation in cosmic-string systems. By contrast, PBH formation from a scaling
monopole network alone has rarely been considered.}

Independently of their formation mechanism, the dynamics of PBHs can be significantly modified if they carry charges under additional gauge symmetries. Charged black holes have been studied both in general relativity and in cosmological contexts, including electrically or magnetically charged PBHs and their impact on binary formation and merger rates \cite{Bekenstein:1971hc,Ruffini:1971bza,Bai:2019zcd,Liu:2020vsy,Liu:2020cds,Bozzola:2020mjx,Maldacena:2020skw}. In particular, long-range forces can alter the formation and evolution of PBH binaries, leading to potentially observable signatures in gravitational wave (GW) experiments.\footnote{In this paper, we however mainly study the GW from the scaling rather than from the PBH/BH binaries.}

In this work, we study PBH formation from scaling monopole networks.  The key
point is that the monopole network does not only contribute to the averaged
energy density, but also generates stochastic Hubble-patch overdensities.  Since
the monopole number in each Hubble patch fluctuates, rare patches can contain
an unusually large amount of defect activity and hence a large total energy
density.  If the symmetry-breaking scale is sufficiently high, these local
overdensities originated from the (would-be) Nambu-Goldstone boson (NGB) gradient energies can exceed the threshold for gravitational collapse and form
PBHs.

We investigate this mechanism by performing classical lattice simulations of
the scalar field dynamics.  We measure the total energy density and the
monopole content in each Hubble patch. From the correlation, we construct the probability
distribution of the Hubble-patch energy densities.  We find that the
high-energy tail of this distribution can be large enough to produce PBHs for the Higgs expectation value $v$ satisfying
$v\gtrsim 0.1M_{\rm pl}$, where $M_{\rm pl}=2.4\times 10^{18}\GEV$ is the reduced Planck scale. Since the monopole network is in a scaling regime, PBHs can be produced over an extended period, leading naturally to a broad mass
spectrum. 
The setup is motivated, for example, by supersymmetric or
moduli-like theories in which light scalars, which are stabilized in the symmetric phase due to dynamical mass, acquire a near-Planckian vacuum expectation value when the Hubble parameter becomes small enough. 
Alternatively, the large expectation value can be naturally realized by thermal effects if the weakness of the $\phi_a$ coupling is controlled by a single large wave function renormalization constant~\cite{Yin:2024txg,Yin:2024pri}. 

The same scaling dynamics also sources GWs because of the large expectation value $v$.  The monopole
network contains relativistic and strongly inhomogeneous scalar gradients, which
generate anisotropic stress during the scaling evolution.  We show that the
resulting GW spectrum is correlated with the PBH mass spectrum,
because both originate from the same Hubble-scale monopole dynamics.  This
correlation provides a characteristic prediction of the scenario.

Finally, we discuss cosmological scenarios in which the monopole network does
not survive until late times. This is necessary because a long-lived global
monopole network is strongly constrained by CMB observations.  We show that the
network can have a finite lifetime due to the population bias and a blue-tilted initial fluctuation if the weakly coupled Higgs is set on the hilltop by the $\O(1)$ Hubble induced mass, which is around the Hubble parameter.
Alternatively, the cosmological problems are solved if the symmetry-breaking scale changes after PBH formation. These scenarios allow the PBHs formed during the scaling era to remain as dark matter while
suppressing the residual monopole or light-particle abundance. 

The rest of this paper is organized as follows.
In Sec.~\ref{chap:1}, we review the basic properties of global monopoles and
their scaling behavior, emphasizing the long-range gradient energy and the
resulting Hubble-scale energy fraction.
In Sec.~\ref{chap:2}, we present the mechanism of PBH formation from stochastic
Hubble-patch overdensities in the scaling monopole network. 
In Sec.~\ref{chap:3}, we study the stochastic GW background
generated by the same scaling dynamics and show how its present spectrum is
correlated with the PBH mass spectrum.
In Sec.~\ref{chap:4}, we summarize our results and discuss implications, 
including the possibility that the PBHs carry hidden magnetic charge. 
Several technical details are collected in the appendices: fitting functions
for the GW spectra, a lower bound on the Hubble-patch gradient
energy from boundary winding, cosmological scenarios in which the monopole
network has a finite lifetime, the gravitational field of a global monopole,
the inefficiency of PBH formation from delayed rolling without monopoles, and
the lattice definitions of Hubble-patch observables.

\section{Global monopoles and their properties}
\lac{1}

We begin by reviewing the basic properties of global monopoles.  They also
provide a useful effective description of gauge monopoles in the small gauge coupling regime, where the gauge screening length is longer than the relevant
cosmological scale. In most cases, we consider global monopole for clarity and we will mention the gauge case when relevant. 

Global monopoles are formed when a global $O(3)$ symmetry is spontaneously
broken to $O(2)$ \cite{Barriola:1989hx,Vilenkin:2000jqa}.  A minimal model is
described by a scalar triplet $\phi^a$ $(a=1,2,3)$ with the Lagrangian
\begin{equation}
\mathcal{L}
=
\frac{1}{2}\partial_\mu \phi^a \partial^\mu \phi^a
-
\frac{\lambda}{4}
\left(\phi^a\phi^a-v^2\right)^2 .
\end{equation}
The vacuum manifold is $S^2$, and monopole configurations are characterized by
\begin{equation}
\pi_2(S^2)=\mathbb{Z}.
\end{equation}

The spherically symmetric monopole solution takes the form
\begin{equation}
\phi^a
=
v\,h(r)\frac{x^a}{r},
\end{equation}
where $h(0)=0$ and $h(r)\to1$ at large $r$.  The core size is
\begin{equation}
r_{\rm core}
\sim
(\sqrt{\lambda}\,v)^{-1}.
\end{equation}

A distinctive feature of global monopoles is their long-range gradient energy.
Outside the core, the radial mode is approximately fixed and the angular
gradient gives
\begin{equation}
\rho_{\rm grad}(r)
\sim
\frac{v^2}{r^2}.
\end{equation}
Therefore the energy enclosed within radius $R$ grows linearly,
\begin{equation}
E(R)
\sim
4\pi v^2 R .
\label{eq:global-mono-energy-frw}
\end{equation}
This linear growth is cut off in cosmology by the distance to neighboring
antimonopoles.  It also implies that a
monopole--antimonopole pair feels an approximately constant attractive force,
\begin{equation}
F
\sim
\frac{\partial E}{\partial R}
\sim
4\pi v^2 .
\end{equation}

Monopoles are produced by the Kibble mechanism when the symmetry is broken
\cite{Kibble:1976sj}.  After formation, the strong monopole--antimonopole
attraction drives annihilation, and the network approaches a scaling regime.
Numerical simulations show that the number of defects per Hubble volume becomes
$\O(1)$ \cite{Bennett:1990xy,Yamaguchi:2001rf,Martins:2008ks}.  We parametrize the total
monopole plus antimonopole number density as
\begin{align}
    n_{\rm def}
    \equiv
    n_M+n_{\bar M}
    =
    \xi_M H^3,
    \qquad
    \xi_M=\O(1),
\end{align}
with  $H$ being the Hubble parameter. 
Thus each Hubble volume contains an order-one, but statistically fluctuating,
number of defects.

Since the energy of each global monopole is cut off at the horizon scale in the
scaling regime, the energy per defect is estimated as
\begin{align}
    E_M(H^{-1})
    \sim
    4\pi v^2 H^{-1}.
\end{align}
The monopole-network energy density is therefore
\begin{align}
    \rho_m
    \sim
    n_{\rm def} E_M(H^{-1})
    \sim
    4\pi \xi_M v^2 H^2 .
\end{align}
Using
$
    \rho_{\rm bg}
    =
    3M_{\rm pl}^2H^2 ,
$
we obtain
\begin{align}
    \frac{\rho_m}{\rho_{\rm bg}}
    \sim
    \frac{4\pi}{3}\xi_M
    \frac{v^2}{M_{\rm pl}^2}.
    \label{eq:monopole-energy-fraction}
\end{align}
Thus, in the scaling regime, the monopole-network energy density follows the
background evolution until the scaling regime ends.

The end of scaling may be due to explicit breaking of the global $O(3)$, the gauge coupling by gauging $\SU(2)\subset O(3) \OR$ an initial population bias.  

We will consider the parameter region $v$ around the Planck scale while the Higgs mass is much smaller. 
Such a weakly interacting slim particle (WISP) can appear naturally in UV
models; see Refs.\,\cite{Jaeckel:2010ni,Ringwald:2012hr,Arias:2012az,Graham:2015ouw,Marsh:2015xka,Irastorza:2018dyq,DiLuzio:2020wdo,Albertus:2026fbe,Arza:2026rsl}
for reviews on WISPs.  For instance, one may consider a supersymmetric 
model in which the scalar field couples to the supersymmetry-breaking field $Z$ through
the Kahler potential
\beq
{\cal K}\supset
c_1\frac{|Z|^2}{M_{\rm pl}^2} |\f_a|^2
-
c_2\frac{|Z|^2}{M_{\rm pl}^4} |\f_a|^4 ,
\eeq
with no superpotential. Alternatively, $\phi_a$ may be an axion-like particle
arising from the compactification of the fifth component of an $\SU(2)$ gauge
field with the compactification radius around the Planck scale. In both cases the nontrivial local minima is naturally around the Planck scale, which spontaneously breaks the $\SU(2)\subset O(3)$.\footnote{The feebly interacting scenario can also realize a time-dependent expectation
value that naturally decreases from the Planck scale~\cite{Yin:2024txg}.  In this
scenario, the scalar potential contains all renormalizable terms with $\O(1)$ parameters, while the kinetic
term of $\f_a$ is very large, as in a weakly coupled gauge theory. The resulting monopole has time dependent $v$. Although the analysis does not directly apply, the PBH formation and GWs should qualitatively hold.} 

\section{PBH formation from the global monopole network}

\lac{2}
%\subsection{PBH formation from the global monopole network}
\subsection{Distribution of Hubble-patch overdensities}

We characterize each Hubble patch by the total number of defects,
\begin{align}
    N_{{\rm tot},H}
    =
    N_M+N_{\bar M},
    \label{eq:hubble-total-defect-number}
\end{align}
where $N_M$ and $N_{\bar M}$ denote the numbers of monopoles and antimonopoles
inside the patch.  This variable should not be interpreted simply as a measure
of static gradient energy.  In particular, a nearby monopole--antimonopole pair
has a partially cancelled far-field configuration and therefore carries less
large-scale gradient energy than two well-separated isolated defects.  Instead,
$N_{{\rm tot},H}$ is used here as an empirical proxy for local defect activity,
including close pairs, nonlinear relaxation, and annihilation events (On the other hand by using the winding number one can derive a lower limit on the overdensities of the Hubble patch, see Appendix \ref{app:2}).

We define the normalized Hubble-patch energy density by
\begin{align}
    R_H
    =
    \frac{\rho_{{\rm tot},H}}
         {\langle \rho_{{\rm tot},H}\rangle}.
    \label{eq:normalized-hubble-patch-energy}
\end{align}
The full distribution is decomposed as a mixture over total-defect-number
sectors,
\begin{align}
    P(R_H)
    =
    \sum_{N=0}^{\infty}
    P_N\,
    P(R_H\mid N_{{\rm tot},H}=N),
    \label{eq:RH-Ntot-mixture}
\end{align}
where $P_N\equiv P(N_{{\rm tot},H}=N)$ is the sector weight.

\begin{figure}[t]
    \centering
    \includegraphics[width=\linewidth]{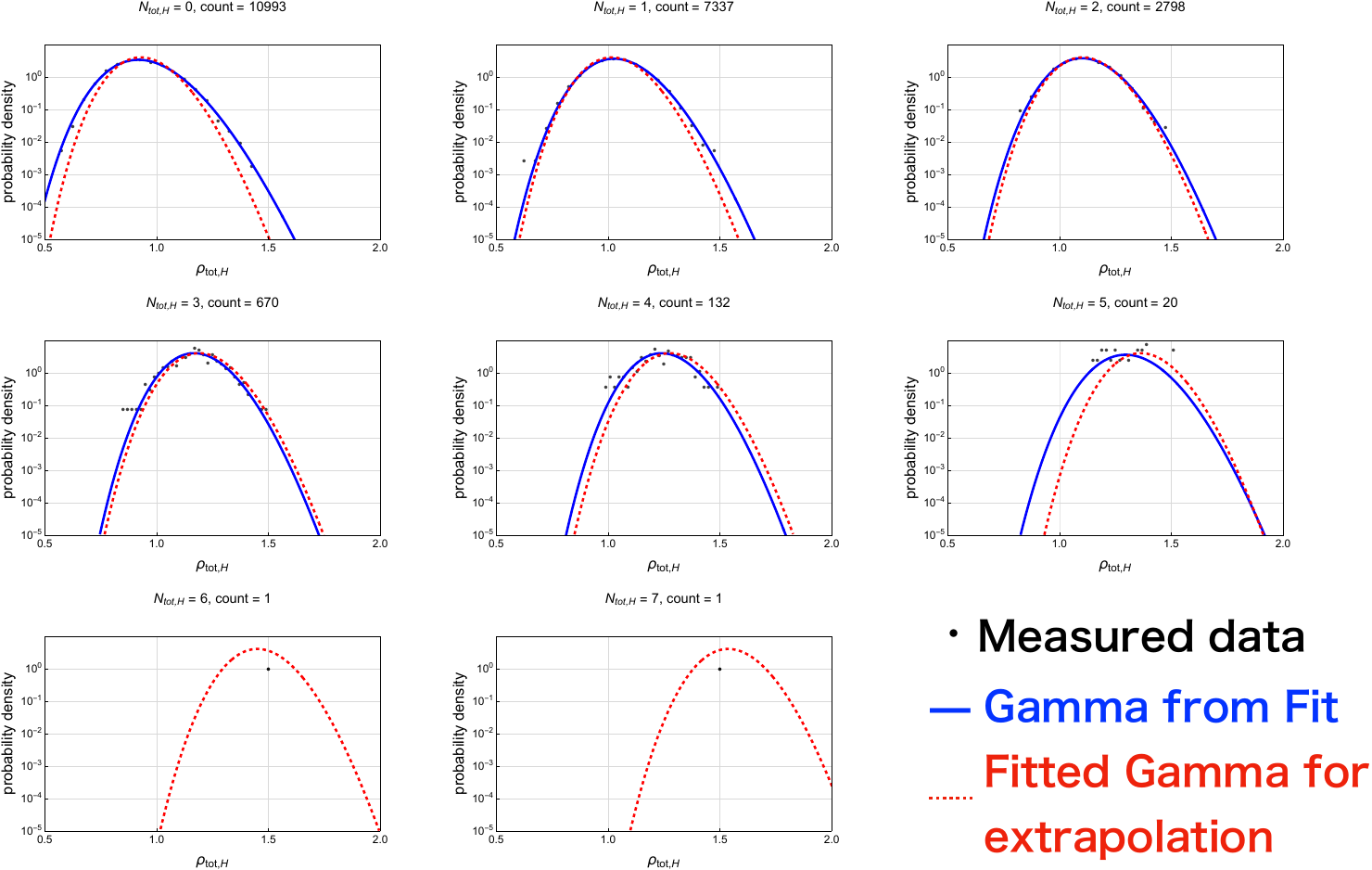}
    \caption{
    Conditional probability density of $\rho_{{\rm tot},H}$ for fixed
    $N_{{\rm tot},H}$.  The black points show the measured histograms, the blue
    solid curves show the Gamma distributions determined by the measured mean
    and variance in each sector, and the red dotted curves show the Gamma
    distributions obtained from the fitted functions $\mu_N$ and $\sigma_N$ used
    for extrapolation.  The sample size of each sector is shown in the panel
    title. Here we showed the result with the radiation-dominated case corresponding to the right panels of Figs. \ref{fig:1} and \ref{fig:2}.
    }
    \label{fig:conditional_gamma}
\end{figure}

For each fixed-$N_{{\rm tot},H}$ sector, we fit the conditional distribution
$P(R_H\mid N_{{\rm tot},H}=N)$ using the lattice data.  Since $R_H$ is positive
and is built from local positive energy contributions, we use a Gamma
distribution as a simple phenomenological model.  Its mean and width,
$\mu_N$ and $\sigma_N$, are calibrated sector by sector. 
The resulting Gamma distributions, together with the measured data points (black points) from lattice simulation in different $N_{\rm tot, H}$, are shown by blue solid lines in Fig.\ref{fig:conditional_gamma}

In the statistically reliable range, both are well described by smooth
functions of $N$, and we keep the leading terms in their expansion,
\begin{align}
    \mu_N
    =
    \mu_0+\mu_1 N,  ~~~ \sigma_N
    =
    \sigma_0
     +\sigma_1 N .
    \label{eq:mean-sigma-Ntot-fit}
\end{align}
This fit allows us to extrapolate the conditional energy distribution to rare
large-$N_{{\rm tot},H}$ sectors.  The measured values of $\mu_N$ increase
monotonically with $N_{{\rm tot},H}$ and are reasonably well described by a
linear approximation.  This trend is consistent with the expectation from the
lower bound on the gradient energy in sectors with finite winding number, as
discussed in Appendix~\ref{app:2}.
However, $\sigma_N$ shows larger scatter, although the variation remains at the
level of $\O(10)\%$.  In some fitting ranges, the fitted slope of $\sigma_N$
changes sign.  In particular, when the fitted slope is negative, a linear
extrapolation would eventually give $\sigma_N<0$ at sufficiently large
$N_{{\rm tot},H}$, which is clearly an artifact of the extrapolation.  Therefore,
for the extrapolation to rare large-$N_{{\rm tot},H}$ sectors, we conservatively
take $\sigma_N$ to be the smallest value measured in the actual data.  
The resulting Gamma distribution is shown by red dashed lines in Fig.\ref{fig:conditional_gamma}.

Once the mean value is fixed, the Poisson distribution reproduces the overall
shape of the measured $P_N$ reasonably well on the logarithmic scale relevant for
the rare large-$N$ tail.  We nevertheless treat this Poisson fit as a simple one-parameter extrapolation model, since the bin-by-bin residuals indicate small
but statistically significant deviations from an ideal Poisson distribution. The extrapolated Hubble-patch energy distribution is then written as
\begin{align}
    P_{\rm ext}(R_H)
    =
    \sum_{N=0}^{N_{\rm max}}
    P_N^{\rm fit}\,
    P_{\rm fit}(R_H\mid N_{{\rm tot},H}=N),
    \label{eq:RH-extrapolated-mixture}
\end{align}
where $P_{\rm fit}(R_H\mid N_{{\rm tot},H}=N)$ denotes the fitted Gamma
distribution in the fixed-$N$ sector.  The PBH formation probability is
therefore
\begin{align}
    P_{\rm form}
    =
    \int_{R_c}^{\infty} dR_H\, P_{\rm ext}(R_H)
    =
    \sum_{N=0}^{N_{\rm max}}
    P_N^{\rm fit}
    \int_{R_c}^{\infty} dR_H\,
    P_{\rm fit}(R_H\mid N_{{\rm tot},H}=N).
    \label{eq:Pform-Ntot-mixture}
\end{align}

\begin{figure}[t]
    \centering
    \includegraphics[width=0.49\textwidth]{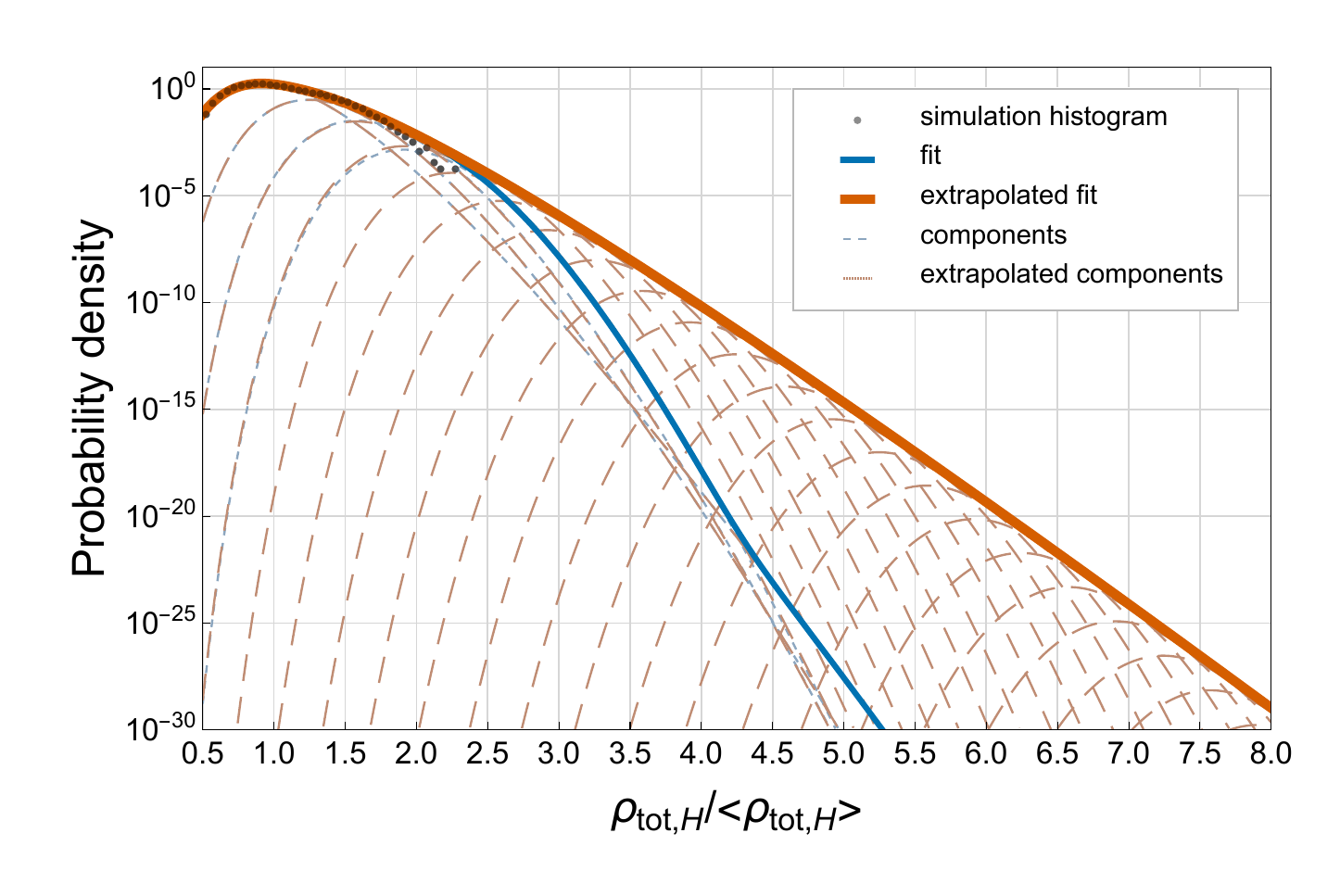}
        \includegraphics[width=0.49\textwidth]{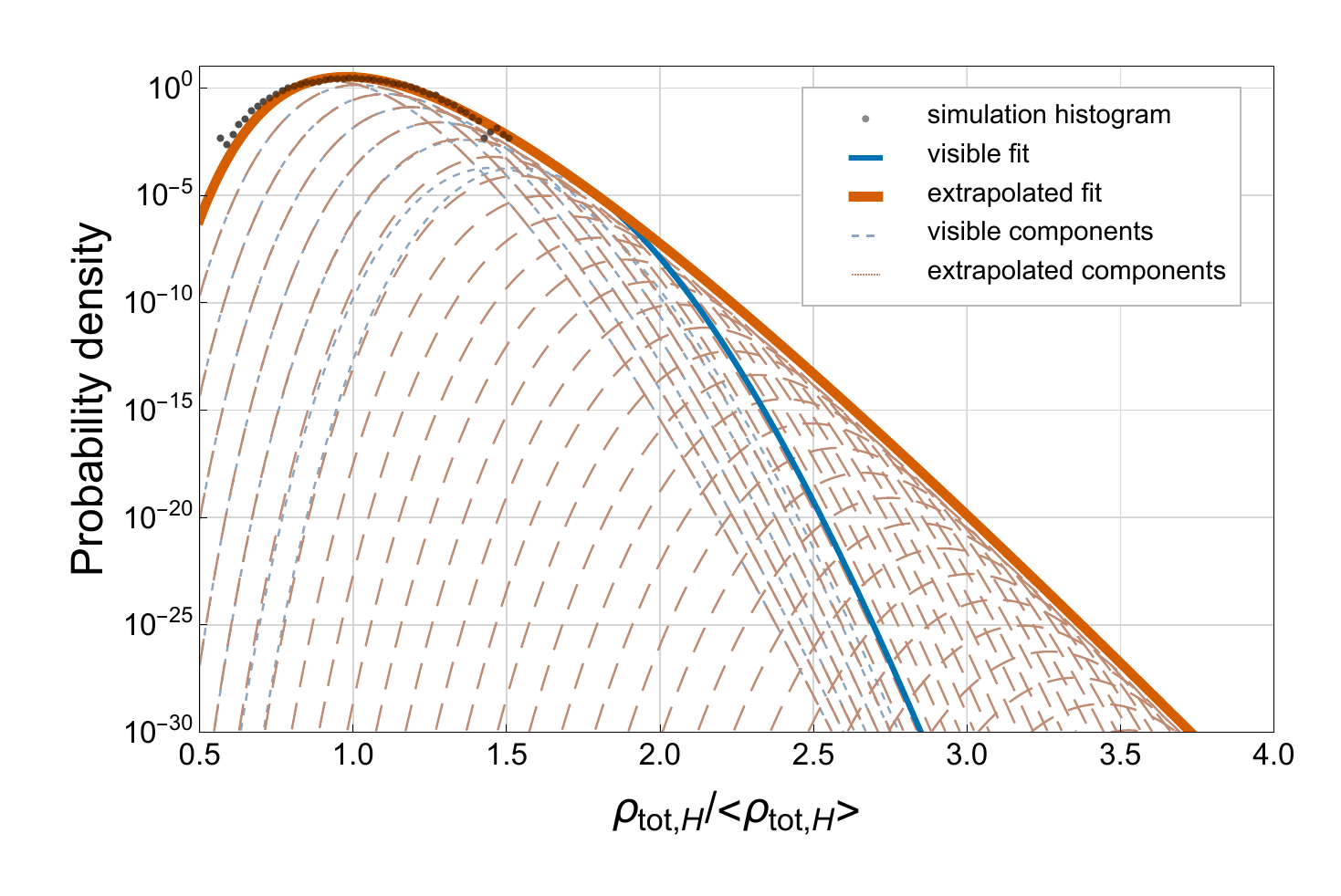}
    \caption{
Hubble-patch energy distribution in the matter-dominated (left) and radiation-dominated (right) at $\tau =10/m_0$.
The black points show the simulation histogram of
$R_H=\rho_{{\rm tot},H}/\langle\rho_{{\rm tot},H}\rangle$.
The blue curve shows the contribution from the Gamma distribution fit obtained using  the observed
$N_{{\rm tot},H}$ sectors, which could be considered as the conservative estimate of the overdensity, while the red curve includes the extrapolated
high-$N_{{\rm tot},H}$ sectors (see the main text). Dashed curves show the individual
fixed-$N_{{\rm tot},H}$ components entering the mixture.
    }
    \label{fig:1}
\end{figure}

\begin{figure}[t]
    \centering
    \includegraphics[width=0.49\textwidth]{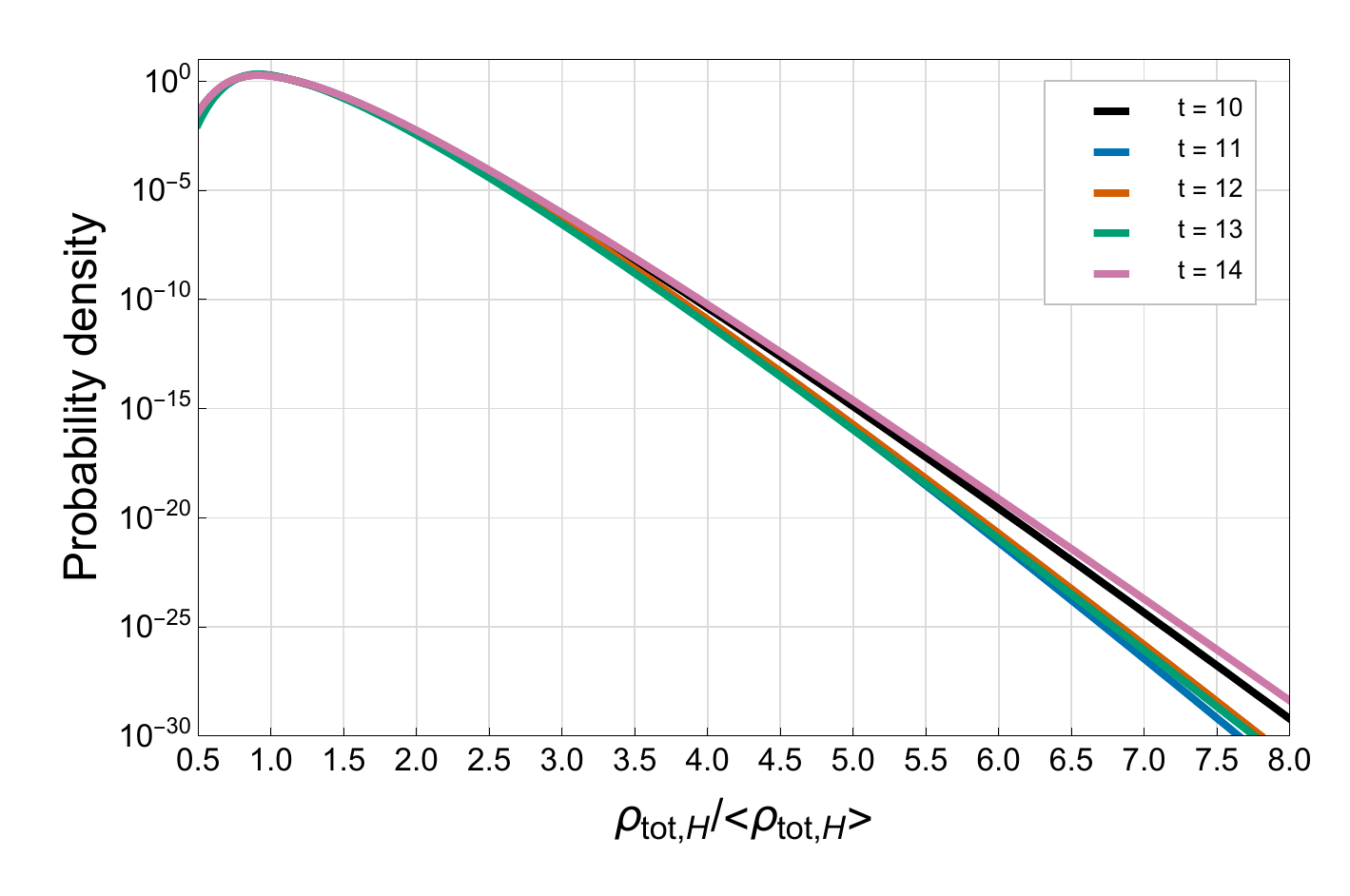}
    \includegraphics[width=0.49\textwidth]{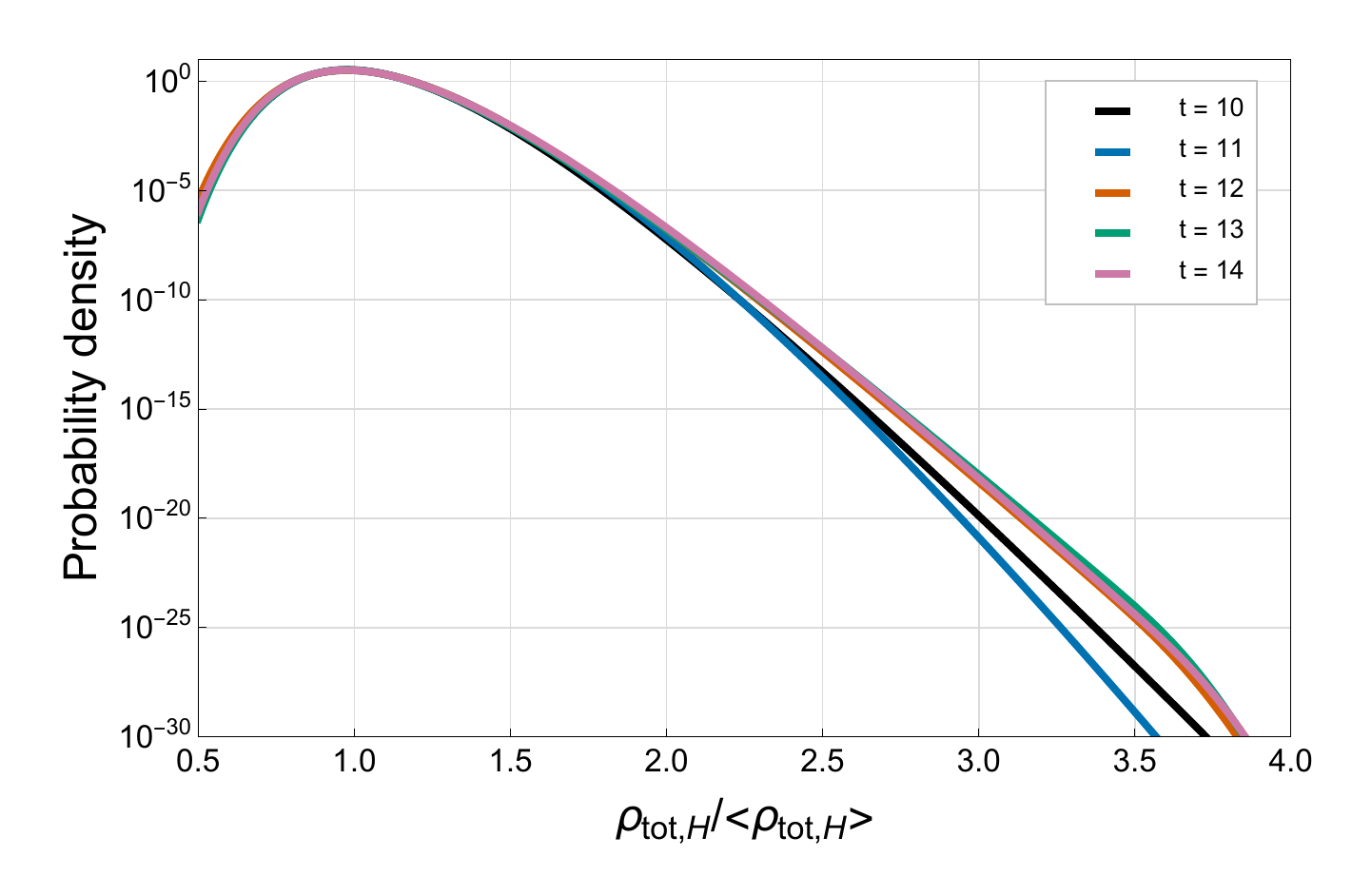}
    \caption{
    Time dependence of the extrapolated Hubble-patch energy distribution for matter- and radiation-dominated case.
    The curves show the fitted and extrapolated mixture distribution for
    $R_H=\rho_{{\rm tot},H}/\langle\rho_{{\rm tot},H}\rangle$ at different
    times.  
    }
    \label{fig:2}
\end{figure}

The numerical results shown in Figs.~\ref{fig:1} and \ref{fig:2} are obtained
using a modified version of {\tt CosmoLattice}~\cite{Figueroa:2020rrl,Figueroa:2021yhd} with Gaussian initial
fluctuations.  We use a fixed expanding background and a lattice with
$N^3=(512)^3$, starting from the conformal time $\tau_0=1/m_0$ and $\sqrt{\lambda} v= H_0=0.5 m_0$, 
with the Gaussian fluctuation. 
The Hubble-patch diagnostics are
evaluated at several times and are used to calibrate the
$N_{{\rm tot},H}$-based mixture model described above. 

In each Hubble patch, we measure the total monopole number 
$N_{{\rm tot},H}$ using the winding-number estimator based on clustering and
gradient flow, together with the total energy density in the same patch. We
then determine the conditional distribution of
$R_H=\rho_{{\rm tot},H}/\langle\rho_{{\rm tot},H}\rangle$ in each
fixed-$N_{{\rm tot},H}$ sector and fit the parameters introduced above.
Details of the lattice implementation and the Hubble-patch measurements are
given in Appendix~\ref{app:lattice}.

In the left panels of Figs. \ref{fig:1} and \ref{fig:2}, we take a matter-dominated (MD) background.  This choice is
motivated by the fact that, around the formation epoch, the energy density of
the radial mode behaves as a matter-like component anyway unless it efficiently decays
or is drained into other fields. 

To assess the latter sensitivity, in the right panel 
we also consider a radiation-dominated (RD) setup in which the effective quartic coupling
is changed after an expansion of factor $3$ as $\lambda_{\rm eff}\propto a^{-2}$, so that the
radial mass redshifts and the scalar sector behaves more closely to radiation.

We find that Hubble patches with $\O(1)$ density fluctuations in the scalar sector can appear with
probabilities larger than $10^{-30}$. In fact achieving $\O(1)$ density fluctuations with the quadratic hilltop potential without those topological defects are difficult due to the Hubble expansion (see Appendix\,\ref{app:delayed-roll-fails}). 
The
approximate overlap of the fitted curves at different times in Fig.~\ref{fig:2}
indicates that the high-energy tail is stable over the scaling regime used in the analysis at least via an order-of-estimate on the prediction of $v$ for the PBH formation as we will discuss soon.

\subsection{PBH formation from monopole-induced overdensities}
\label{subsec:PBH-formation}

We now discuss the formation of PBHs from the Hubble-patch overdensities
estimated above. We define 
\begin{align}
    f_\phi
    \equiv
    \frac{\langle \rho_{\rm tot,H}\rangle}{\rho_{\rm bg}}
    % \sim  \frac{4\pi}{3}\xi_M
    % \frac{v^2}{M_{\rm pl}^2}.
    \label{eq:scalar-sector-fraction}
\end{align}
In the last estimate, we assume that the total energy density in the scalar sector, $\rho_{\rm tot,H}$, is dominated by the monopole energy density. 
From the numerical simulation in the interval $8/m_0<\tau<15/m_0$, we fit
\beq \laq{fit}
f_\phi= \(9.2 \pm 0.1\) \frac{v^2}{M^2_{\rm pl}} [{\rm MD}],~~~~~~f_\phi= \(16.9 \pm 0.5\)\frac{v^2}{M^2_{\rm pl}} [{\rm RD}].
\eeq 
This value is consistent with the previous studies for the monopole scaling $f_\f \sim  \frac{4\pi}{3}\xi_M
     \frac{v^2}{M_{\rm pl}^2}$ with $\x_H=\O(1)$~\cite{Bennett:1990xy,Yamaguchi:2001rf}, by assuming the monopole dominating the scalar sector energy density.
Thus the actual overdensity relative to the total background is
\begin{align}
    \delta_H
    \simeq
    f_\phi (R_H-1).
    \label{eq:actual-overdensity-from-RH}
\end{align}
The PBH formation probability at the epoch with Hubble scale $H$ is therefore
\begin{align}
    P_{\rm form}(H)
    =
    \int_{\delta_c/f_\phi(H)+1}^\infty dR_H\,
    P_{\rm ext}(R_H;H),
    \label{eq:Pform-with-fphi}
\end{align}
where $\delta_c$ is the collapse threshold.
For the radiation-dominated case we take $\delta_c=0.4$ as a benchmark~\cite{Harada:2013epa,Escriva:2019phb}.
For the matter-dominated case, we use the same value as a prompt-collapse criterion.\footnote{
In the present case, the overdensities are generated by an active
scaling monopole network rather than by passive dust perturbations, which would grow to enhance the PBH formation~\cite{Khlopov:1980mg,Harada:2016mhb,Harada:2017fjm}.  Monopole
motion, annihilation, and the redistribution of energy into scalar gradients and
radiation can erase or rearrange subhorizon overdensities.  We therefore count
only prompt collapse of Hubble patches with order-one overdensity and leave a
dedicated treatment of possible delayed collapse for future work.}
From Fig.\,\ref{fig:1} one can see by approximating $\int_{\delta_c/f_\phi(H)+1}^\infty dR_H\,
    P_{\rm ext}(R_H;H)\sim  R_H P_{\rm ext}(R_H;H)
   |_{R_H=\delta_c/f_\phi(H)+1}$, and using $f_\f=1$, that 
   $P_{\rm form}\lesssim 0.5 \AND 0.09$ for matter and radiation-dominated Universe, respectively. 
Using the fitted data~\Eq{fit} and by taking $R_H<7.5, \AND 3.5$ respectively (see Fig.\ref{fig:1}), we get the condition for PBH formation with non-negligible amount ($P_{\rm form}>10^{-30}$)
$
v>0.08 M_{\rm pl} [{\rm MD}],\AND 0.1 M_{\rm pl} [{\rm RD}].
$\footnote{If the mass is time varying, which is naturally realized with the feebly interacting scenario~\cite{Yin:2024txg,Yin:2024pri}, this limit does not apply because the core of the monopole is no longer cosmological constant. }
 As a consequence, we have the successful parameter region for the PBH formation:
 \beq 
 \boxed{v\gtrsim 0.1 M_{\rm pl}  }
 \eeq 
 For  $v\gg 0.2 M_{\rm pl}$, the PBHs can be overproduced, which can set limits on the monopole formation as a function of formation era. This constraint, however, can be evaded by considering an efficient late-time entropy dilution. For our numerical estimation, assuming a fixed background, we need $f_\f<1$ at the formation. Then Eq.~\eq{fit} gives $v/M_{\rm pl}\lesssim$  0.33(MD) and 0.24(RD). 
 $v\lesssim 1.65 M_{\rm pl}$ may also be needed to avoid topological inflation~\cite{Sakai:1995nh}. 
Violating those limits, we need a simulation with general relativity to take into account of the backreaction or topological inflation, which is beyond our scope. However, even in those case, we expect efficient PBH formation because the overdensities are more significant. 

The initial PBH energy fraction is then estimated as
\begin{align}
    \beta_{\rm PBH}(H)
    \simeq
    \gamma_{\rm col} P_{\rm form}(H),
    \label{eq:beta-PBH-with-fphi}
\end{align}
where $\gamma_{\rm col}=\O(0.1-1)$ parametrizes the fraction of the horizon
mass ending up in the PBH.  The PBH mass is
\begin{align}
    M_{\rm PBH}(H)
    =
    4\pi\gamma_{\rm col}
    \frac{M_{\rm pl}^2}{H}.
    \label{eq:PBH-mass-H}
\end{align}
If PBHs are formed over an interval $H_{\rm i}>H>H_{\rm f}$, their masses span
$
    M_{\rm min}
    =
    4\pi\gamma_{\rm col}\frac{M_{\rm pl}^2}{H_{\rm i}},
%    \qquad
    M_{\rm max}
    =
    4\pi\gamma_{\rm col}\frac{M_{\rm pl}^2}{H_{\rm f}} .
$
The end of the formation may be because the end of the monopole scaling or the suppression of the overdensities. The latter can be simply achieved by changing the background equation of state (see Fig.\ref{fig:1}).

Strictly speaking, PBHs formed from patches with nonzero winding can initially
carry scalar hair.  During the subsequent scaling evolution, they can also
interact and merge with monopoles or antimonopoles.  However, neutralization occurs within an $\O(1)$ Hubble time in the scaling regime. Notice that the PBH remains by eating other antimonopoles rather than annihilating~\cite{Stojkovic:2004hz}. 
Therefore, the mass of the neutralized PBH is expected to remain of the same
order as the horizon-mass estimate above. On the other hand, the PBH with the hair can also remain if it is screened by the pseudo-NGB mass or gauge boson mass, which also means that the scaling is terminated.

Although other relics, such as monopoles and Higgs particles, may in
principle affect late-time cosmology, we assume scenarios in which they do not
cause any cosmological problem.  In particular, after the epochs relevant for PBH production, the universe is assumed to follow the standard cosmological history.
Several simple and concrete possibilities realizing this assumption are discussed
in Appendix~\ref{app:latercosmo}.

Under these assumptions, we consider two natural cases for PBH formation: PBH
production during a matter-dominated reheating phase and PBH production during a
radiation-dominated phase.

\paragraph{PBH mass spectrum in the matter-dominated era} 
Now let us discuss the PBH spectrum in the present Universe.
First, consider the case in which PBHs form during an early matter-dominated era
and the Universe reheats at temperature $T_{\rm RH}$.  Since both PBHs and the dominant matter component redshift as
$a^{-3}$ before reheating, the PBH fraction is unchanged until reheating.
After reheating, entropy conservation gives
\begin{align}
    \frac{d\Omega_{\rm PBH,0}^{\rm MD}}{d\ln M_{\rm PBH}} h^2
    =
    \frac{3}{4}
    \frac{s_0 T_{\rm RH}}{\rho_{c,0}/h^2}
    \beta_{\rm PBH}
    \approx
    0.02
    \left(
        \frac{T_{\rm RH}}{1\,{\rm GeV}}
    \right)
    \left(
        \frac{\beta_{\rm PBH}}{10^{-10}}
    \right).
    \label{eq:PBH-abundance-MD-num}
\end{align}
From the previous subsection, 
\begin{align}
    \beta_{\rm PBH}
    \sim
    \gamma_{\rm col} \frac{4\pi}{3} \xi_H
    \frac{v^2}{M_{\rm pl}^2}
    P_{\rm ext}
    \left(
        \frac{\delta_c}
             {4\pi \xi_H v^2/\left(3M_{\rm pl}^2\right)}
        +1,\,
        H
    \right),
\end{align}
where we approximated
\begin{align}
    P_{\rm form}
    \sim
    P_{\rm ext}
    \left(
        \frac{\delta_c}
             {4\pi \xi_H v^2/\left(3M_{\rm pl}^2\right)}
        +1,\,
        H
    \right),
    \qquad
    f_\phi
    \sim
    \frac{4\pi}{3} \xi_H\frac{v^2}{M_{\rm pl}^2},
\end{align}
for the parametric discussion, while using the previous numerical results gives more precise conclusions.
Given $\xi_H=\O(1)$, the required abundance can be obtained for
$v=\O(0.1)M_{\rm pl}$, as suggested by Fig.~\ref{fig:1}.  If
$P_{\rm form}(H)$ is approximately flat during the scaling regime, as indicated
by Fig.~\ref{fig:2}, the resulting PBH spectrum is also approximately flat. 

The spectra are shown in the lower panel of Fig.\ref{fig:MD} together with the various reaches and limits of the PBH searches.  
Here, we consider that the end of reheating matches with the end of scaling for simplicity for having this figure. Such a possibility can be obtained for the scenario in Appendix.\,\ref{app:tran}.
Although these limits and reaches are derived for monochromatic PBH mass functions, we have verified that our extended mass function also satisfies the corresponding constraints using the criterion of Ref.~\cite{Carr:2017jsz}.
Depending on the duration of the scaling regime, this scenario can also account for the hints reported by Subaru Hyper Suprime-Cam~\cite{Sugiyama:2026kpv}, the Optical Gravitational Lensing Experiment (OGLE)~\cite{Niikura:2019kqi}, and Ca-rich gap transients~\cite{Smirnov:2022zip}. These hints may even be explained simultaneously, if other constraints are ignored.

\paragraph{PBH mass spectrum in the radiation-dominated era}
If the Higgs decays with Bose enhancement or parametric resonance, the energy can soon be transferred into radiation. Then we can have the formation and scaling in the radiation-dominated Universe. 

Let the entropy density at the 
formation 
$
    s_*
    =
    \frac{2\pi^2}{45}g_{*s}T^3,
    T
    =
    \left(
        \frac{90}{\pi^2 g_*}
    \right)^{1/4}
    \sqrt{H M_{\rm pl}} .
$
Here $g_{*s},\AND g_*$ are the relativistic degrees of freedom for entropy density and energy density, respectively.
The present PBH abundance per logarithmic mass interval is then
\begin{align}
    \frac{d\Omega^{\rm RD}_{\rm PBH,0}}{d\ln M_{\rm PBH}} h^2
    &=
    \frac{s_0}{\rho_{c,0}/h^2}\,
    \frac{3M_{\rm pl}^2H^2}{s_*}\,
    \beta_{\rm PBH}(4\pi\gamma_{\rm col}M_{\rm pl}^2/M_{\rm PBH}).\\
   &\approx
    0.009
    \left(
        \frac{\gamma_{\rm col}}{0.2}
    \right)^{1/2}
    \left(
        \frac{106.75}{g_{*s}}
    \right)^{1/4}
    \left(
        \frac{10^{18}{\rm g}}{M_{\rm PBH}}
    \right)^{1/2}
    \frac{\beta_{\rm PBH}}{10^{-17}}
    \label{eq:PBH-abundance-RD-num}
\end{align}
Thus, in the radiation-dominated case, lighter PBHs are enhanced more strongly
in the present abundance, because PBHs redshift as matter whereas the background
redshifts as radiation. Again, it is possible to have the correct abundance.

The spectra are shown in the lower panel of Fig.~\ref{fig:RD}.
It may also explain the GW signal data from LVK collaboration~\cite{Andres-Carcasona:2024wqk}.

\section{Gravitational waves}
\lac{3}
For PBH formation we require $v$ to be close to the Planck scale,
typically $v=\O(0.1)M_{\rm pl}$.
Since the stress tensor during monopole formation and the subsequent scaling
evolution is strongly inhomogeneous and anisotropic, and since the monopole
energy density is large enough for PBH formation, a sizable stochastic
GW background can be generated.

A useful estimate follows from dimensional analysis.  Assuming that the Higgs
field is integrated out, the GWs are sourced by the NGB sector.
The equation of motion of the NGBs is controlled by the gradient time scale and
the Hubble time scale, which are of the same order in the scaling solution.
Therefore, after rescaling time by the Hubble scale and normalizing the NGB
fields by $v$, the equations of motion become independent of the overall
energy scales.  In these dimensionless variables, the anisotropic stress generated by the NGB
dynamics is an order-one quantity.  Restoring the dimensions, each derivative
brings a factor of $H$ and each NGB field brings a factor of $v$.  Therefore the
physical anisotropic stress scales as
$
    \Pi \sim v^2 H^2 .
$
Since the GWs are produced over one Hubble time, the tensor
amplitude is estimated as
$
    h \sim \frac{\Pi}{M_{\rm pl}^2 H^2}
      \sim \frac{v^2}{M_{\rm pl}^2}.
$
Thus the GW energy density scales as
$
    \rho_{\rm GW}
    \sim
    M_{\rm pl}^2 H^2 h^2
    \sim
    \frac{v^4 H^2}{M_{\rm pl}^2}.
$
Dividing by the total energy density,
$\rho_{\rm tot}=3M_{\rm pl}^2H^2$, one arrives at the formula,
\begin{align}
  \Omega_{\rm GW}[H]
  \equiv
  \frac{H^{-1}d\rho_{\rm GW}/dt}{\rho_{\rm tot}}
  \simeq
  C_{\rm GW}
  \left(
    \frac{v}{M_{\rm pl}}
  \right)^4 ,
  \label{eq:OmegaGW-peak-estimate}
\end{align}
where $C_{\rm GW}$ is a dimensionless efficiency factor determined by the
nonlinear dynamics of the transition.  From the numerical simulation, we find
$C_{\rm GW}=\O(1)$.  This argument holds as long as the scaling regime is maintained.  The GWs are mostly produced at frequencies of order $H$ at the
time of production.  After production, the GWs are redshifted as
radiation.

This order-estimate argument also applies to other scaling topological defects,
such as global cosmic string networks, when their dynamics is controlled by NGB
gradients. Our argument gives consistent result with the general result for the scaling sources in Refs.\,~\cite{Figueroa:2012kw,Figueroa:2020rrl, Fenu:2009qf, Figueroa:2020lvo}. 
In the following, we discuss the GWs produced during radiation-dominated and matter-dominated epochs, which is in the opposite order from
before, for clarity.

\paragraph{Radiation-dominated Universe}
If the monopole-producing transition takes place during radiation
domination, the frequency today is estimated as
\begin{align}
  f_0^{\rm RD}
  \simeq
  1.6\times 10^{-4}{\rm Hz}\,
  \left(
    \frac{T_*}{1{\rm TeV}}
  \right)
  \left(
    \frac{g_{*,s}}{106.75}
  \right)^{1/6},
  \label{eq:GW-peak-frequency-RD}
\end{align}
where $T_*$ is the temperature at the production for the corresponding frequency. Note that $T_*$ is a free variable during the scaling of the monopole network. 
The present GW
abundance is estimated as
\begin{align}
  \Omega_{\rm GW,0}^{\rm RD} h^2
  \simeq
  2\times 10^{-5}
  \left(
    \frac{100}{g_*}
  \right)^{1/3}
  C_{\rm GW}
  \left(
    \frac{v}{M_{\rm pl}}
  \right)^4 .
  \label{eq:GW-present-estimate-RD}
\end{align}
Thus, for $v$ close to the Planck scale, the monopole-producing transition can
lead to an observable stochastic GW background. This almost scale-invariant spectrum is kept in the range between the formation (corresponding to $f_{\rm form}=f_0^{\rm RD}[T_{\rm form}]$) and end ($f_{\rm end}=f_0^{\rm RD}[T_{\rm end}]$)) of the network, with $f_{\rm end}<f_{\rm form}.$ We find numerically that $(f_{\rm form}/f)^{4.2}$ and $(f/f_{\rm end})^{2.7}$ from the numerical data, the latter of which is consistent with causality.
The fitted GW spectra are shown in the top panel of Fig.\ref{fig:RD} together with various future reaches and current limits.

\paragraph{Matter-dominated Universe}
If the monopole-producing transition takes place during an early
matter-dominated era, the frequency today is estimated as
\begin{align}
  f_0^{\rm MD}
  \simeq
  1.6\times 10^{-4}{\rm Hz}\,
  \left(
    \frac{T_{\rm RH}}{1{\rm TeV}}
  \right)
  \left(
    \frac{g_{*,s,{\rm RH}}}{106.75}
  \right)^{1/6}
  \left(
    \frac{H_*}{H_{\rm RH}}
  \right)^{1/3},
  \label{eq:GW-peak-frequency-MD}
\end{align}
where $H_{\rm RH}$ and
$T_{\rm RH}$, respectively, are the Hubble scale and temperature at reheating.  The factor
$(H_*/H_{\rm RH})^{1/3}$ comes from the redshifting of the frequency during the
early matter-dominated era.  The GWs are also diluted relative
to the dominant matter component until reheating.  The present abundance is
\begin{align}
  \Omega_{\rm GW,0}^{\rm MD} h^2
  \simeq
  2\times 10^{-5}
  \left(
    \frac{100}{g_{*,{\rm RH}}}
  \right)^{1/3}
  C_{\rm GW}
  \left(
    \frac{v}{M_{\rm pl}}
  \right)^4
  \left(
    \frac{H_{\rm RH}}{H_*}
  \right)^{2/3}.
  \label{eq:GW-present-estimate-MD}
\end{align}
Here the factor $(H_{\rm RH}/H_*)^{2/3}$ comes from the redshifting of
GWs as radiation during the early matter-dominated era. 
Therefore, 
the GW
abundance is suppressed if reheating occurs much later than the end of the scaling network.
Given the scale invariance in the radiation-dominated case, one can derive
\beq\Omega_{\rm GW,0} \propto   (f_0^{\rm MD})^{-2},
\eeq  
for the GWs emitted during the scaling regime. This behavior is consistent with the numerical simulation. At much larger $f_0^{\rm MD}$, the spectrum would enter the formation regime, which is not shown in the figure.
Thus the peak spectrum corresponds to the Hubble parameter at the end of the scaling. Below we should have $\Omega_{\rm GW,0 }\propto (f_0^{\rm MD})^3$ from the causality. This is consistent with our fit $\propto k^{3.3}$. 

\begin{figure}[t]
  \centering
  \includegraphics[width=0.7\textwidth]{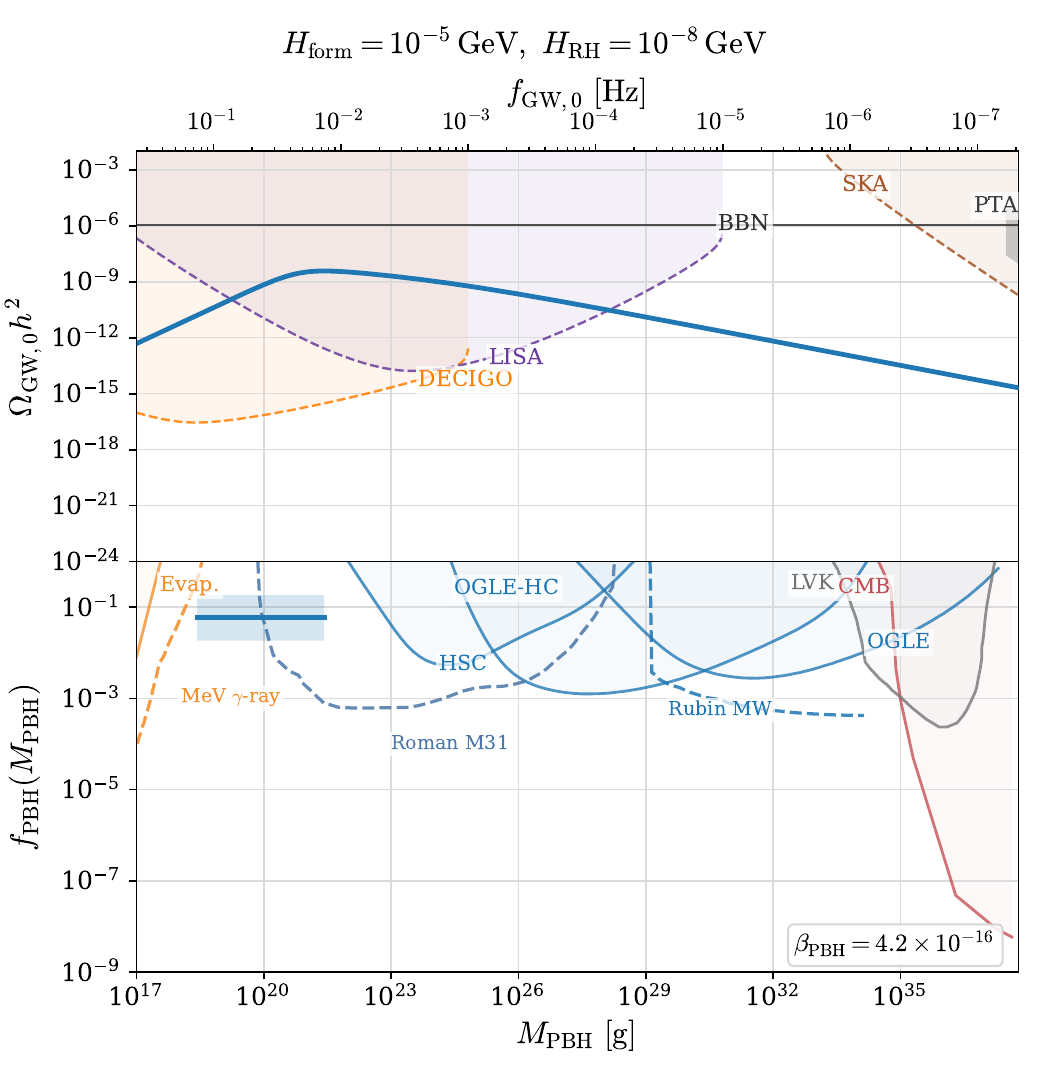}
  \caption{
  The upper panel shows the present-day GW spectrum $\Omega_{\rm GW,0}h^2$ for an early matter-dominated (MD) phase by taking $v/M_{\rm pl}=0.105$. The black line shows the BBN constraint~\cite{Pagano:2015hma}. The GW reaches (dashed lines) include SKA~\cite{Janssen:2014dka}, LISA~\cite{Robson:2018ifk}, DECIGO~\cite{Yagi:2011wg,Kuroyanagi:2014qza}. 
  The bottom panel shows the PBH dark matter fraction corresponding to the top panel cosmology. $\beta_{\rm PBH}$ is obtained by integrating the extrapolated tail of $P_{\rm ext}(R_H)$ from $1+\delta_c/f_\phi$. The blue curve is applied from the MD fitting result (see Appendix~\ref{app:fit}). 
  The blue band indicates a one-order-of magnitude fluctuation in $\beta_{\rm PBH}[H]$. 
  The shaded and labeled curves indicate Subaru-HSC\cite{Croon:2020ouk}, OGLE-hc~\cite{Mroz:2024wia}, QGLE~\cite{Mroz:2024mse}, CMB anisotropy~\cite{Serpico:2020ehh}, LIGO-Virgo-KAGRA (LVK) O3~\cite{Andres-Carcasona:2024wqk}, evaporation constraints from extragalactic photon background~\cite{Carr:2009jm}.
The dashed lines are projected sensitivities of MeV gamma-ray~\cite{Coogan:2020tuf}, Roman M31 and Rubin Milky-Way microlensing~\cite{LSSTDarkMatterGroup:2019mwo}. The data points are taken from Ref.~\cite{Schmitz:2020syl,Kavanagh_PBHbounds_2019}. 
  }
  \label{fig:MD}
\end{figure}

\begin{figure}[t]
  \centering
  \includegraphics[width=0.7\linewidth]{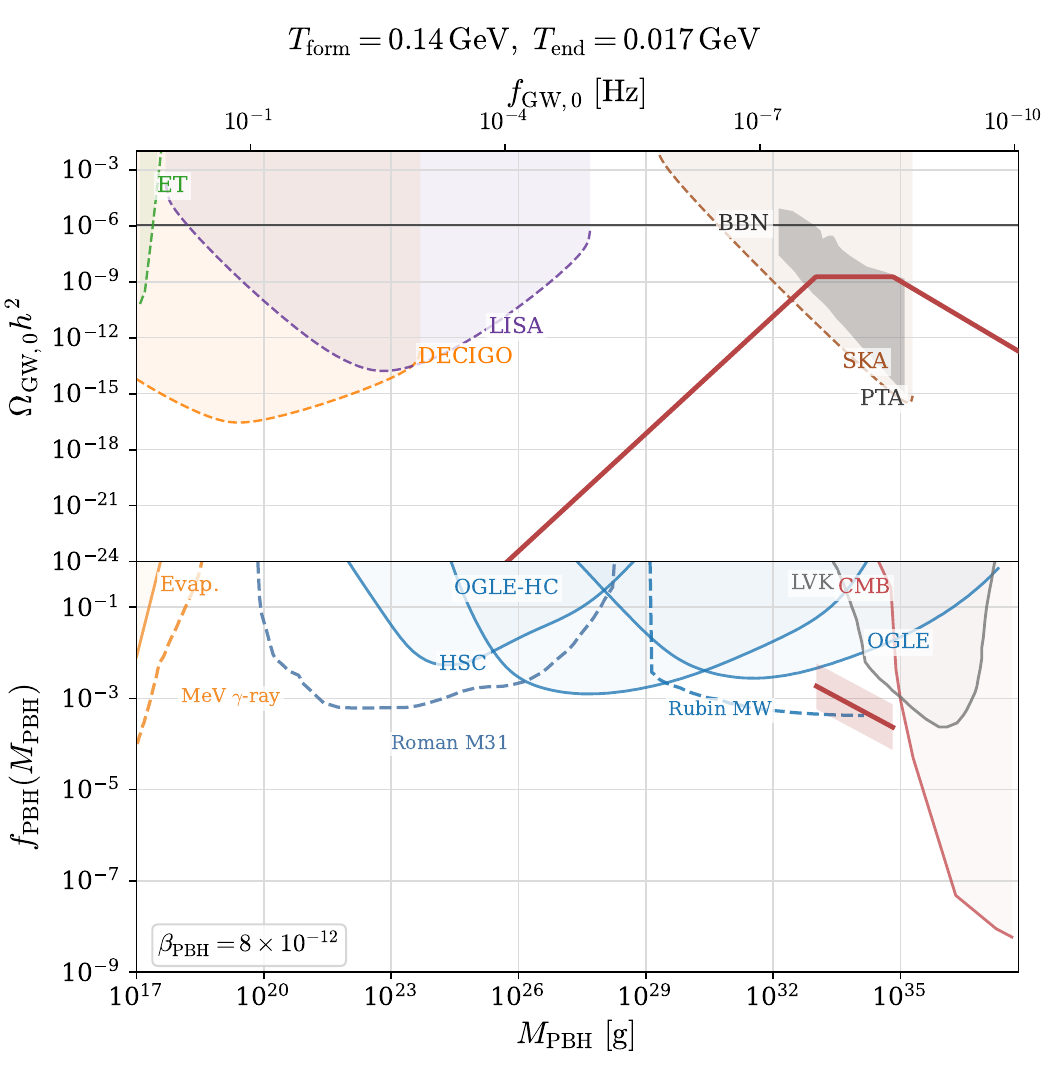}
  \caption{
  The same figure as Fig.\ref{fig:MD} but in the RD case with $v/M_{\rm pl}\approx0.14$. 
  For GW spectrum function, IR and UV dependence are taken from the fitting function and the duration of middle scaling region is controlled by $T_{\rm form}$ and $T_{\rm end}$. The gray shaded region is the signal region reported by the PTA~\cite{NANOGrav:2023gor,Antoniadis:2023ott,Reardon:2023gzh,Xu:2023wog}. The GW reach also includes ET~\cite{Punturo:2010zz,Hild:2010id,ET:2019dnz} in addition to the ones in Fig.~\ref{fig:MD}.
  }
  \label{fig:RD}
\end{figure}

The fitted GW spectrum are shown in the top panel of Fig.\ref{fig:MD}.

%%%%%%%%%%%%%%%%%%%%%%%%%%%%%%%%%%%%%%%%%%%%%

%%%%%%%%%%%%%%%%%%%%%%%%%%%%%%%%%%%%%%%%%%%%%
%%%%%%%%%%%%%%%%%%%%%%%%%%%%%%%%%%%%%%%%%%%%%

\section{Conclusions and discussion}
\lac{4}
In this paper, we have studied overdensities in scaling monopole networks.  This
regime is realized when the monopoles are global, or when they are local
monopoles with sufficiently small gauge coupling.  In this case, stochastic
Hubble-patch overdensities sourced by the monopole network can lead to primordial blackhole (PBH)
formation.  Even if the monopole network disappears due to the later cosmological
evolution, the PBHs formed during the scaling era can remain and constitute the
dominant component of dark matter.  This scenario is also linked to a stochastic
gravitational-wave (GW) background generated by the nonlinear monopole dynamics,
which can be probed by future GW detectors such as SKA~\cite{Janssen:2014dka,Weltman:2018zrl}, LISA~\cite{LISA:2017pwj}, DECIGO~\cite{Kawamura:2011zz}, ET~\cite{Punturo:2010zz,Sathyaprakash:2012jk}, and CE~\cite{Evans:2021gyd} (see also Refs.\,~\cite{NANOGrav:2023gor,Antoniadis:2023ott,Reardon:2023gzh,Xu:2023wog} for the recent signal in pulsar timing arrays). 
Although we focus on the particular region for the PBH formation, our mechanism works in wider mass parameter region. The future limits and constraints can be found in, e.g.,
Ref.\cite{Carr:2026hot}. 

An interesting possibility is that the monopoles originate from a hidden gauge symmetry and are therefore hidden magnetic monopoles.  In this case, PBHs formed from monopole-induced overdensities can inherit hidden magnetic charge.  Such PBHs may initially carry scalar hair before gauge screening becomes effective, and can subsequently interact with ambient monopoles and antimonopoles during the scaling regime.  These interactions can neutralize the scalar hair or magnetic charge through the absorption of oppositely charged defects, while the PBHs themselves remain.  However, as shown in \Eqs{PBH-abundance-MD-num} and \eq{PBH-abundance-RD-num}, the PBH component produced near the end of the scaling regime can already make a sizable contribution, especially in the matter-dominated case.  We find that an $\O(10)\%$ fraction of Hubble patches carries a nonzero winding number.  Therefore, PBHs formed from such patches are magnetically charged, giving an $\O(10)\%$ charged fraction of the PBH population.
The scenario, e.g., in Appendix \ref{app:tran}, makes the magnetically charged PBHs dominant dark matter component. 

This does not immediately imply a conflict with large-scale structure
formation.  Since the long-range hidden magnetic force is screened due to antimagnetically charged PBH on sufficiently
large scales, and the large-scale dark matter distribution can remain
effectively neutral.  Nevertheless, the hidden magnetic force can still be
important on smaller scales, especially for PBH binaries.

Let us estimate the size of this effect.  The charged gauge-boson mass is
$
    m_W \sim gv .
$
We focus on the epoch when the charged gauge-boson mass becomes comparable to
the Hubble scale,
$
    m_W \sim H .
$
This gives
$
    g \sim \frac{H}{v}.
$
The magnetic charge is then of order
$
    q_m \sim \frac{4\pi}{g}
    \sim
    4\pi \frac{v}{H}.
$
On the other hand, the PBH mass formed at this epoch is
$
    M_{\rm PBH}
    \sim
    4\pi\gamma_{\rm col}\frac{M_{\rm pl}^2}{H}.
$
Therefore the ratio between the hidden magnetic Coulomb force and the
gravitational force is estimated as
$
    \frac{F_{\rm mag}}{F_{\rm grav}}
    \sim
    \left(
        \frac{q_m M_{\rm pl}}{M_{\rm PBH}}
    \right)^2
    \sim
    \frac{1}{\gamma_{\rm col}^2}
    \left(
        \frac{v}{M_{\rm pl}}
    \right)^2 .
$
Thus, in the parameter region relevant for PBH formation,
$v=\O(0.1)M_{\rm pl}$, the hidden magnetic force can be comparable to gravity on unscreened scales. Note that regions with charge-to-mass ratios above the extremal bound cannot collapse into charged black holes carrying the full charge due to the repulsive force from the magnetic interaction. In other words, we cannot have over-extremal PBH.

Oppositely charged PBHs feel an enhanced attraction, while equally charged PBHs
feel a reduced effective attraction.  Hence, magnetically charged PBHs can have binary formation rates, merger times,
merger-rate distributions, and GW signals that differ from
those of ordinary neutral PBHs~\cite{Liebling:2016orx,Liu:2020vsy,Bozzola:2020mjx}.
In addition, if radiation into the hidden gauge sector is kinematically allowed, the hidden
charge can provide an extra channel for energy loss and thereby affect the inspiral and merger.  Future measurements of the PBH mass
spectrum, merger-rate distribution, and GW waveforms can
therefore further probe this scenario.  If the residual $\U(1)$ is subsequently broken in
such a way that the magnetic flux is confined into flux tubes, the charged PBHs
may instead become confined PBHs connected by hidden strings.  The dynamics of such confined PBHs warrant a separate study.
%%%%%%%%%%%%%%%%%%%%%%%%%%%%%%%%%%%%%%%%%%%%%%%%%%%%%%%

\appendix

%%%%%%%%%%%%%%%%%%%%%%%%%%%%%%%%%%%%%%%%%%%%%%%%%%
\section{Fitting functions for GW spectra}
\label{app:fit}

\paragraph{GW spectrum in the matter-dominated era}

The generated spectrum from the lattice simulation is fitted by the following function as 
\begin{align}
    \Omega_{\rm GW}^{\rm MD}(k) (v/M_{\rm pl})^{-2} f_\f^{-1}= \frac{A}{c_l(k/k_0)^{p_l}+c_m(k/k_0)^{p_m}+c_h(k/k_0)^{p_h}}, 
\end{align}
where the coefficients are $(A, k_0, c_l, c_m, p_l, p_m, p_h)^{\rm MD}=(0.040, 0.56m_0, 0.013, 0.14, -3.3, -2.4, 1.2)$ and $c_h=1-c_l-c_m$(See Ref.\,\cite{Kitajima:2023cek}). 

Since $f_\phi\propto v^2/M_{\rm pl}^2$, this machine normalization is equivalent to extracting the expected $(v/M_{\rm pl})^4$ scaling.

\paragraph{GW spectrum in the radiation-dominated era}

The generated spectrum is fitted as 
\begin{align}
    \Omega_{\rm GW}^{\rm RD}(k)(v/M_{\rm pl})^{-2} f_\f^{-1}= \frac{A}{c_l(k/k_0)^{p_l}+c_m(k/k_0)^{p_m}+c_h(k/k_0)^{p_h}}, 
\end{align}
where $(A, k_0, c_l, c_m, p_l, p_m, p_h)^{\rm RD}=(0.040, 0.96m_0, 0.020, 0.93, -2.7, 0.067, 4.2)$. 

For the radiation-dominated case, we only use the tails of $\propto k^{2.7}$ and $\propto k^{-4.2}$ and we connect the interval by a scale invariant line for a generic interval $H_{\rm form}$ and $H_{\rm end}$. For the mapping to the GW spectrum in the current Universe (see e.g. Ref.\,\cite{Kitajima:2023cek,Vanvlasselaer:2026fay})

\begin{figure}[t]
    \centering
    \begin{minipage}[t]{0.32\linewidth}
        \centering
        \includegraphics[width=\linewidth]{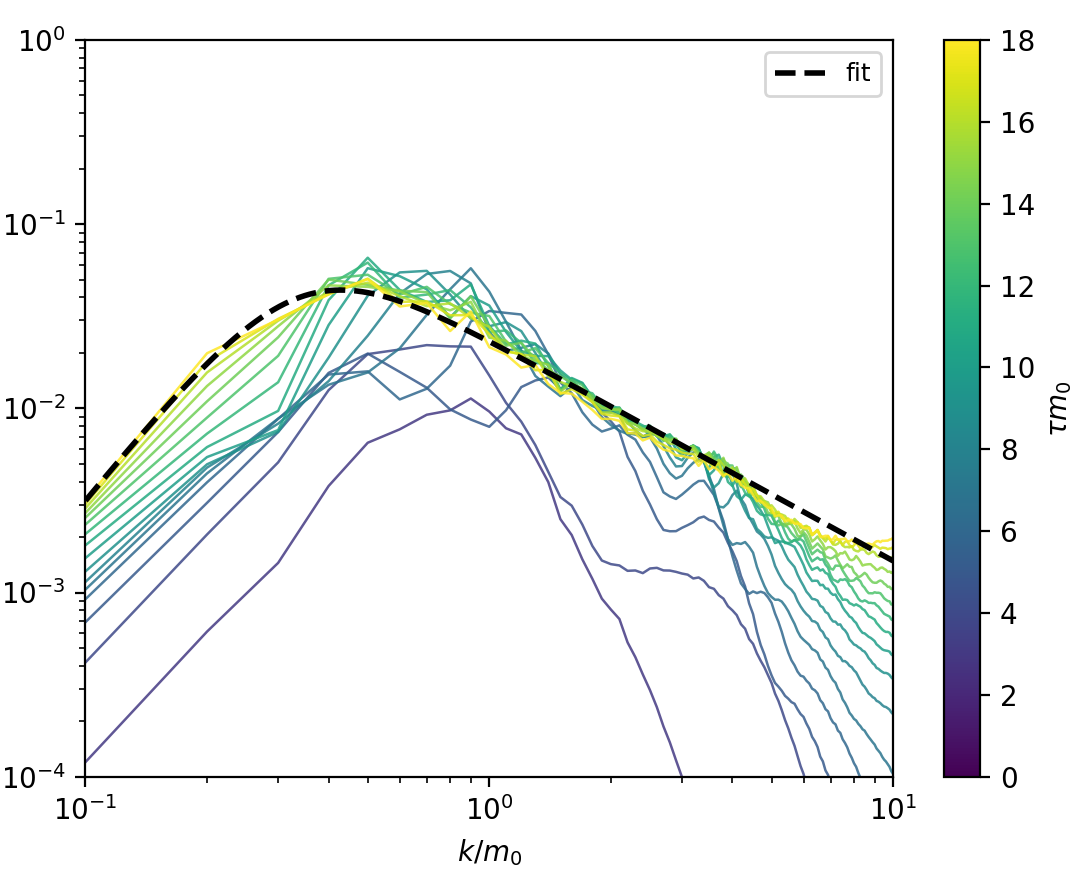}
        \vspace{-2mm}
        \centerline{(a) MD}
    \end{minipage}
    \hfill
    \begin{minipage}[t]{0.32\linewidth}
        \centering
        \includegraphics[width=\linewidth]{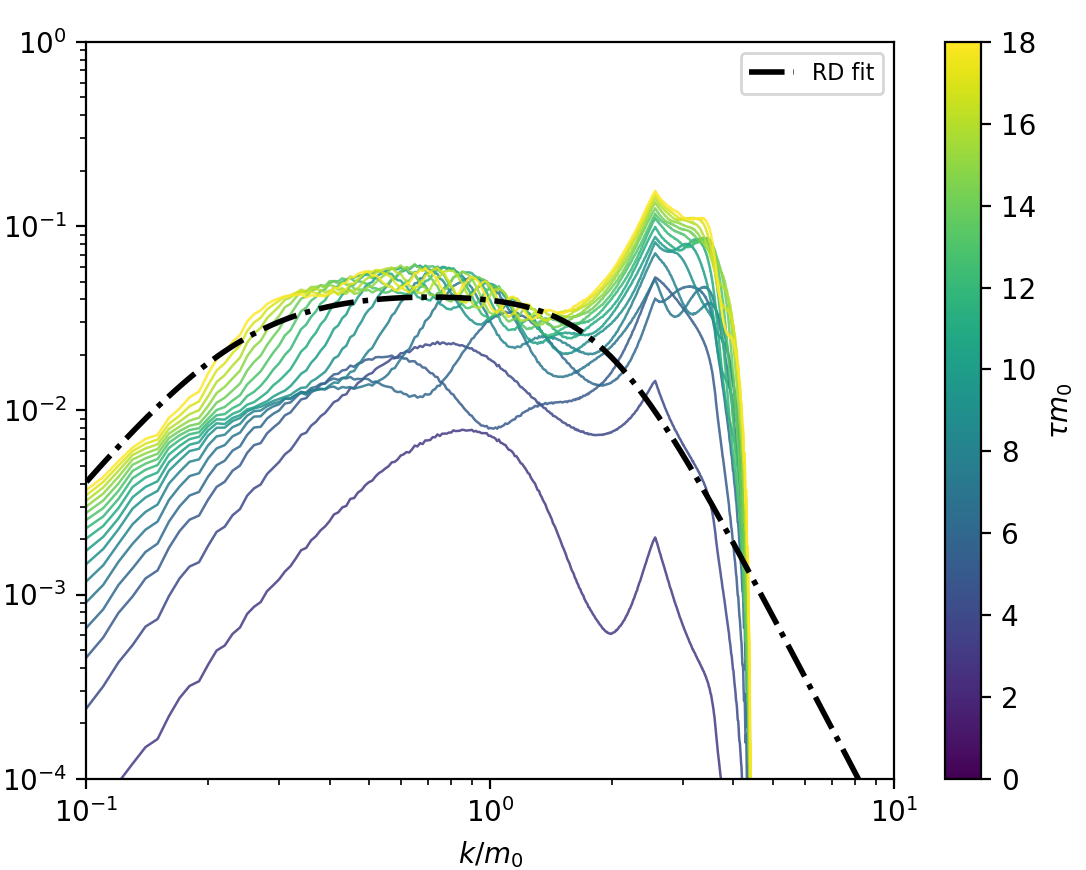}
        \vspace{-2mm}
        \centerline{(b) RD, smaller $k$}
    \end{minipage}
    \hfill
    \begin{minipage}[t]{0.31\linewidth}
        \centering
        \includegraphics[width=\linewidth]{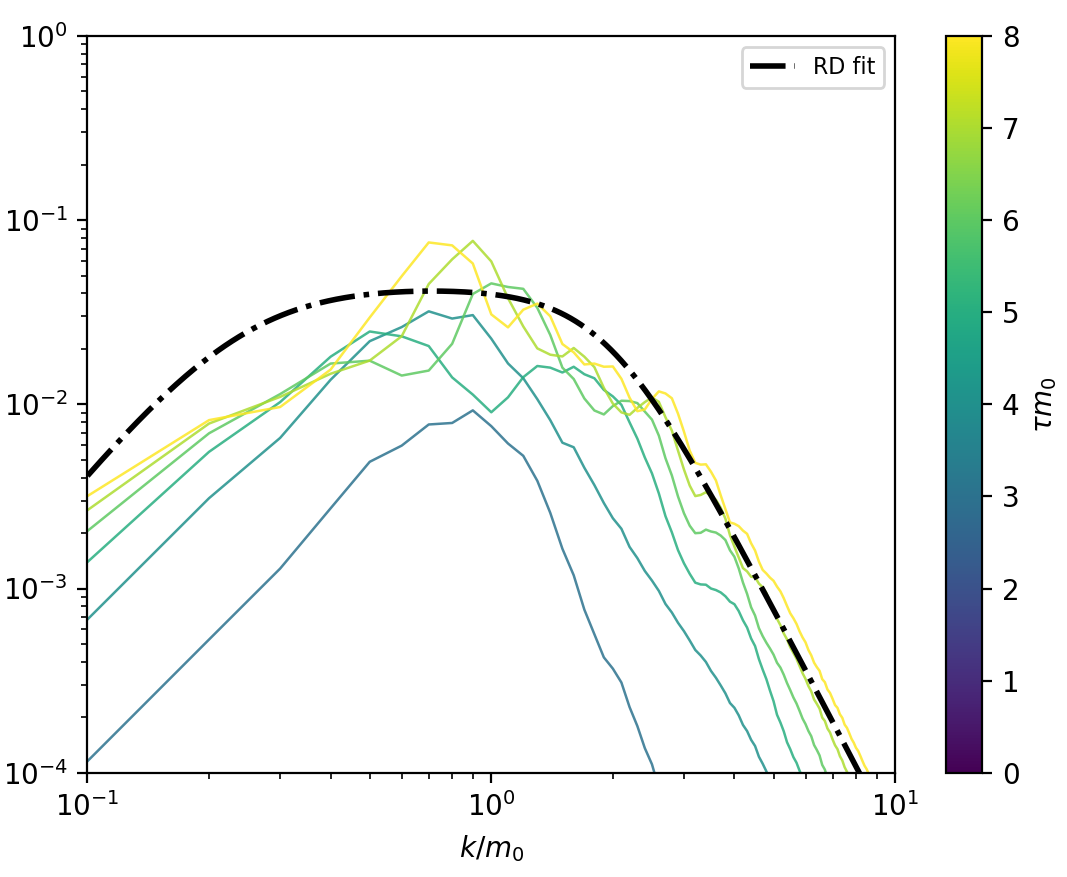}
        \vspace{-2mm}
        \centerline{(c) RD, larger $k$}
    \end{minipage}
   \caption{
Time evolution of the GW spectra $\Omega_{\rm GW} \times (v/M_{\rm pl})^{-2}\times f_\f^{-1}$ obtained from the lattice simulations.
Panel (a) shows the matter-dominated case, while panels (b) and (c) show the
radiation-dominated case.
The black curves show the fitting functions.
For the radiation-dominated case, the fitting function is constructed by
superposing the low-$k$ tail, the approximately scale-invariant intermediate
component, and the high-$k$ tail.
}
    \label{fig:gw_spectrum_fit}
\end{figure}

% \WYC{Check the form use $k_{\rm form}$ and $k_{\rm end}$. }

\section{A generic lower gradient bound from the boundary winding}
\label{app:2}

In this appendix we record a simple lower bound associated with the net
winding number in a Hubble patch.  The result should be regarded as a
topological lower bound on the large-scale gradient energy, not as a complete
description of the Hubble-patch energy distribution.

We define the boundary winding number by
\begin{align}
    Q_H
    =
    \frac{1}{8\pi}
    \int_{\partial V_H} dS_i\,
    \epsilon^{ijk}\epsilon^{abc}
    \hat\phi^a
    \partial_j \hat\phi^b
    \partial_k \hat\phi^c ,
    \qquad
    \hat\phi^a \equiv \frac{\phi^a}{|\phi|}.
    \label{eq:app-hubble-winding-number}
\end{align}
When monopoles are well separated and no defect lies on the boundary, this
winding agrees with the net number of defects inside the Hubble patch,
\begin{align}
    Q_H \simeq N_M-N_{\bar M}.
\end{align}

Outside the monopole cores, the radial mode is approximately fixed,
\begin{align}
    \phi^a = v n^a,
    \qquad
    n^a n^a = 1 .
\end{align}
On a sphere surrounding the patch, the angular field $n^a$ defines a map from
the spatial two-sphere to the vacuum manifold $S^2$.  For a configuration of
degree $Q$, the two-dimensional sigma-model energy obeys the bound
\begin{align}
    E_{S^2}
    \ge
    4\pi |Q| .
    \label{eq:app-sigma-model-bound}
\end{align}
Applying this bound shell by shell over a region of physical size $R$, the
gradient energy satisfies the parametric lower bound
\begin{align}
    E(R)
    \gtrsim
    4\pi v^2 |Q| R ,
    \label{eq:app-global-monopole-energy-bound}
\end{align}
up to an order-one geometric factor.

For a Hubble patch, $R\simeq H^{-1}$, and hence a patch with boundary winding
$Q_H$ contains at least
\begin{align}
    E_H(Q_H)
    \gtrsim
    4\pi v^2 |Q_H| H^{-1}
    \label{eq:app-hubble-patch-energy-bound}
\end{align}
of large-scale gradient energy.  The corresponding estimate for the horizon-scale overdensity is
\begin{align}
    \delta_H(Q_H)
    \equiv
    \frac{E_H(Q_H) H^3}{\rho_{\rm bg}} .
\end{align}
Using
\begin{align}
    \rho_{\rm bg}
    =
    3 M_{\rm pl}^2 H^2 ,
\end{align}
we obtain
\begin{align}
    \delta_H(Q_H)
    \gtrsim
    \frac{4\pi}{3}
    \frac{v^2}{M_{\rm pl}^2}
    |Q_H| .
    \label{eq:app-delta-from-QH}
\end{align}

This lower bound is useful for showing that nonzero net winding necessarily
carries large-scale gradient energy.  However, it does not determine the full
energy distribution.  In particular, a monopole--antimonopole pair with small
separation has a partially cancelled far-field configuration, and can carry
less large-scale gradient energy than two isolated defects.  Conversely, such a
nearby pair can still contribute to the local energy through core energy,
localized gradients, kinetic energy, and radiation during annihilation.  This
is why the main text uses $N_{{\rm tot},H}$ as an empirical proxy for local
defect activity rather than relying only on the net winding $Q_H$.

\section{Cosmology for PBH dark matter}
\label{app:latercosmo}
In the present scenario, the monopole network and the particles produced during
its subsequent evolution can also contribute to dark matter.  These include the
monopoles themselves, gauge bosons, radial-mode excitations, and pNGBs produced
during annihilation or collapse of the network.  A detailed study of these
components will be performed elsewhere.  Here we point out the simple scenarios
under which PBHs can become the dominant dark matter component.

In particular, if the global monopole network survives until recombination, the
symmetry-breaking scale at that epoch is constrained by the CMB.  Existing CMB
analyses of global monopole networks constrain their contribution to the
temperature power spectrum to be small, giving $f_{10}<0.024$ and corresponding
to $v\lesssim 6.4\times 10^{15}\,{\rm GeV}$ for global monopoles
\cite{Lopez-Eiguren:2017dmc}.

\subsection{Population bias scenario}
We consider a simple population-bias scenario in which the monopole network has a finite lifetime. 
The essential ingredient is that the scalar field is not exactly massless during inflation. 
For example, let us introduce the non-minimal coupling
\beq 
{\cal L}\supset -{1\over 2}\x \f_a \f^a R ,
\eeq 
where $R$ is the Ricci scalar. During quasi-de Sitter inflation, $R\simeq 12H_{\rm inf}^2$, with $H_{\rm inf}$ being the Hubble parameter during inflation. Hence, this term induces the Hubble mass
\beq 
m^2_{\rm Hubble}=12\x H_{\rm inf}^2 .
\eeq 
For
\beq 
{1\over 12}\lesssim \x < {3\over 16},
\eeq 
we have
\beq 
H_{\rm inf}\lesssim m_{\rm Hubble}< {3\over 2}H_{\rm inf}.
\eeq 
In this regime the superhorizon fluctuation is not conserved. Instead, it decays after horizon exit as
\beq 
\d\f_k \propto a^{-\alpha}, 
\qquad
\alpha={3\over 2}-\nu,
\qquad
\nu=\sqrt{ {9\over 4}-{m^2_{\rm Hubble}\over H_{\rm inf}^2} } .
\eeq 
Equivalently, the fluctuation spectrum is blue-tilted,
\beq 
{\cal P}_{\d\f}(k)\simeq 
\left({H_{\rm inf}\over 2\pi}\right)^2
C_\nu
\left({k\over aH_{\rm inf}}\right)^{3-2\nu},
\eeq 
where $C_\nu$ is an ${\cal O}(1)$ numerical factor. The corresponding spectral index is
\beq 
n_\f-1=3-2\nu .
\eeq 
Thus the long-wavelength correlation is suppressed compared with the scale-invariant case.

The stochastic distribution of the coarse-grained field is also modified. For a quadratic potential, the equilibrium variance is approximately
\beq 
\vev{\f^2}\simeq {3H_{\rm inf}^4\over 8\pi^2 m^2_{\rm Hubble}}
={H_{\rm inf}^2\over 32\pi^2\x}.
\eeq 
Therefore, for $m_{\rm Hubble}\sim H_{\rm inf}$, the typical coarse-grained field value in one inflationary Hubble patch is displaced from the origin
\beq 
|\phi|_{\rm typ}\sim \sqrt{\langle \phi^a\phi^a\rangle}\sim 0.1-0.2\,H_{\rm inf}.
\eeq 
The ensemble average vanishes by the $O(3)$ symmetry, but the observable Universe corresponds to one stochastic realization.  This provides a population bias. The same inflationary fluctuations in smaller scales (e.g. Horizon scale at the formation) are $\sim H_{\rm inf}$. 
When the tachyonic mass scale becomes comparable to the Hubble scale, 
 \beq \sqrt{\l v^2}\sim H,\eeq 
the field rolls toward the symmetry-breaking minimum and monopoles are formed by the usual Kibble mechanism. The evolution at the early stage is like the ordinary symmetry breaking and it forms the monopole network. 

As a result, we obtain the following two features:
\begin{itemize}
\item The coarse-grained field value in each Hubble patch is not exactly at the origin, giving a population bias among patches.
\item The fluctuation spectrum is blue-tilted rather than scale invariant, so the large-scale correlation is reduced and the scaling regime of the monopole network does not persist for a very long time; see, e.g., Refs.\,\cite{Gonzalez:2022mcx,Kitajima:2023kzu,2911336} for domain wall and \cite{Yin:2024pri} also for cosmic string. This is checked within the simulation time of $\tau<15$ by using the lattice simulation in the case of global monopole by taking the initial power spectrum for $2\nu\gtrsim 0.1$. 
\end{itemize}
These facts imply that the monopole network can collapse at a finite time after its formation, without requiring an additional assumption tied to PBH formation. After the collapse, the energy stored in the network can be transferred into light degrees of freedom. In particular, sufficiently light pseudo-NGBs or gauge fields may contribute to the radiation component. We also noticed that when $2\nu \lesssim 1$, the probability distribution gets broader. 

One discussion we should make is the fate of the Higgs radial mode, which should be weakly coupled so as not to erase the initial condition, set by the inflationary phase. When the monopoles annhilate, harmless NGB dark radiation are mostly produced since the monopoles are dominated by the NGB gradient energy. However, at the formation of the monopole the semi-relativistic Higgs bosons are significantly produced contributing to the matter, which should be suppressed. 
One simple scenario is stimulated decays~\cite{Moroi:2020has,Moroi:2020bkq}, or
tachyonic/parametric resonance to have the energy of the scalar sector
into radiation, e.g. by coupling the Higgs to the Standard Model-like Higgs boson \cite{Kofman:1994rk,Kofman:1997yn,Amin:2018kkg}.

\subsection{Transient $v$ scenario}
\label{app:tran}

Another way to be consistent with the monopole limit is to have
$v=\O(0.1)M_{\rm pl}$ during PBH formation, while reducing it after PBH
formation. In this case we may even have the thermal mass to the Higgs field driving it to the origin as the usual phase transition and there is no population bias. 

Since the monopole-network energy density scales as
$\rho_M/\rho_{\rm bg}\propto v^2/M_{\rm pl}^2$, a later decrease of $v$
suppresses the remaining monopole energy density while leaving the already
formed PBHs intact. Such a time dependence can be realized if the scalar
potential receives a late-time contribution from another sector. 
In the matter-dominated scenario, one may consider the coupling of $\rho(s)(\f_a \f^a -v^2)^2$ in addition to the vacuum potential $\propto (\f_a \f^a -v_{\rm today}^2)^2$ with $v_{\rm today}$ being the vacuum expectation value today. 
Here we assume $s$ is the matter field that dominates the Universe. Then $\phi$ will gets an effective potential during the era, to induce the large expecation value around the Planck scale while it is suppressed today.\footnote{Alternatively, for example,
$\phi$ may couple through a portal interaction to a hidden non-Abelian gauge
sector, whose dynamical scale generates an additional late-time potential, in a
way analogous to the QCD axion \cite{Peccei:1977hh,Weinberg:1977ma,Wilczek:1977pj}.
Also, a simple possibility is to consider the tachyonic thermal mass. In the feebly interacting model, the transient Planckian expectation value is predicted~\cite{Yin:2024txg}.} This implies that at the end of the reheating phase, the induced potential eventually disappears and we get the smaller vacuum expectation value. Although we still have the scaling network of the monopole until the screening scale becomes relevant, the tension is highly reduced and the formation of the PBH is highly suppressed.

The light NGBs (or very weakly coupled light gauge bosons) contribute to the dark radiation. The monopole remains until today but the abundance is suppressed $v_{\rm today}^2/M_{\rm pl}^2$, which were formed when $v\sim M_{\rm pl}$. The Higgs mass gets suppressed by a factor $v_{\rm today}/M_{\rm pl}$ and thus does the final Higgs matter density today given the number conservation. The Higgs becomes subdominant dark matter if $v_{\rm today} T_{\rm RH}/ M_{\rm pl}\ll 10^{-10}\GEV$. Alternatively, it may just dissipate the energy to the Standard Model particles. 
\\
\\
The two scenarios should also work for the PBH formation by other topological defects such as cosmic strings, which also require a near-Planckian Higgs expectation, value to evade the limits from the late time cosmology.

\section{Gravitational field of global monopoles}

The linearly growing energy in Eq.~\eqref{eq:global-mono-energy-frw} does not
appear as an ordinary localized Arnowitt-Deser-Misner (ADM) mass.  Instead, its leading gravitational
effect is a solid-angle deficit.  Indeed, since
\begin{equation}
  E(r)\sim 4\pi v^2 r ,
\end{equation}
the dimensionless quantity controlling the metric perturbation is
\begin{equation}
  \frac{2G E(r)}{r}\sim 8\pi Gv^2 .
\end{equation}
This quantity is independent of $r$.  Thus the long-range NGB gradient
energy does not generate a conventional $1/r$ Newtonian potential. Rather, it
changes the asymptotic angular geometry.

The exterior metric of a global monopole is approximately given by the
Barriola-Vilenkin form~\cite{Barriola:1989hx}
\begin{equation}
  ds^2
  =
  -B(r)dt^2+B(r)^{-1}dr^2+r^2d\Omega^2,
  \qquad
  B(r)=1-\Delta-\frac{2GM}{r},
  \laq{metric}
\end{equation}
where
\begin{equation}
  \Delta=8\pi Gv^2 .
\end{equation}
The parameter $\Delta$ measures the solid-angle deficit produced by the
long-range scalar gradients.  Equivalently, after an appropriate rescaling of
the time and radial coordinates, the constant term in $B(r)$ can be interpreted
as a deficit in the asymptotic solid angle.  The parameter $M$ denotes the
coefficient of the finite $1/r$ term in the metric.  For an isolated regular
global monopole, the linearly growing NGB gradient energy is
encoded in the deficit-angle term $\Delta$, while $M$ represents only the residual
finite contribution after this linear piece has been separated off.  This
residual contribution is of order the monopole core energy and can be negative
\cite{Harari:1990cz}.  Therefore, for an isolated monopole and at distances
$r\gg r_{\rm core}\sim 1/(\sqrt{\l}v)$,  this core-scale
contribution is not the dominant gravitational effect.

This interpretation changes if the winding is neutralized at a finite radius,
for example by antimonopoles.  In that case, the NGB gradient energy
does not continue to grow linearly to infinity.  At distances much larger than
the neutralization scale, the spacetime can approach an ordinary
asymptotically flat geometry, whose leading far-field term is characterized by
an ADM mass, up to multipole corrections.  Thus, when a configuration contains
both monopoles and antimonopoles, which is the case for  PBH formation, one expects to use the usual criterion for the overdensities when studying PBH formation.

\section{PBH formation from the stochastic delayed-rolling: difficulties and loopholes}
\label{app:delayed-roll-fails}

In the main text, PBHs are produced by overdensities associated with monopole
formation and the subsequent scaling network.  One may ask whether PBHs can
instead be produced more simply from delayed rolling of the scalar field with initial Hubble scale correlation length from
the symmetric hilltop without forming a topological defect, since the field experiences a tachyonic instability after
the transition~\cite{Felder:2000hj,Felder:2001kt}. 
Then the stochastic realization of the field may make the overdensity and generate PBH. 
In this appendix, we show
quantitatively that this delayed-roll mechanism is inefficient in the quadratic hilltop potential. The monopole scaling dynamics is
therefore essential.

When 
\begin{align}
    \phi=\phi_{\rm end}\sim v ,
\end{align}
one cannot use the quadratic hilltop to study the system and the tachyonic growth stops. 
Let the typical initial displacement at the onset of the tachyonic instability
be
\begin{align}
    \phi_{\rm typ}=A\sigma_\phi ,
    \qquad A=O(1).
\end{align}
For avoiding the topological formation we may take $A>1$.
The required growth factor for a typical patch to reach the nonlinear regime is
\begin{align}
    Y
    \equiv
    {\phi_{\rm end}\over \phi_{\rm typ}}
    \simeq
    {v\over A\sigma_\phi}.
    \label{eq:Y_def_app}
\end{align}
For $\sigma_\phi\simeq H_*/(2\pi)$, where $H_*=m_\F$, this gives
\begin{align}
    Y
    \simeq
    {2\pi v\over A H_*}.
    \label{eq:Y_Hstar_app}
\end{align}
For PBH formation at small horizon mass, $H_*$ is much smaller than $v$, while
$v$ should be close to the Planck scale in order for the scalar sector to carry
an order-one fraction of the total energy density.  For example,
\begin{align}
    v=10^{18}{\rm GeV},
    \qquad
    H_*=10^{-3}{\rm GeV},
    \qquad
    A=1
\end{align}
gives
\begin{align}
    Y\simeq 6\times 10^{21}.
    \label{eq:Y_numeric_app}
\end{align}
Thus the typical patch must be amplified by more than twenty orders of
magnitude before it leaves the hilltop regime.

We now estimate the probability that a patch is delayed by $\Delta N$ relative
to a typical patch. For $\phi<\phi_{\rm end}$, the linear approximation holds. Let
$U(N;N_*)$ denote the linear growth factor of the
hilltop field,
\begin{align}
    \phi(N)
    =
    U(N;N_*)\phi_i,
    \qquad
    U(N_*;N_*)=1 .
    \label{eq:U_def_app}
\end{align}
The typical patch with $\f_i=\f_{\rm typ}$ reaches $\phi_{\rm end}$ at $N=N_{\rm typ}$, where
\begin{align}
    U(N_{\rm typ};N_*)=Y .
    \label{eq:Ntyp_def_app}
\end{align}
A patch is delayed by at least $\Delta N$ if it has not reached $\f_{\rm end}$
regime at $N=N_{\rm typ}+\Delta N$.  This requires
\begin{align}
    U(N_{\rm typ}+\Delta N;N_*)|\phi_i|
    <
    \phi_{\rm end}.
\end{align}
Using \Eq{Ntyp_def_app}, this condition becomes
\begin{align}
    |\phi_i|
    <
    \phi_{\rm typ}
    {U(N_{\rm typ};N_*)\over U(N_{\rm typ}+\Delta N;N_*)}.
    \label{eq:delay_condition_app}
\end{align}
Therefore the one-patch delay probability is
\begin{align}
    P_{\rm delay}(\Delta N;Y)
    =
    {\rm erf}
    \left[
    {A\over \sqrt{2}}
    {U(N_{\rm typ};N_*)\over U(N_{\rm typ}+\Delta N;N_*)}
    \right].
    \label{eq:Pdelay_exact_app}
\end{align}

To see the parametric behavior, we allow the tachyonic mass to scale as
\begin{align}
    m_\F\propto a^{-q},
    \qquad
    H\propto a^{-p/2}.
\end{align}
The case relevant for our setup is $q=0$, but keeping $q$ is useful for showing
how the degree of tuning depends on the time dependence of the mass.  Then
\begin{align}
    {m_\F\over H}
    \propto
    e^{\alpha (N-N_*)/2},
    \qquad
    \alpha\equiv p-2q .
    \label{eq:alpha_def_app}
\end{align}
For $\alpha>0$ and large $Y$, the extra growth over the delay interval is
approximately
\begin{align}
    \log {U(N_{\rm typ}+\Delta N;N_*)\over U(N_{\rm typ};N_*)}
    \simeq
    \log Y
    \left(
    e^{\alpha\Delta N/2}-1
    \right),
    \label{eq:extra_growth_app}
\end{align}
up to terms that are not enhanced by $\log Y$.  Hence
\begin{align}
    P_{\rm delay}(\Delta N;Y)
    \simeq
    {\rm erf}
    \left[
    {A\over \sqrt{2}}
    \exp\left\{
    -\log Y
    \left(
    e^{\alpha\Delta N/2}-1
    \right)
    \right\}
    \right].
    \label{eq:Pdelay_scaling_app}
\end{align}
This expression shows that the tuning is controlled by $\alpha=p-2q$.  A
constant mass has $\alpha=p$ and gives the strongest suppression.  A mass
scaling as $m_\F\propto a^{-1}$ has $\alpha=p-2$ and reduces the tuning.  If
$m_\F\propto H$, then $\alpha=0$, and the large hierarchy $Y$ no longer
exponentially suppresses the delay probability.

The PBH probability is further suppressed because a whole region corresponding
to the collapse horizon must be delayed coherently.  If the relevant delay is
$\Delta N$, the number of independent initial Hubble patches inside the
collapse region is estimated as
\begin{align}
    N_{\rm patch}
    =
    e^{3(p/2-1)\Delta N}.
    \label{eq:Npatch_app}
\end{align}
Thus
\begin{align}
    \beta_{\rm PBH}
    \simeq
    \left[
    P_{\rm delay}(\Delta N;Y)
    \right]^{N_{\rm patch}} .
    \label{eq:beta_delay_app}
\end{align}
The typical patch must have rolled down, entered the oscillatory regime, and
redshifted relative to the delayed patch; otherwise, the contrast in the sum of
the kinetic and potential energies remains negligible.  Therefore, the delayed
patch has to remain near the hilltop and avoid oscillation for longer than an
order-one Hubble time after a typical patch has started oscillating.  In the
following estimate, we take\footnote{This condition is more severe than the one to have overdensity $\d_c>0.4$ for the most conservative case $f_\f=1$.}
\begin{align}
    \Delta N = 1
\end{align}
as the relevant benchmark delay.

For $\Delta N=1$, the coherence factors are
\begin{align}
    N_{\rm patch}
    =
    e^{3/2}
    \simeq
    4.5
    \qquad
    (p=3),
    \label{eq:Npatch_p3_eff_app}
\end{align}
and
\begin{align}
    N_{\rm patch}
    =
    e^{3}
    \simeq
    20
    \qquad
    (p=4).
    \label{eq:Npatch_p4_eff_app}
\end{align}
Thus even if a single patch has a nonzero probability to be delayed, the PBH
probability is further suppressed as in \Eq{beta_delay_app}.

For the representative hierarchy $Y=6\times 10^{21}$ and $A=1$, one obtains
the following order-of-magnitude estimates:
\begin{center}
\begin{tabular}{c|c|c|c|c}
\hline
background & mass scaling & $\alpha$ & $\log_{10}P_{\rm delay}$ & $\log_{10}\beta_{\rm PBH}$ \\
\hline
$p=4$ & $m_\F={\rm const}$ & $4$ & $-139$ & $-2.8\times 10^3$ \\
$p=4$ & $m_\F\propto a^{-1}$ & $2$ & $-38$ & $-7.6\times 10^2$ \\
$p=4$ & $m_\F\propto H$ & $0$ & $-0.39$ & $-7.8$ \\
\hline
$p=3$ & $m_\F={\rm const}$ & $3$ & $-76$ & $-3.4\times 10^2$ \\
$p=3$ & $m_\F\propto a^{-1}$ & $1$ & $-14$ & $-64$ \\
$p=3$ & $m_\F\propto H$ & $0$ & $-0.34$ & $-1.5$ \\
\hline
\end{tabular}
\end{center}
The constant-mass case and the $m_\F\propto a^{-1}$ case are therefore highly
suppressed.

We have also checked this conclusion numerically by evolving the scalar field
with Hubble-scale initial fluctuations and with initial
$\vev{|\f|}\sim H$, but without monopole formation.  Even for a modest
hierarchy between the symmetry-breaking scale and the Hubble scale, the
Hubble-patch distribution of the total energy density remains sharply localized
near the mean and does not develop a sizable high-density tail.

\paragraph{Loopholes}
The only case in the table where the delay probability itself is not strongly
suppressed is $m_\F\propto H$.   A simple possibility is to
introduce a nonminimal coupling between $\F$ and the Hubble parameter, which
generates a Hubble-induced mass and naturally gives a tachyonic mass
proportional to $H$, at least during a late-time matter-dominated era.  

Another loophole is to consider a non-quadratic hilltop.  This may be realized
in low-scale inflation, where an explosive tachyonic instability phase was
recently found~\cite{Masubuchi:2026eau}.  A further study is important for
constraining such low-scale inflation scenarios.

\section{Hubble-patch observables}
\label{app:lattice}

We measure the monopole number, the winding charge, and the energy density on a
Hubble-patch basis.  

At each output time, we first determine the size of a Hubble patch in lattice
units,
\begin{equation}
  L_H^{\rm lat}
  =
  \frac{a/\dot a}{dx},
\end{equation}
where $dx$ is the lattice spacing. We use the data when this is larger than one
lattice spacing.  The number of Hubble blocks per side is then
\begin{equation}
  N_H
  =
  \left\lfloor
  \frac{N_{\rm lat}}{L_H^{\rm lat}}
  \right\rfloor ,
\end{equation}
and the total number of Hubble blocks is
\begin{equation}
  N_{\rm block}=N_H^3 .
\end{equation}
The number of lattice cells used per side is
\begin{equation}
  N_{\rm used}
  =
  \left\lfloor
  N_H L_H^{\rm lat}
  \right\rfloor .
\end{equation}
Cells outside this used volume are discarded.  This prescription avoids mixing
incomplete Hubble patches with complete ones.

The local monopole charge is computed on each elementary lattice cube from the
solid angle swept by the normalized scalar field direction
\begin{equation}
  \hat n
  =
  \frac{\boldsymbol{\phi}}{|\boldsymbol{\phi}|}.
\end{equation}
For a triangular face with vertices $\hat n_1,\hat n_2,\hat n_3$, we use the
oriented solid angle
\begin{equation}
  \Omega(\hat n_1,\hat n_2,\hat n_3)
  =
  2\tan^{-1}
  \frac{
    \hat n_1\cdot(\hat n_2\times \hat n_3)
  }{
    1+\hat n_1\cdot\hat n_2
     +\hat n_2\cdot\hat n_3
     +\hat n_3\cdot\hat n_1
  } .
\end{equation}
This is the standard geometrical definition of the lattice winding number based
on the image area on the vacuum manifold~\cite{Berg:1981er}.  The charge of an
elementary cube is then
\begin{equation}
  Q_{\rm cube}
  =
  \frac{1}{4\pi}
  \sum_{\Delta\in\partial{\rm cube}}
  \Omega_\Delta ,
\end{equation}
where the cube surface is divided into twelve oriented triangles.

For the Hubble-patch monopole count, we do not count the raw cube charges
directly.  Instead, we first apply a short flow to the eight normalized field
directions on each cube and then compute the corresponding flowed cube charge.
The flow is a local smoothing on the unit sphere.  At each flow step, each
vertex is shifted along the tangent projection of the graph Laplacian,
\begin{equation}
  \hat n_i
  \to
  {\cal N}\left[
  \hat n_i
  +
  \epsilon
  \left(
  \sum_{j\in{\rm n.n.}(i)}(\hat n_j-\hat n_i)
  -
  \hat n_i
  \hat n_i\cdot
  \sum_{j\in{\rm n.n.}(i)}(\hat n_j-\hat n_i)
  \right)
  \right],
\end{equation}
where ${\cal N}$ denotes normalization to unit length. 
${\rm n.n.}(i)$ denotes the set of nearest-neighbor vertices connected to 
vertex $i$ by an edge of the cube. 
In the analysis, we use
four flow steps with $\epsilon=0.02$.  This short flow removes lattice-scale
angular noise while keeping the field on the vacuum manifold.

After the flow, cubes with nonzero rounded charge are grouped into connected
clusters using nearest-neighbor connectivity.  The charge of a cluster $C$ is
defined by summing the integer flowed charges of the cubes in the cluster,
\begin{equation}
  Q_C
  =
  \sum_{{\rm cubes}\in C}
  {\rm round}(Q_{\rm cube}^{\rm flow}) .
\end{equation}
If a charged cluster crosses a Hubble-block boundary, it is not assigned to a
single Hubble patch.  Otherwise, its charge is assigned to the Hubble block that
contains the cluster.  The monopole and antimonopole numbers in a Hubble patch
are then defined by
\begin{equation}
  N_{{\rm mon},H}
  =
  \sum_{C\subset H}
  \max(Q_C,0),
  \qquad
  N_{{\rm anti},H}
  =
  \sum_{C\subset H}
  \max(-Q_C,0),
\end{equation}
and
\begin{equation}
  N_{{\rm tot},H}
  =
  N_{{\rm mon},H}
  +
  N_{{\rm anti},H}.
\end{equation}
Thus the monopole count used in the analysis is the cube-flow-cluster count.

The gradient and total energy densities in each Hubble patch are computed by
summing the local energy density over all lattice sites inside the patch and
dividing by the number of sites in the patch,
\begin{equation}
  \rho_{{\rm grad},H}
  =
  \frac{1}{N_{{\rm site},H}}
  \sum_{\boldsymbol{x}\in H}
  \rho_{\rm grad}(\boldsymbol{x}),
\end{equation}
and
\begin{equation}
  \rho_{{\rm tot},H}
  =
  \frac{1}{N_{{\rm site},H}}
  \sum_{\boldsymbol{x}\in H}
  \left[
    \rho_{\rm kin}(\boldsymbol{x})
    +
    \rho_{\rm grad}(\boldsymbol{x})
    +
    \rho_{\rm pot}(\boldsymbol{x})
  \right].
\end{equation}
The normalized Hubble-patch energy densities are defined by
\begin{equation}
  x_{{\rm grad},H}
  =
  \frac{\rho_{{\rm grad},H}}
       {\langle \rho_{{\rm grad},H} \rangle_H},
  \qquad
  x_{{\rm tot},H}
  =
  \frac{\rho_{{\rm tot},H}}
       {\langle \rho_{{\rm tot},H} \rangle_H},
\end{equation}
where $\langle\cdots\rangle_H$ denotes an average over all non-empty Hubble
patches at the same output time.

\bibliography{GenericALPDMbound.bib}

\begin{thebibliography}{109}%
\makeatletter
\providecommand \@ifxundefined [1]{%
 \@ifx{#1\undefined}
}%
\providecommand \@ifnum [1]{%
 \ifnum #1\expandafter \@firstoftwo
 \else \expandafter \@secondoftwo
 \fi
}%
\providecommand \@ifx [1]{%
 \ifx #1\expandafter \@firstoftwo
 \else \expandafter \@secondoftwo
 \fi
}%
\providecommand \natexlab [1]{#1}%
\providecommand \enquote  [1]{``#1''}%
\providecommand \bibnamefont  [1]{#1}%
\providecommand \bibfnamefont [1]{#1}%
\providecommand \citenamefont [1]{#1}%
\providecommand \href@noop [0]{\@secondoftwo}%
\providecommand \href [0]{\begingroup \@sanitize@url \@href}%
\providecommand \@href[1]{\@@startlink{#1}\@@href}%
\providecommand \@@href[1]{\endgroup#1\@@endlink}%
\providecommand \@sanitize@url [0]{\catcode `\\12\catcode `\$12\catcode
  `\&12\catcode `\#12\catcode `\^12\catcode `\_12\catcode `\%12\relax}%
\providecommand \@@startlink[1]{}%
\providecommand \@@endlink[0]{}%
\providecommand \url  [0]{\begingroup\@sanitize@url \@url }%
\providecommand \@url [1]{\endgroup\@href {#1}{\urlprefix }}%
\providecommand \urlprefix  [0]{URL }%
\providecommand \Eprint [0]{\href }%
\providecommand \doibase [0]{http://dx.doi.org/}%
\providecommand \selectlanguage [0]{\@gobble}%
\providecommand \bibinfo  [0]{\@secondoftwo}%
\providecommand \bibfield  [0]{\@secondoftwo}%
\providecommand \translation [1]{[#1]}%
\providecommand \BibitemOpen [0]{}%
\providecommand \bibitemStop [0]{}%
\providecommand \bibitemNoStop [0]{.\EOS\space}%
\providecommand \EOS [0]{\spacefactor3000\relax}%
\providecommand \BibitemShut  [1]{\csname bibitem#1\endcsname}%
\let\auto@bib@innerbib\@empty
%</preamble>
\bibitem [{\citenamefont {Carr}\ and\ \citenamefont
  {Hawking}(1974)}]{Carr:1974nx}%
  \BibitemOpen
  \bibfield  {author} {\bibinfo {author} {\bibfnamefont {B.~J.}\ \bibnamefont
  {Carr}}\ and\ \bibinfo {author} {\bibfnamefont {S.~W.}\ \bibnamefont
  {Hawking}},\ }\href {\doibase 10.1093/mnras/168.2.399} {\bibfield  {journal}
  {\bibinfo  {journal} {Mon. Not. Roy. Astron. Soc.}\ }\textbf {\bibinfo
  {volume} {168}},\ \bibinfo {pages} {399} (\bibinfo {year}
  {1974})}\BibitemShut {NoStop}%
\bibitem [{\citenamefont {Carr}(1975)}]{Carr:1975qj}%
  \BibitemOpen
  \bibfield  {author} {\bibinfo {author} {\bibfnamefont {B.~J.}\ \bibnamefont
  {Carr}},\ }\href {\doibase 10.1086/153853} {\bibfield  {journal} {\bibinfo
  {journal} {Astrophys. J.}\ }\textbf {\bibinfo {volume} {201}},\ \bibinfo
  {pages} {1} (\bibinfo {year} {1975})}\BibitemShut {NoStop}%
\bibitem [{\citenamefont {Sasaki}\ \emph {et~al.}(2018)\citenamefont {Sasaki},
  \citenamefont {Suyama}, \citenamefont {Tanaka},\ and\ \citenamefont
  {Yokoyama}}]{Sasaki:2018dmp}%
  \BibitemOpen
  \bibfield  {author} {\bibinfo {author} {\bibfnamefont {M.}~\bibnamefont
  {Sasaki}}, \bibinfo {author} {\bibfnamefont {T.}~\bibnamefont {Suyama}},
  \bibinfo {author} {\bibfnamefont {T.}~\bibnamefont {Tanaka}}, \ and\ \bibinfo
  {author} {\bibfnamefont {S.}~\bibnamefont {Yokoyama}},\ }\href {\doibase
  10.1088/1361-6382/aaa7b4} {\bibfield  {journal} {\bibinfo  {journal} {Class.
  Quant. Grav.}\ }\textbf {\bibinfo {volume} {35}},\ \bibinfo {pages} {063001}
  (\bibinfo {year} {2018})},\ \Eprint {http://arxiv.org/abs/1801.05235}
  {arXiv:1801.05235 [astro-ph.CO]} \BibitemShut {NoStop}%
\bibitem [{\citenamefont {Carr}\ \emph {et~al.}(2021)\citenamefont {Carr},
  \citenamefont {Kohri}, \citenamefont {Sendouda},\ and\ \citenamefont
  {Yokoyama}}]{Carr:2020gox}%
  \BibitemOpen
  \bibfield  {author} {\bibinfo {author} {\bibfnamefont {B.}~\bibnamefont
  {Carr}}, \bibinfo {author} {\bibfnamefont {K.}~\bibnamefont {Kohri}},
  \bibinfo {author} {\bibfnamefont {Y.}~\bibnamefont {Sendouda}}, \ and\
  \bibinfo {author} {\bibfnamefont {J.}~\bibnamefont {Yokoyama}},\ }\href
  {\doibase 10.1088/1361-6633/ac1e31} {\bibfield  {journal} {\bibinfo
  {journal} {Rept. Prog. Phys.}\ }\textbf {\bibinfo {volume} {84}},\ \bibinfo
  {pages} {116902} (\bibinfo {year} {2021})},\ \Eprint
  {http://arxiv.org/abs/2002.12778} {arXiv:2002.12778 [astro-ph.CO]}
  \BibitemShut {NoStop}%
\bibitem [{\citenamefont {Villanueva-Domingo}\ \emph
  {et~al.}(2021)\citenamefont {Villanueva-Domingo}, \citenamefont {Mena},\ and\
  \citenamefont {Palomares-Ruiz}}]{Villanueva-Domingo:2021spv}%
  \BibitemOpen
  \bibfield  {author} {\bibinfo {author} {\bibfnamefont {P.}~\bibnamefont
  {Villanueva-Domingo}}, \bibinfo {author} {\bibfnamefont {O.}~\bibnamefont
  {Mena}}, \ and\ \bibinfo {author} {\bibfnamefont {S.}~\bibnamefont
  {Palomares-Ruiz}},\ }\href {\doibase 10.3389/fspas.2021.681084} {\bibfield
  {journal} {\bibinfo  {journal} {Front. Astron. Space Sci.}\ }\textbf
  {\bibinfo {volume} {8}},\ \bibinfo {pages} {87} (\bibinfo {year} {2021})},\
  \Eprint {http://arxiv.org/abs/2103.12087} {arXiv:2103.12087 [astro-ph.CO]}
  \BibitemShut {NoStop}%
\bibitem [{\citenamefont {Green}\ and\ \citenamefont
  {Kavanagh}(2021)}]{Green:2020jor}%
  \BibitemOpen
  \bibfield  {author} {\bibinfo {author} {\bibfnamefont {A.~M.}\ \bibnamefont
  {Green}}\ and\ \bibinfo {author} {\bibfnamefont {B.~J.}\ \bibnamefont
  {Kavanagh}},\ }\href {\doibase 10.1088/1361-6471/abc534} {\bibfield
  {journal} {\bibinfo  {journal} {J. Phys. G}\ }\textbf {\bibinfo {volume}
  {48}},\ \bibinfo {pages} {043001} (\bibinfo {year} {2021})},\ \Eprint
  {http://arxiv.org/abs/2007.10722} {arXiv:2007.10722 [astro-ph.CO]}
  \BibitemShut {NoStop}%
\bibitem [{\citenamefont {Carr}\ \emph {et~al.}(2026)\citenamefont {Carr},
  \citenamefont {Iovino}, \citenamefont {Perna}, \citenamefont {Vaskonen},\
  and\ \citenamefont {Veerm{\"a}e}}]{Carr:2026hot}%
  \BibitemOpen
  \bibfield  {author} {\bibinfo {author} {\bibfnamefont {B.}~\bibnamefont
  {Carr}}, \bibinfo {author} {\bibfnamefont {A.~J.}\ \bibnamefont {Iovino}},
  \bibinfo {author} {\bibfnamefont {G.}~\bibnamefont {Perna}}, \bibinfo
  {author} {\bibfnamefont {V.}~\bibnamefont {Vaskonen}}, \ and\ \bibinfo
  {author} {\bibfnamefont {H.}~\bibnamefont {Veerm{\"a}e}},\ }\href {\doibase
  10.1007/s40766-026-00080-z} {\bibfield  {journal} {\bibinfo  {journal} {Riv.
  Nuovo Cim.}\ }\textbf {\bibinfo {volume} {49}},\ \bibinfo {pages} {225}
  (\bibinfo {year} {2026})},\ \Eprint {http://arxiv.org/abs/2601.06024}
  {arXiv:2601.06024 [astro-ph.CO]} \BibitemShut {NoStop}%
\bibitem [{\citenamefont {Hawking}(1989)}]{Hawking:1987bn}%
  \BibitemOpen
  \bibfield  {author} {\bibinfo {author} {\bibfnamefont {S.~W.}\ \bibnamefont
  {Hawking}},\ }\href {\doibase 10.1016/0370-2693(89)90206-2} {\bibfield
  {journal} {\bibinfo  {journal} {Phys. Lett. B}\ }\textbf {\bibinfo {volume}
  {231}},\ \bibinfo {pages} {237} (\bibinfo {year} {1989})}\BibitemShut
  {NoStop}%
\bibitem [{\citenamefont {Polnarev}\ and\ \citenamefont
  {Zembowicz}(1991)}]{Polnarev:1988dh}%
  \BibitemOpen
  \bibfield  {author} {\bibinfo {author} {\bibfnamefont {A.}~\bibnamefont
  {Polnarev}}\ and\ \bibinfo {author} {\bibfnamefont {R.}~\bibnamefont
  {Zembowicz}},\ }\href {\doibase 10.1103/PhysRevD.43.1106} {\bibfield
  {journal} {\bibinfo  {journal} {Phys. Rev. D}\ }\textbf {\bibinfo {volume}
  {43}},\ \bibinfo {pages} {1106} (\bibinfo {year} {1991})}\BibitemShut
  {NoStop}%
\bibitem [{\citenamefont {Caldwell}\ and\ \citenamefont
  {Casper}(1996)}]{Caldwell:1995fu}%
  \BibitemOpen
  \bibfield  {author} {\bibinfo {author} {\bibfnamefont {R.~R.}\ \bibnamefont
  {Caldwell}}\ and\ \bibinfo {author} {\bibfnamefont {P.}~\bibnamefont
  {Casper}},\ }\href {\doibase 10.1103/PhysRevD.53.3002} {\bibfield  {journal}
  {\bibinfo  {journal} {Phys. Rev. D}\ }\textbf {\bibinfo {volume} {53}},\
  \bibinfo {pages} {3002} (\bibinfo {year} {1996})},\ \Eprint
  {http://arxiv.org/abs/gr-qc/9509012} {arXiv:gr-qc/9509012} \BibitemShut
  {NoStop}%
\bibitem [{\citenamefont {MacGibbon}\ \emph {et~al.}(1998)\citenamefont
  {MacGibbon}, \citenamefont {Brandenberger},\ and\ \citenamefont
  {Wichoski}}]{MacGibbon:1997pu}%
  \BibitemOpen
  \bibfield  {author} {\bibinfo {author} {\bibfnamefont {J.~H.}\ \bibnamefont
  {MacGibbon}}, \bibinfo {author} {\bibfnamefont {R.~H.}\ \bibnamefont
  {Brandenberger}}, \ and\ \bibinfo {author} {\bibfnamefont {U.~F.}\
  \bibnamefont {Wichoski}},\ }\href {\doibase 10.1103/PhysRevD.57.2158}
  {\bibfield  {journal} {\bibinfo  {journal} {Phys. Rev. D}\ }\textbf {\bibinfo
  {volume} {57}},\ \bibinfo {pages} {2158} (\bibinfo {year} {1998})},\ \Eprint
  {http://arxiv.org/abs/astro-ph/9707146} {arXiv:astro-ph/9707146} \BibitemShut
  {NoStop}%
\bibitem [{\citenamefont {Helfer}\ \emph {et~al.}(2019)\citenamefont {Helfer},
  \citenamefont {Aurrekoetxea},\ and\ \citenamefont {Lim}}]{Helfer:2018qgv}%
  \BibitemOpen
  \bibfield  {author} {\bibinfo {author} {\bibfnamefont {T.}~\bibnamefont
  {Helfer}}, \bibinfo {author} {\bibfnamefont {J.~C.}\ \bibnamefont
  {Aurrekoetxea}}, \ and\ \bibinfo {author} {\bibfnamefont {E.~A.}\
  \bibnamefont {Lim}},\ }\href {\doibase 10.1103/PhysRevD.99.104028} {\bibfield
   {journal} {\bibinfo  {journal} {Phys. Rev. D}\ }\textbf {\bibinfo {volume}
  {99}},\ \bibinfo {pages} {104028} (\bibinfo {year} {2019})},\ \Eprint
  {http://arxiv.org/abs/1808.06678} {arXiv:1808.06678 [gr-qc]} \BibitemShut
  {NoStop}%
\bibitem [{\citenamefont {Garriga}\ and\ \citenamefont
  {Sakellariadou}(1993)}]{Garriga:1993gj}%
  \BibitemOpen
  \bibfield  {author} {\bibinfo {author} {\bibfnamefont {J.}~\bibnamefont
  {Garriga}}\ and\ \bibinfo {author} {\bibfnamefont {M.}~\bibnamefont
  {Sakellariadou}},\ }\href {\doibase 10.1103/PhysRevD.48.2502} {\bibfield
  {journal} {\bibinfo  {journal} {Phys. Rev. D}\ }\textbf {\bibinfo {volume}
  {48}},\ \bibinfo {pages} {2502} (\bibinfo {year} {1993})},\ \Eprint
  {http://arxiv.org/abs/hep-th/9303024} {arXiv:hep-th/9303024} \BibitemShut
  {NoStop}%
\bibitem [{\citenamefont {James-Turner}\ \emph {et~al.}(2020)\citenamefont
  {James-Turner}, \citenamefont {Weil}, \citenamefont {Green},\ and\
  \citenamefont {Copeland}}]{James-Turner:2019ssu}%
  \BibitemOpen
  \bibfield  {author} {\bibinfo {author} {\bibfnamefont {C.}~\bibnamefont
  {James-Turner}}, \bibinfo {author} {\bibfnamefont {D.~P.~B.}\ \bibnamefont
  {Weil}}, \bibinfo {author} {\bibfnamefont {A.~M.}\ \bibnamefont {Green}}, \
  and\ \bibinfo {author} {\bibfnamefont {E.~J.}\ \bibnamefont {Copeland}},\
  }\href {\doibase 10.1103/PhysRevD.101.123526} {\bibfield  {journal} {\bibinfo
   {journal} {Phys. Rev. D}\ }\textbf {\bibinfo {volume} {101}},\ \bibinfo
  {pages} {123526} (\bibinfo {year} {2020})},\ \Eprint
  {http://arxiv.org/abs/1911.12658} {arXiv:1911.12658 [astro-ph.CO]}
  \BibitemShut {NoStop}%
\bibitem [{\citenamefont {Ferrer}\ \emph {et~al.}(2019)\citenamefont {Ferrer},
  \citenamefont {Masso}, \citenamefont {Panico}, \citenamefont {Pujolas},\ and\
  \citenamefont {Rompineve}}]{Ferrer:2018uiu}%
  \BibitemOpen
  \bibfield  {author} {\bibinfo {author} {\bibfnamefont {F.}~\bibnamefont
  {Ferrer}}, \bibinfo {author} {\bibfnamefont {E.}~\bibnamefont {Masso}},
  \bibinfo {author} {\bibfnamefont {G.}~\bibnamefont {Panico}}, \bibinfo
  {author} {\bibfnamefont {O.}~\bibnamefont {Pujolas}}, \ and\ \bibinfo
  {author} {\bibfnamefont {F.}~\bibnamefont {Rompineve}},\ }\href {\doibase
  10.1103/PhysRevLett.122.101301} {\bibfield  {journal} {\bibinfo  {journal}
  {Phys. Rev. Lett.}\ }\textbf {\bibinfo {volume} {122}},\ \bibinfo {pages}
  {101301} (\bibinfo {year} {2019})},\ \Eprint
  {http://arxiv.org/abs/1807.01707} {arXiv:1807.01707 [hep-ph]} \BibitemShut
  {NoStop}%
\bibitem [{\citenamefont {Liu}\ \emph {et~al.}(2020{\natexlab{a}})\citenamefont
  {Liu}, \citenamefont {Guo},\ and\ \citenamefont {Cai}}]{Liu:2019lul}%
  \BibitemOpen
  \bibfield  {author} {\bibinfo {author} {\bibfnamefont {J.}~\bibnamefont
  {Liu}}, \bibinfo {author} {\bibfnamefont {Z.-K.}\ \bibnamefont {Guo}}, \ and\
  \bibinfo {author} {\bibfnamefont {R.-G.}\ \bibnamefont {Cai}},\ }\href
  {\doibase 10.1103/PhysRevD.101.023513} {\bibfield  {journal} {\bibinfo
  {journal} {Phys. Rev. D}\ }\textbf {\bibinfo {volume} {101}},\ \bibinfo
  {pages} {023513} (\bibinfo {year} {2020}{\natexlab{a}})},\ \Eprint
  {http://arxiv.org/abs/1908.02662} {arXiv:1908.02662 [astro-ph.CO]}
  \BibitemShut {NoStop}%
\bibitem [{\citenamefont {Gouttenoire}\ and\ \citenamefont
  {Volansky}(2024)}]{Gouttenoire:2023naa}%
  \BibitemOpen
  \bibfield  {author} {\bibinfo {author} {\bibfnamefont {Y.}~\bibnamefont
  {Gouttenoire}}\ and\ \bibinfo {author} {\bibfnamefont {T.}~\bibnamefont
  {Volansky}},\ }\href {\doibase 10.1103/PhysRevD.110.043514} {\bibfield
  {journal} {\bibinfo  {journal} {Phys. Rev. D}\ }\textbf {\bibinfo {volume}
  {110}},\ \bibinfo {pages} {043514} (\bibinfo {year} {2024})},\ \Eprint
  {http://arxiv.org/abs/2305.04942} {arXiv:2305.04942 [hep-ph]} \BibitemShut
  {NoStop}%
\bibitem [{\citenamefont {Kitajima}\ \emph {et~al.}(2024)\citenamefont
  {Kitajima}, \citenamefont {Lee}, \citenamefont {Murai}, \citenamefont
  {Takahashi},\ and\ \citenamefont {Yin}}]{Kitajima:2023cek}%
  \BibitemOpen
  \bibfield  {author} {\bibinfo {author} {\bibfnamefont {N.}~\bibnamefont
  {Kitajima}}, \bibinfo {author} {\bibfnamefont {J.}~\bibnamefont {Lee}},
  \bibinfo {author} {\bibfnamefont {K.}~\bibnamefont {Murai}}, \bibinfo
  {author} {\bibfnamefont {F.}~\bibnamefont {Takahashi}}, \ and\ \bibinfo
  {author} {\bibfnamefont {W.}~\bibnamefont {Yin}},\ }\href {\doibase
  10.1016/j.physletb.2024.138586} {\bibfield  {journal} {\bibinfo  {journal}
  {Phys. Lett. B}\ }\textbf {\bibinfo {volume} {851}},\ \bibinfo {pages}
  {138586} (\bibinfo {year} {2024})},\ \Eprint
  {http://arxiv.org/abs/2306.17146} {arXiv:2306.17146 [hep-ph]} \BibitemShut
  {NoStop}%
\bibitem [{\citenamefont {Gouttenoire}\ and\ \citenamefont
  {Vitagliano}(2024)}]{Gouttenoire:2023ftk}%
  \BibitemOpen
  \bibfield  {author} {\bibinfo {author} {\bibfnamefont {Y.}~\bibnamefont
  {Gouttenoire}}\ and\ \bibinfo {author} {\bibfnamefont {E.}~\bibnamefont
  {Vitagliano}},\ }\href {\doibase 10.1103/PhysRevD.110.L061306} {\bibfield
  {journal} {\bibinfo  {journal} {Phys. Rev. D}\ }\textbf {\bibinfo {volume}
  {110}},\ \bibinfo {pages} {L061306} (\bibinfo {year} {2024})},\ \Eprint
  {http://arxiv.org/abs/2306.17841} {arXiv:2306.17841 [gr-qc]} \BibitemShut
  {NoStop}%
\bibitem [{\citenamefont {Ferreira}\ \emph {et~al.}(2024)\citenamefont
  {Ferreira}, \citenamefont {Notari}, \citenamefont {Pujol{\`a}s},\ and\
  \citenamefont {Rompineve}}]{Ferreira:2024eru}%
  \BibitemOpen
  \bibfield  {author} {\bibinfo {author} {\bibfnamefont {R.~Z.}\ \bibnamefont
  {Ferreira}}, \bibinfo {author} {\bibfnamefont {A.}~\bibnamefont {Notari}},
  \bibinfo {author} {\bibfnamefont {O.}~\bibnamefont {Pujol{\`a}s}}, \ and\
  \bibinfo {author} {\bibfnamefont {F.}~\bibnamefont {Rompineve}},\ }\href
  {\doibase 10.1088/1475-7516/2024/06/020} {\bibfield  {journal} {\bibinfo
  {journal} {JCAP}\ }\textbf {\bibinfo {volume} {06}},\ \bibinfo {pages} {020}
  (\bibinfo {year} {2024})},\ \Eprint {http://arxiv.org/abs/2401.14331}
  {arXiv:2401.14331 [astro-ph.CO]} \BibitemShut {NoStop}%
\bibitem [{\citenamefont {Masubuchi}\ \emph {et~al.}(2026)\citenamefont
  {Masubuchi}, \citenamefont {Narita},\ and\ \citenamefont
  {Yin}}]{Masubuchi:2026eau}%
  \BibitemOpen
  \bibfield  {author} {\bibinfo {author} {\bibfnamefont {H.}~\bibnamefont
  {Masubuchi}}, \bibinfo {author} {\bibfnamefont {Y.}~\bibnamefont {Narita}}, \
  and\ \bibinfo {author} {\bibfnamefont {W.}~\bibnamefont {Yin}},\ }\href@noop
  {} {\  (\bibinfo {year} {2026})},\ \Eprint {http://arxiv.org/abs/2602.15825}
  {arXiv:2602.15825 [hep-ph]} \BibitemShut {NoStop}%
\bibitem [{\citenamefont {Sugeno}\ and\ \citenamefont
  {Yin}(2026)}]{Sugeno:2025kwx}%
  \BibitemOpen
  \bibfield  {author} {\bibinfo {author} {\bibfnamefont {T.}~\bibnamefont
  {Sugeno}}\ and\ \bibinfo {author} {\bibfnamefont {W.}~\bibnamefont {Yin}},\
  }\href {\doibase 10.1007/JHEP04(2026)108} {\bibfield  {journal} {\bibinfo
  {journal} {JHEP}\ }\textbf {\bibinfo {volume} {04}},\ \bibinfo {pages} {108}
  (\bibinfo {year} {2026})},\ \Eprint {http://arxiv.org/abs/2511.19429}
  {arXiv:2511.19429 [hep-ph]} \BibitemShut {NoStop}%
\bibitem [{\citenamefont {Miyazaki}\ \emph {et~al.}(2026)\citenamefont
  {Miyazaki}, \citenamefont {Narita}, \citenamefont {Song}, \citenamefont
  {Yaginuma},\ and\ \citenamefont {Yin}}]{Miyazaki:2025tvq}%
  \BibitemOpen
  \bibfield  {author} {\bibinfo {author} {\bibfnamefont {M.}~\bibnamefont
  {Miyazaki}}, \bibinfo {author} {\bibfnamefont {Y.}~\bibnamefont {Narita}},
  \bibinfo {author} {\bibfnamefont {D.}~\bibnamefont {Song}}, \bibinfo {author}
  {\bibfnamefont {N.}~\bibnamefont {Yaginuma}}, \ and\ \bibinfo {author}
  {\bibfnamefont {W.}~\bibnamefont {Yin}},\ }\href {\doibase
  10.1007/JHEP05(2026)119} {\bibfield  {journal} {\bibinfo  {journal} {JHEP}\
  }\textbf {\bibinfo {volume} {05}},\ \bibinfo {pages} {119} (\bibinfo {year}
  {2026})},\ \Eprint {http://arxiv.org/abs/2509.13292} {arXiv:2509.13292
  [hep-ph]} \BibitemShut {NoStop}%
\bibitem [{\citenamefont {Matsuda}(2006)}]{Matsuda:2005ez}%
  \BibitemOpen
  \bibfield  {author} {\bibinfo {author} {\bibfnamefont {T.}~\bibnamefont
  {Matsuda}},\ }\href {\doibase 10.1088/1126-6708/2006/04/017} {\bibfield
  {journal} {\bibinfo  {journal} {JHEP}\ }\textbf {\bibinfo {volume} {04}},\
  \bibinfo {pages} {017} (\bibinfo {year} {2006})},\ \Eprint
  {http://arxiv.org/abs/hep-ph/0509062} {arXiv:hep-ph/0509062} \BibitemShut
  {NoStop}%
\bibitem [{\citenamefont {Bekenstein}(1972)}]{Bekenstein:1971hc}%
  \BibitemOpen
  \bibfield  {author} {\bibinfo {author} {\bibfnamefont {J.~D.}\ \bibnamefont
  {Bekenstein}},\ }\href {\doibase 10.1103/PhysRevD.5.1239} {\bibfield
  {journal} {\bibinfo  {journal} {Phys. Rev. D}\ }\textbf {\bibinfo {volume}
  {5}},\ \bibinfo {pages} {1239} (\bibinfo {year} {1972})}\BibitemShut
  {NoStop}%
\bibitem [{\citenamefont {Ruffini}\ and\ \citenamefont
  {Wheeler}(1971)}]{Ruffini:1971bza}%
  \BibitemOpen
  \bibfield  {author} {\bibinfo {author} {\bibfnamefont {R.}~\bibnamefont
  {Ruffini}}\ and\ \bibinfo {author} {\bibfnamefont {J.~A.}\ \bibnamefont
  {Wheeler}},\ }\href {\doibase 10.1063/1.3022513} {\bibfield  {journal}
  {\bibinfo  {journal} {Phys. Today}\ }\textbf {\bibinfo {volume} {24}},\
  \bibinfo {pages} {30} (\bibinfo {year} {1971})}\BibitemShut {NoStop}%
\bibitem [{\citenamefont {Bai}\ and\ \citenamefont
  {Orlofsky}(2020)}]{Bai:2019zcd}%
  \BibitemOpen
  \bibfield  {author} {\bibinfo {author} {\bibfnamefont {Y.}~\bibnamefont
  {Bai}}\ and\ \bibinfo {author} {\bibfnamefont {N.}~\bibnamefont {Orlofsky}},\
  }\href {\doibase 10.1103/PhysRevD.101.055006} {\bibfield  {journal} {\bibinfo
   {journal} {Phys. Rev. D}\ }\textbf {\bibinfo {volume} {101}},\ \bibinfo
  {pages} {055006} (\bibinfo {year} {2020})},\ \Eprint
  {http://arxiv.org/abs/1906.04858} {arXiv:1906.04858 [hep-ph]} \BibitemShut
  {NoStop}%
\bibitem [{\citenamefont {Liu}\ \emph {et~al.}(2020{\natexlab{b}})\citenamefont
  {Liu}, \citenamefont {Christiansen}, \citenamefont {Guo}, \citenamefont
  {Cai},\ and\ \citenamefont {Kim}}]{Liu:2020vsy}%
  \BibitemOpen
  \bibfield  {author} {\bibinfo {author} {\bibfnamefont {L.}~\bibnamefont
  {Liu}}, \bibinfo {author} {\bibfnamefont {O.}~\bibnamefont {Christiansen}},
  \bibinfo {author} {\bibfnamefont {Z.-K.}\ \bibnamefont {Guo}}, \bibinfo
  {author} {\bibfnamefont {R.-G.}\ \bibnamefont {Cai}}, \ and\ \bibinfo
  {author} {\bibfnamefont {S.~P.}\ \bibnamefont {Kim}},\ }\href {\doibase
  10.1103/PhysRevD.102.103520} {\bibfield  {journal} {\bibinfo  {journal}
  {Phys. Rev. D}\ }\textbf {\bibinfo {volume} {102}},\ \bibinfo {pages}
  {103520} (\bibinfo {year} {2020}{\natexlab{b}})},\ \Eprint
  {http://arxiv.org/abs/2008.02326} {arXiv:2008.02326 [gr-qc]} \BibitemShut
  {NoStop}%
\bibitem [{\citenamefont {Liu}\ \emph {et~al.}(2020{\natexlab{c}})\citenamefont
  {Liu}, \citenamefont {Guo}, \citenamefont {Cai},\ and\ \citenamefont
  {Kim}}]{Liu:2020cds}%
  \BibitemOpen
  \bibfield  {author} {\bibinfo {author} {\bibfnamefont {L.}~\bibnamefont
  {Liu}}, \bibinfo {author} {\bibfnamefont {Z.-K.}\ \bibnamefont {Guo}},
  \bibinfo {author} {\bibfnamefont {R.-G.}\ \bibnamefont {Cai}}, \ and\
  \bibinfo {author} {\bibfnamefont {S.~P.}\ \bibnamefont {Kim}},\ }\href
  {\doibase 10.1103/PhysRevD.102.043508} {\bibfield  {journal} {\bibinfo
  {journal} {Phys. Rev. D}\ }\textbf {\bibinfo {volume} {102}},\ \bibinfo
  {pages} {043508} (\bibinfo {year} {2020}{\natexlab{c}})},\ \Eprint
  {http://arxiv.org/abs/2001.02984} {arXiv:2001.02984 [astro-ph.CO]}
  \BibitemShut {NoStop}%
\bibitem [{\citenamefont {Bozzola}\ and\ \citenamefont
  {Paschalidis}(2021)}]{Bozzola:2020mjx}%
  \BibitemOpen
  \bibfield  {author} {\bibinfo {author} {\bibfnamefont {G.}~\bibnamefont
  {Bozzola}}\ and\ \bibinfo {author} {\bibfnamefont {V.}~\bibnamefont
  {Paschalidis}},\ }\href {\doibase 10.1103/PhysRevLett.126.041103} {\bibfield
  {journal} {\bibinfo  {journal} {Phys. Rev. Lett.}\ }\textbf {\bibinfo
  {volume} {126}},\ \bibinfo {pages} {041103} (\bibinfo {year} {2021})},\
  \Eprint {http://arxiv.org/abs/2006.15764} {arXiv:2006.15764 [gr-qc]}
  \BibitemShut {NoStop}%
\bibitem [{\citenamefont {Maldacena}(2021)}]{Maldacena:2020skw}%
  \BibitemOpen
  \bibfield  {author} {\bibinfo {author} {\bibfnamefont {J.}~\bibnamefont
  {Maldacena}},\ }\href {\doibase 10.1007/JHEP04(2021)079} {\bibfield
  {journal} {\bibinfo  {journal} {JHEP}\ }\textbf {\bibinfo {volume} {04}},\
  \bibinfo {pages} {079} (\bibinfo {year} {2021})},\ \Eprint
  {http://arxiv.org/abs/2004.06084} {arXiv:2004.06084 [hep-th]} \BibitemShut
  {NoStop}%
\bibitem [{\citenamefont {Yin}(2025)}]{Yin:2024txg}%
  \BibitemOpen
  \bibfield  {author} {\bibinfo {author} {\bibfnamefont {W.}~\bibnamefont
  {Yin}},\ }\href {\doibase 10.1007/JHEP10(2025)177} {\bibfield  {journal}
  {\bibinfo  {journal} {JHEP}\ }\textbf {\bibinfo {volume} {10}},\ \bibinfo
  {pages} {177} (\bibinfo {year} {2025})},\ \Eprint
  {http://arxiv.org/abs/2412.17802} {arXiv:2412.17802 [hep-ph]} \BibitemShut
  {NoStop}%
\bibitem [{\citenamefont {Yin}(2024)}]{Yin:2024pri}%
  \BibitemOpen
  \bibfield  {author} {\bibinfo {author} {\bibfnamefont {W.}~\bibnamefont
  {Yin}},\ }\href {\doibase 10.1093/ptep/ptaf053} {\  (\bibinfo {year}
  {2024}),\ 10.1093/ptep/ptaf053},\ \Eprint {http://arxiv.org/abs/2412.19798}
  {arXiv:2412.19798 [hep-ph]} \BibitemShut {NoStop}%
\bibitem [{\citenamefont {Barriola}\ and\ \citenamefont
  {Vilenkin}(1989)}]{Barriola:1989hx}%
  \BibitemOpen
  \bibfield  {author} {\bibinfo {author} {\bibfnamefont {M.}~\bibnamefont
  {Barriola}}\ and\ \bibinfo {author} {\bibfnamefont {A.}~\bibnamefont
  {Vilenkin}},\ }\href {\doibase 10.1103/PhysRevLett.63.341} {\bibfield
  {journal} {\bibinfo  {journal} {Phys. Rev. Lett.}\ }\textbf {\bibinfo
  {volume} {63}},\ \bibinfo {pages} {341} (\bibinfo {year} {1989})}\BibitemShut
  {NoStop}%
\bibitem [{\citenamefont {Vilenkin}\ and\ \citenamefont
  {Shellard}(2000)}]{Vilenkin:2000jqa}%
  \BibitemOpen
  \bibfield  {author} {\bibinfo {author} {\bibfnamefont {A.}~\bibnamefont
  {Vilenkin}}\ and\ \bibinfo {author} {\bibfnamefont {E.~P.~S.}\ \bibnamefont
  {Shellard}},\ }\href@noop {} {\emph {\bibinfo {title} {{Cosmic Strings and
  Other Topological Defects}}}}\ (\bibinfo  {publisher} {Cambridge University
  Press},\ \bibinfo {year} {2000})\BibitemShut {NoStop}%
\bibitem [{\citenamefont {Kibble}(1976)}]{Kibble:1976sj}%
  \BibitemOpen
  \bibfield  {author} {\bibinfo {author} {\bibfnamefont {T.~W.~B.}\
  \bibnamefont {Kibble}},\ }\href {\doibase 10.1088/0305-4470/9/8/029}
  {\bibfield  {journal} {\bibinfo  {journal} {J. Phys. A}\ }\textbf {\bibinfo
  {volume} {9}},\ \bibinfo {pages} {1387} (\bibinfo {year} {1976})}\BibitemShut
  {NoStop}%
\bibitem [{\citenamefont {Bennett}\ and\ \citenamefont
  {Rhie}(1990)}]{Bennett:1990xy}%
  \BibitemOpen
  \bibfield  {author} {\bibinfo {author} {\bibfnamefont {D.~P.}\ \bibnamefont
  {Bennett}}\ and\ \bibinfo {author} {\bibfnamefont {S.~H.}\ \bibnamefont
  {Rhie}},\ }\href {\doibase 10.1103/PhysRevLett.65.1709} {\bibfield  {journal}
  {\bibinfo  {journal} {Phys. Rev. Lett.}\ }\textbf {\bibinfo {volume} {65}},\
  \bibinfo {pages} {1709} (\bibinfo {year} {1990})}\BibitemShut {NoStop}%
\bibitem [{\citenamefont {Yamaguchi}(2001)}]{Yamaguchi:2001rf}%
  \BibitemOpen
  \bibfield  {author} {\bibinfo {author} {\bibfnamefont {M.}~\bibnamefont
  {Yamaguchi}},\ }\href {\doibase 10.1103/PhysRevD.64.081301} {\bibfield
  {journal} {\bibinfo  {journal} {Phys. Rev. D}\ }\textbf {\bibinfo {volume}
  {64}},\ \bibinfo {pages} {081301} (\bibinfo {year} {2001})},\ \Eprint
  {http://arxiv.org/abs/hep-ph/0103130} {arXiv:hep-ph/0103130} \BibitemShut
  {NoStop}%
\bibitem [{\citenamefont {Martins}\ and\ \citenamefont
  {Achucarro}(2008)}]{Martins:2008ks}%
  \BibitemOpen
  \bibfield  {author} {\bibinfo {author} {\bibfnamefont {C.~J. A.~P.}\
  \bibnamefont {Martins}}\ and\ \bibinfo {author} {\bibfnamefont
  {A.}~\bibnamefont {Achucarro}},\ }\href {\doibase 10.1103/PhysRevD.78.083541}
  {\bibfield  {journal} {\bibinfo  {journal} {Phys. Rev. D}\ }\textbf {\bibinfo
  {volume} {78}},\ \bibinfo {pages} {083541} (\bibinfo {year} {2008})},\
  \Eprint {http://arxiv.org/abs/0806.2671} {arXiv:0806.2671 [hep-ph]}
  \BibitemShut {NoStop}%
\bibitem [{\citenamefont {Jaeckel}\ and\ \citenamefont
  {Ringwald}(2010)}]{Jaeckel:2010ni}%
  \BibitemOpen
  \bibfield  {author} {\bibinfo {author} {\bibfnamefont {J.}~\bibnamefont
  {Jaeckel}}\ and\ \bibinfo {author} {\bibfnamefont {A.}~\bibnamefont
  {Ringwald}},\ }\href {\doibase 10.1146/annurev.nucl.012809.104433} {\bibfield
   {journal} {\bibinfo  {journal} {Ann. Rev. Nucl. Part. Sci.}\ }\textbf
  {\bibinfo {volume} {60}},\ \bibinfo {pages} {405} (\bibinfo {year} {2010})},\
  \Eprint {http://arxiv.org/abs/1002.0329} {arXiv:1002.0329 [hep-ph]}
  \BibitemShut {NoStop}%
\bibitem [{\citenamefont {Ringwald}(2012)}]{Ringwald:2012hr}%
  \BibitemOpen
  \bibfield  {author} {\bibinfo {author} {\bibfnamefont {A.}~\bibnamefont
  {Ringwald}},\ }\href {\doibase 10.1016/j.dark.2012.10.008} {\bibfield
  {journal} {\bibinfo  {journal} {Phys. Dark Univ.}\ }\textbf {\bibinfo
  {volume} {1}},\ \bibinfo {pages} {116} (\bibinfo {year} {2012})},\ \Eprint
  {http://arxiv.org/abs/1210.5081} {arXiv:1210.5081 [hep-ph]} \BibitemShut
  {NoStop}%
\bibitem [{\citenamefont {Arias}\ \emph {et~al.}(2012)\citenamefont {Arias},
  \citenamefont {Cadamuro}, \citenamefont {Goodsell}, \citenamefont {Jaeckel},
  \citenamefont {Redondo},\ and\ \citenamefont {Ringwald}}]{Arias:2012az}%
  \BibitemOpen
  \bibfield  {author} {\bibinfo {author} {\bibfnamefont {P.}~\bibnamefont
  {Arias}}, \bibinfo {author} {\bibfnamefont {D.}~\bibnamefont {Cadamuro}},
  \bibinfo {author} {\bibfnamefont {M.}~\bibnamefont {Goodsell}}, \bibinfo
  {author} {\bibfnamefont {J.}~\bibnamefont {Jaeckel}}, \bibinfo {author}
  {\bibfnamefont {J.}~\bibnamefont {Redondo}}, \ and\ \bibinfo {author}
  {\bibfnamefont {A.}~\bibnamefont {Ringwald}},\ }\href {\doibase
  10.1088/1475-7516/2012/06/013} {\bibfield  {journal} {\bibinfo  {journal}
  {JCAP}\ }\textbf {\bibinfo {volume} {06}},\ \bibinfo {pages} {013} (\bibinfo
  {year} {2012})},\ \Eprint {http://arxiv.org/abs/1201.5902} {arXiv:1201.5902
  [hep-ph]} \BibitemShut {NoStop}%
\bibitem [{\citenamefont {Graham}\ \emph {et~al.}(2015)\citenamefont {Graham},
  \citenamefont {Irastorza}, \citenamefont {Lamoreaux}, \citenamefont
  {Lindner},\ and\ \citenamefont {van Bibber}}]{Graham:2015ouw}%
  \BibitemOpen
  \bibfield  {author} {\bibinfo {author} {\bibfnamefont {P.~W.}\ \bibnamefont
  {Graham}}, \bibinfo {author} {\bibfnamefont {I.~G.}\ \bibnamefont
  {Irastorza}}, \bibinfo {author} {\bibfnamefont {S.~K.}\ \bibnamefont
  {Lamoreaux}}, \bibinfo {author} {\bibfnamefont {A.}~\bibnamefont {Lindner}},
  \ and\ \bibinfo {author} {\bibfnamefont {K.~A.}\ \bibnamefont {van Bibber}},\
  }\href {\doibase 10.1146/annurev-nucl-102014-022120} {\bibfield  {journal}
  {\bibinfo  {journal} {Ann. Rev. Nucl. Part. Sci.}\ }\textbf {\bibinfo
  {volume} {65}},\ \bibinfo {pages} {485} (\bibinfo {year} {2015})},\ \Eprint
  {http://arxiv.org/abs/1602.00039} {arXiv:1602.00039 [hep-ex]} \BibitemShut
  {NoStop}%
\bibitem [{\citenamefont {Marsh}(2016)}]{Marsh:2015xka}%
  \BibitemOpen
  \bibfield  {author} {\bibinfo {author} {\bibfnamefont {D.~J.~E.}\
  \bibnamefont {Marsh}},\ }\href {\doibase 10.1016/j.physrep.2016.06.005}
  {\bibfield  {journal} {\bibinfo  {journal} {Phys. Rept.}\ }\textbf {\bibinfo
  {volume} {643}},\ \bibinfo {pages} {1} (\bibinfo {year} {2016})},\ \Eprint
  {http://arxiv.org/abs/1510.07633} {arXiv:1510.07633 [astro-ph.CO]}
  \BibitemShut {NoStop}%
\bibitem [{\citenamefont {Irastorza}\ and\ \citenamefont
  {Redondo}(2018)}]{Irastorza:2018dyq}%
  \BibitemOpen
  \bibfield  {author} {\bibinfo {author} {\bibfnamefont {I.~G.}\ \bibnamefont
  {Irastorza}}\ and\ \bibinfo {author} {\bibfnamefont {J.}~\bibnamefont
  {Redondo}},\ }\href {\doibase 10.1016/j.ppnp.2018.05.003} {\bibfield
  {journal} {\bibinfo  {journal} {Prog. Part. Nucl. Phys.}\ }\textbf {\bibinfo
  {volume} {102}},\ \bibinfo {pages} {89} (\bibinfo {year} {2018})},\ \Eprint
  {http://arxiv.org/abs/1801.08127} {arXiv:1801.08127 [hep-ph]} \BibitemShut
  {NoStop}%
\bibitem [{\citenamefont {Di~Luzio}\ \emph {et~al.}(2020)\citenamefont
  {Di~Luzio}, \citenamefont {Giannotti}, \citenamefont {Nardi},\ and\
  \citenamefont {Visinelli}}]{DiLuzio:2020wdo}%
  \BibitemOpen
  \bibfield  {author} {\bibinfo {author} {\bibfnamefont {L.}~\bibnamefont
  {Di~Luzio}}, \bibinfo {author} {\bibfnamefont {M.}~\bibnamefont {Giannotti}},
  \bibinfo {author} {\bibfnamefont {E.}~\bibnamefont {Nardi}}, \ and\ \bibinfo
  {author} {\bibfnamefont {L.}~\bibnamefont {Visinelli}},\ }\href {\doibase
  10.1016/j.physrep.2020.06.002} {\bibfield  {journal} {\bibinfo  {journal}
  {Phys. Rept.}\ }\textbf {\bibinfo {volume} {870}},\ \bibinfo {pages} {1}
  (\bibinfo {year} {2020})},\ \Eprint {http://arxiv.org/abs/2003.01100}
  {arXiv:2003.01100 [hep-ph]} \BibitemShut {NoStop}%
\bibitem [{\citenamefont {Albertus}\ \emph {et~al.}(2026)\citenamefont
  {Albertus} \emph {et~al.}}]{Albertus:2026fbe}%
  \BibitemOpen
  \bibfield  {author} {\bibinfo {author} {\bibfnamefont {C.}~\bibnamefont
  {Albertus}} \emph {et~al.},\ }\href@noop {} {\  (\bibinfo {year} {2026})},\
  \Eprint {http://arxiv.org/abs/2602.09089} {arXiv:2602.09089 [hep-ph]}
  \BibitemShut {NoStop}%
\bibitem [{\citenamefont {Arza}\ \emph {et~al.}(2026)\citenamefont {Arza} \emph
  {et~al.}}]{Arza:2026rsl}%
  \BibitemOpen
  \bibfield  {author} {\bibinfo {author} {\bibfnamefont {A.}~\bibnamefont
  {Arza}} \emph {et~al.},\ }\href@noop {} {\  (\bibinfo {year} {2026})},\
  \Eprint {http://arxiv.org/abs/2603.03433} {arXiv:2603.03433 [hep-ph]}
  \BibitemShut {NoStop}%
\bibitem [{\citenamefont {Figueroa}\ \emph {et~al.}(2021)\citenamefont
  {Figueroa}, \citenamefont {Florio}, \citenamefont {Torrenti},\ and\
  \citenamefont {Valkenburg}}]{Figueroa:2020rrl}%
  \BibitemOpen
  \bibfield  {author} {\bibinfo {author} {\bibfnamefont {D.~G.}\ \bibnamefont
  {Figueroa}}, \bibinfo {author} {\bibfnamefont {A.}~\bibnamefont {Florio}},
  \bibinfo {author} {\bibfnamefont {F.}~\bibnamefont {Torrenti}}, \ and\
  \bibinfo {author} {\bibfnamefont {W.}~\bibnamefont {Valkenburg}},\ }\href
  {\doibase 10.1088/1475-7516/2021/04/035} {\bibfield  {journal} {\bibinfo
  {journal} {JCAP}\ }\textbf {\bibinfo {volume} {04}},\ \bibinfo {pages} {035}
  (\bibinfo {year} {2021})},\ \Eprint {http://arxiv.org/abs/2006.15122}
  {arXiv:2006.15122 [astro-ph.CO]} \BibitemShut {NoStop}%
\bibitem [{\citenamefont {Figueroa}\ \emph {et~al.}(2023)\citenamefont
  {Figueroa}, \citenamefont {Florio}, \citenamefont {Torrenti},\ and\
  \citenamefont {Valkenburg}}]{Figueroa:2021yhd}%
  \BibitemOpen
  \bibfield  {author} {\bibinfo {author} {\bibfnamefont {D.~G.}\ \bibnamefont
  {Figueroa}}, \bibinfo {author} {\bibfnamefont {A.}~\bibnamefont {Florio}},
  \bibinfo {author} {\bibfnamefont {F.}~\bibnamefont {Torrenti}}, \ and\
  \bibinfo {author} {\bibfnamefont {W.}~\bibnamefont {Valkenburg}},\ }\href
  {\doibase 10.1016/j.cpc.2022.108586} {\bibfield  {journal} {\bibinfo
  {journal} {Comput. Phys. Commun.}\ }\textbf {\bibinfo {volume} {283}},\
  \bibinfo {pages} {108586} (\bibinfo {year} {2023})},\ \Eprint
  {http://arxiv.org/abs/2102.01031} {arXiv:2102.01031 [astro-ph.CO]}
  \BibitemShut {NoStop}%
\bibitem [{\citenamefont {Harada}\ \emph {et~al.}(2013)\citenamefont {Harada},
  \citenamefont {Yoo},\ and\ \citenamefont {Kohri}}]{Harada:2013epa}%
  \BibitemOpen
  \bibfield  {author} {\bibinfo {author} {\bibfnamefont {T.}~\bibnamefont
  {Harada}}, \bibinfo {author} {\bibfnamefont {C.-M.}\ \bibnamefont {Yoo}}, \
  and\ \bibinfo {author} {\bibfnamefont {K.}~\bibnamefont {Kohri}},\ }\href
  {\doibase 10.1103/PhysRevD.88.084051} {\bibfield  {journal} {\bibinfo
  {journal} {Phys. Rev. D}\ }\textbf {\bibinfo {volume} {88}},\ \bibinfo
  {pages} {084051} (\bibinfo {year} {2013})},\ \bibinfo {note} {[Erratum:
  Phys.Rev.D 89, 029903 (2014)]},\ \Eprint {http://arxiv.org/abs/1309.4201}
  {arXiv:1309.4201 [astro-ph.CO]} \BibitemShut {NoStop}%
\bibitem [{\citenamefont {Escriv{\`a}}\ \emph {et~al.}(2020)\citenamefont
  {Escriv{\`a}}, \citenamefont {Germani},\ and\ \citenamefont
  {Sheth}}]{Escriva:2019phb}%
  \BibitemOpen
  \bibfield  {author} {\bibinfo {author} {\bibfnamefont {A.}~\bibnamefont
  {Escriv{\`a}}}, \bibinfo {author} {\bibfnamefont {C.}~\bibnamefont
  {Germani}}, \ and\ \bibinfo {author} {\bibfnamefont {R.~K.}\ \bibnamefont
  {Sheth}},\ }\href {\doibase 10.1103/PhysRevD.101.044022} {\bibfield
  {journal} {\bibinfo  {journal} {Phys. Rev. D}\ }\textbf {\bibinfo {volume}
  {101}},\ \bibinfo {pages} {044022} (\bibinfo {year} {2020})},\ \Eprint
  {http://arxiv.org/abs/1907.13311} {arXiv:1907.13311 [gr-qc]} \BibitemShut
  {NoStop}%
\bibitem [{\citenamefont {Khlopov}\ and\ \citenamefont
  {Polnarev}(1980)}]{Khlopov:1980mg}%
  \BibitemOpen
  \bibfield  {author} {\bibinfo {author} {\bibfnamefont {M.~Y.}\ \bibnamefont
  {Khlopov}}\ and\ \bibinfo {author} {\bibfnamefont {A.~G.}\ \bibnamefont
  {Polnarev}},\ }\href {\doibase 10.1016/0370-2693(80)90624-3} {\bibfield
  {journal} {\bibinfo  {journal} {Phys. Lett. B}\ }\textbf {\bibinfo {volume}
  {97}},\ \bibinfo {pages} {383} (\bibinfo {year} {1980})}\BibitemShut
  {NoStop}%
\bibitem [{\citenamefont {Harada}\ \emph {et~al.}(2016)\citenamefont {Harada},
  \citenamefont {Yoo}, \citenamefont {Kohri}, \citenamefont {Nakao},\ and\
  \citenamefont {Jhingan}}]{Harada:2016mhb}%
  \BibitemOpen
  \bibfield  {author} {\bibinfo {author} {\bibfnamefont {T.}~\bibnamefont
  {Harada}}, \bibinfo {author} {\bibfnamefont {C.-M.}\ \bibnamefont {Yoo}},
  \bibinfo {author} {\bibfnamefont {K.}~\bibnamefont {Kohri}}, \bibinfo
  {author} {\bibfnamefont {K.-i.}\ \bibnamefont {Nakao}}, \ and\ \bibinfo
  {author} {\bibfnamefont {S.}~\bibnamefont {Jhingan}},\ }\href {\doibase
  10.3847/1538-4357/833/1/61} {\bibfield  {journal} {\bibinfo  {journal}
  {Astrophys. J.}\ }\textbf {\bibinfo {volume} {833}},\ \bibinfo {pages} {61}
  (\bibinfo {year} {2016})},\ \Eprint {http://arxiv.org/abs/1609.01588}
  {arXiv:1609.01588 [astro-ph.CO]} \BibitemShut {NoStop}%
\bibitem [{\citenamefont {Harada}\ \emph {et~al.}(2017)\citenamefont {Harada},
  \citenamefont {Yoo}, \citenamefont {Kohri},\ and\ \citenamefont
  {Nakao}}]{Harada:2017fjm}%
  \BibitemOpen
  \bibfield  {author} {\bibinfo {author} {\bibfnamefont {T.}~\bibnamefont
  {Harada}}, \bibinfo {author} {\bibfnamefont {C.-M.}\ \bibnamefont {Yoo}},
  \bibinfo {author} {\bibfnamefont {K.}~\bibnamefont {Kohri}}, \ and\ \bibinfo
  {author} {\bibfnamefont {K.-I.}\ \bibnamefont {Nakao}},\ }\href {\doibase
  10.1103/PhysRevD.96.083517} {\bibfield  {journal} {\bibinfo  {journal} {Phys.
  Rev. D}\ }\textbf {\bibinfo {volume} {96}},\ \bibinfo {pages} {083517}
  (\bibinfo {year} {2017})},\ \bibinfo {note} {[Erratum: Phys.Rev.D 99, 069904
  (2019)]},\ \Eprint {http://arxiv.org/abs/1707.03595} {arXiv:1707.03595
  [gr-qc]} \BibitemShut {NoStop}%
\bibitem [{\citenamefont {Sakai}\ \emph {et~al.}(1996)\citenamefont {Sakai},
  \citenamefont {Shinkai}, \citenamefont {Tachizawa},\ and\ \citenamefont
  {Maeda}}]{Sakai:1995nh}%
  \BibitemOpen
  \bibfield  {author} {\bibinfo {author} {\bibfnamefont {N.}~\bibnamefont
  {Sakai}}, \bibinfo {author} {\bibfnamefont {H.-A.}\ \bibnamefont {Shinkai}},
  \bibinfo {author} {\bibfnamefont {T.}~\bibnamefont {Tachizawa}}, \ and\
  \bibinfo {author} {\bibfnamefont {K.-i.}\ \bibnamefont {Maeda}},\ }\href
  {\doibase 10.1103/PhysRevD.53.655} {\bibfield  {journal} {\bibinfo  {journal}
  {Phys. Rev. D}\ }\textbf {\bibinfo {volume} {53}},\ \bibinfo {pages} {655}
  (\bibinfo {year} {1996})},\ \bibinfo {note} {[Erratum: Phys.Rev.D 54, 2981
  (1996)]},\ \Eprint {http://arxiv.org/abs/gr-qc/9506068} {arXiv:gr-qc/9506068}
  \BibitemShut {NoStop}%
\bibitem [{\citenamefont {Stojkovic}\ and\ \citenamefont
  {Freese}(2005)}]{Stojkovic:2004hz}%
  \BibitemOpen
  \bibfield  {author} {\bibinfo {author} {\bibfnamefont {D.}~\bibnamefont
  {Stojkovic}}\ and\ \bibinfo {author} {\bibfnamefont {K.}~\bibnamefont
  {Freese}},\ }\href {\doibase 10.1016/j.physletb.2004.12.019} {\bibfield
  {journal} {\bibinfo  {journal} {Phys. Lett. B}\ }\textbf {\bibinfo {volume}
  {606}},\ \bibinfo {pages} {251} (\bibinfo {year} {2005})},\ \Eprint
  {http://arxiv.org/abs/hep-ph/0403248} {arXiv:hep-ph/0403248} \BibitemShut
  {NoStop}%
\bibitem [{\citenamefont {Carr}\ \emph {et~al.}(2017)\citenamefont {Carr},
  \citenamefont {Raidal}, \citenamefont {Tenkanen}, \citenamefont {Vaskonen},\
  and\ \citenamefont {Veerm{\"a}e}}]{Carr:2017jsz}%
  \BibitemOpen
  \bibfield  {author} {\bibinfo {author} {\bibfnamefont {B.}~\bibnamefont
  {Carr}}, \bibinfo {author} {\bibfnamefont {M.}~\bibnamefont {Raidal}},
  \bibinfo {author} {\bibfnamefont {T.}~\bibnamefont {Tenkanen}}, \bibinfo
  {author} {\bibfnamefont {V.}~\bibnamefont {Vaskonen}}, \ and\ \bibinfo
  {author} {\bibfnamefont {H.}~\bibnamefont {Veerm{\"a}e}},\ }\href {\doibase
  10.1103/PhysRevD.96.023514} {\bibfield  {journal} {\bibinfo  {journal} {Phys.
  Rev. D}\ }\textbf {\bibinfo {volume} {96}},\ \bibinfo {pages} {023514}
  (\bibinfo {year} {2017})},\ \Eprint {http://arxiv.org/abs/1705.05567}
  {arXiv:1705.05567 [astro-ph.CO]} \BibitemShut {NoStop}%
\bibitem [{\citenamefont {Sugiyama}\ \emph {et~al.}(2026)\citenamefont
  {Sugiyama}, \citenamefont {Takada}, \citenamefont {Yasuda},\ and\
  \citenamefont {Tominaga}}]{Sugiyama:2026kpv}%
  \BibitemOpen
  \bibfield  {author} {\bibinfo {author} {\bibfnamefont {S.}~\bibnamefont
  {Sugiyama}}, \bibinfo {author} {\bibfnamefont {M.}~\bibnamefont {Takada}},
  \bibinfo {author} {\bibfnamefont {N.}~\bibnamefont {Yasuda}}, \ and\ \bibinfo
  {author} {\bibfnamefont {N.}~\bibnamefont {Tominaga}},\ }\href@noop {} {\
  (\bibinfo {year} {2026})},\ \Eprint {http://arxiv.org/abs/2602.05840}
  {arXiv:2602.05840 [astro-ph.CO]} \BibitemShut {NoStop}%
\bibitem [{\citenamefont {Niikura}\ \emph {et~al.}(2019)\citenamefont
  {Niikura}, \citenamefont {Takada}, \citenamefont {Yokoyama}, \citenamefont
  {Sumi},\ and\ \citenamefont {Masaki}}]{Niikura:2019kqi}%
  \BibitemOpen
  \bibfield  {author} {\bibinfo {author} {\bibfnamefont {H.}~\bibnamefont
  {Niikura}}, \bibinfo {author} {\bibfnamefont {M.}~\bibnamefont {Takada}},
  \bibinfo {author} {\bibfnamefont {S.}~\bibnamefont {Yokoyama}}, \bibinfo
  {author} {\bibfnamefont {T.}~\bibnamefont {Sumi}}, \ and\ \bibinfo {author}
  {\bibfnamefont {S.}~\bibnamefont {Masaki}},\ }\href {\doibase
  10.1103/PhysRevD.99.083503} {\bibfield  {journal} {\bibinfo  {journal} {Phys.
  Rev. D}\ }\textbf {\bibinfo {volume} {99}},\ \bibinfo {pages} {083503}
  (\bibinfo {year} {2019})},\ \Eprint {http://arxiv.org/abs/1901.07120}
  {arXiv:1901.07120 [astro-ph.CO]} \BibitemShut {NoStop}%
\bibitem [{\citenamefont {Smirnov}\ \emph {et~al.}(2024)\citenamefont
  {Smirnov}, \citenamefont {Goobar}, \citenamefont {Linden},\ and\
  \citenamefont {M{\"o}rtsell}}]{Smirnov:2022zip}%
  \BibitemOpen
  \bibfield  {author} {\bibinfo {author} {\bibfnamefont {J.}~\bibnamefont
  {Smirnov}}, \bibinfo {author} {\bibfnamefont {A.}~\bibnamefont {Goobar}},
  \bibinfo {author} {\bibfnamefont {T.}~\bibnamefont {Linden}}, \ and\ \bibinfo
  {author} {\bibfnamefont {E.}~\bibnamefont {M{\"o}rtsell}},\ }\href {\doibase
  10.1103/PhysRevLett.132.151401} {\bibfield  {journal} {\bibinfo  {journal}
  {Phys. Rev. Lett.}\ }\textbf {\bibinfo {volume} {132}},\ \bibinfo {pages}
  {151401} (\bibinfo {year} {2024})},\ \Eprint
  {http://arxiv.org/abs/2211.00013} {arXiv:2211.00013 [astro-ph.CO]}
  \BibitemShut {NoStop}%
\bibitem [{\citenamefont {Andr{\'e}s-Carcasona}\ \emph
  {et~al.}(2024)\citenamefont {Andr{\'e}s-Carcasona}, \citenamefont {Iovino},
  \citenamefont {Vaskonen}, \citenamefont {Veerm{\"a}e}, \citenamefont
  {Mart{\'\i}nez}, \citenamefont {Pujol{\`a}s},\ and\ \citenamefont
  {Mir}}]{Andres-Carcasona:2024wqk}%
  \BibitemOpen
  \bibfield  {author} {\bibinfo {author} {\bibfnamefont {M.}~\bibnamefont
  {Andr{\'e}s-Carcasona}}, \bibinfo {author} {\bibfnamefont {A.~J.}\
  \bibnamefont {Iovino}}, \bibinfo {author} {\bibfnamefont {V.}~\bibnamefont
  {Vaskonen}}, \bibinfo {author} {\bibfnamefont {H.}~\bibnamefont
  {Veerm{\"a}e}}, \bibinfo {author} {\bibfnamefont {M.}~\bibnamefont
  {Mart{\'\i}nez}}, \bibinfo {author} {\bibfnamefont {O.}~\bibnamefont
  {Pujol{\`a}s}}, \ and\ \bibinfo {author} {\bibfnamefont {L.~M.}\ \bibnamefont
  {Mir}},\ }\href {\doibase 10.1103/PhysRevD.110.023040} {\bibfield  {journal}
  {\bibinfo  {journal} {Phys. Rev. D}\ }\textbf {\bibinfo {volume} {110}},\
  \bibinfo {pages} {023040} (\bibinfo {year} {2024})},\ \Eprint
  {http://arxiv.org/abs/2405.05732} {arXiv:2405.05732 [astro-ph.CO]}
  \BibitemShut {NoStop}%
\bibitem [{\citenamefont {Figueroa}\ \emph {et~al.}(2013)\citenamefont
  {Figueroa}, \citenamefont {Hindmarsh},\ and\ \citenamefont
  {Urrestilla}}]{Figueroa:2012kw}%
  \BibitemOpen
  \bibfield  {author} {\bibinfo {author} {\bibfnamefont {D.~G.}\ \bibnamefont
  {Figueroa}}, \bibinfo {author} {\bibfnamefont {M.}~\bibnamefont {Hindmarsh}},
  \ and\ \bibinfo {author} {\bibfnamefont {J.}~\bibnamefont {Urrestilla}},\
  }\href {\doibase 10.1103/PhysRevLett.110.101302} {\bibfield  {journal}
  {\bibinfo  {journal} {Phys. Rev. Lett.}\ }\textbf {\bibinfo {volume} {110}},\
  \bibinfo {pages} {101302} (\bibinfo {year} {2013})},\ \Eprint
  {http://arxiv.org/abs/1212.5458} {arXiv:1212.5458 [astro-ph.CO]} \BibitemShut
  {NoStop}%
\bibitem [{\citenamefont {Fenu}\ \emph {et~al.}(2009)\citenamefont {Fenu},
  \citenamefont {Figueroa}, \citenamefont {Durrer},\ and\ \citenamefont
  {Garcia-Bellido}}]{Fenu:2009qf}%
  \BibitemOpen
  \bibfield  {author} {\bibinfo {author} {\bibfnamefont {E.}~\bibnamefont
  {Fenu}}, \bibinfo {author} {\bibfnamefont {D.~G.}\ \bibnamefont {Figueroa}},
  \bibinfo {author} {\bibfnamefont {R.}~\bibnamefont {Durrer}}, \ and\ \bibinfo
  {author} {\bibfnamefont {J.}~\bibnamefont {Garcia-Bellido}},\ }\href
  {\doibase 10.1088/1475-7516/2009/10/005} {\bibfield  {journal} {\bibinfo
  {journal} {JCAP}\ }\textbf {\bibinfo {volume} {10}},\ \bibinfo {pages} {005}
  (\bibinfo {year} {2009})},\ \Eprint {http://arxiv.org/abs/0908.0425}
  {arXiv:0908.0425 [astro-ph.CO]} \BibitemShut {NoStop}%
\bibitem [{\citenamefont {Figueroa}\ \emph {et~al.}(2020)\citenamefont
  {Figueroa}, \citenamefont {Hindmarsh}, \citenamefont {Lizarraga},\ and\
  \citenamefont {Urrestilla}}]{Figueroa:2020lvo}%
  \BibitemOpen
  \bibfield  {author} {\bibinfo {author} {\bibfnamefont {D.~G.}\ \bibnamefont
  {Figueroa}}, \bibinfo {author} {\bibfnamefont {M.}~\bibnamefont {Hindmarsh}},
  \bibinfo {author} {\bibfnamefont {J.}~\bibnamefont {Lizarraga}}, \ and\
  \bibinfo {author} {\bibfnamefont {J.}~\bibnamefont {Urrestilla}},\ }\href
  {\doibase 10.1103/PhysRevD.102.103516} {\bibfield  {journal} {\bibinfo
  {journal} {Phys. Rev. D}\ }\textbf {\bibinfo {volume} {102}},\ \bibinfo
  {pages} {103516} (\bibinfo {year} {2020})},\ \Eprint
  {http://arxiv.org/abs/2007.03337} {arXiv:2007.03337 [astro-ph.CO]}
  \BibitemShut {NoStop}%
\bibitem [{\citenamefont {Pagano}\ \emph {et~al.}(2016)\citenamefont {Pagano},
  \citenamefont {Salvati},\ and\ \citenamefont {Melchiorri}}]{Pagano:2015hma}%
  \BibitemOpen
  \bibfield  {author} {\bibinfo {author} {\bibfnamefont {L.}~\bibnamefont
  {Pagano}}, \bibinfo {author} {\bibfnamefont {L.}~\bibnamefont {Salvati}}, \
  and\ \bibinfo {author} {\bibfnamefont {A.}~\bibnamefont {Melchiorri}},\
  }\href {\doibase 10.1016/j.physletb.2016.07.078} {\bibfield  {journal}
  {\bibinfo  {journal} {Phys. Lett. B}\ }\textbf {\bibinfo {volume} {760}},\
  \bibinfo {pages} {823} (\bibinfo {year} {2016})},\ \Eprint
  {http://arxiv.org/abs/1508.02393} {arXiv:1508.02393 [astro-ph.CO]}
  \BibitemShut {NoStop}%
\bibitem [{\citenamefont {Janssen}\ \emph {et~al.}(2015)\citenamefont {Janssen}
  \emph {et~al.}}]{Janssen:2014dka}%
  \BibitemOpen
  \bibfield  {author} {\bibinfo {author} {\bibfnamefont {G.}~\bibnamefont
  {Janssen}} \emph {et~al.},\ }\href {\doibase 10.22323/1.215.0037} {\bibfield
  {journal} {\bibinfo  {journal} {PoS}\ }\textbf {\bibinfo {volume}
  {AASKA14}},\ \bibinfo {pages} {037} (\bibinfo {year} {2015})},\ \Eprint
  {http://arxiv.org/abs/1501.00127} {arXiv:1501.00127 [astro-ph.IM]}
  \BibitemShut {NoStop}%
\bibitem [{\citenamefont {Robson}\ \emph {et~al.}(2019)\citenamefont {Robson},
  \citenamefont {Cornish},\ and\ \citenamefont {Liu}}]{Robson:2018ifk}%
  \BibitemOpen
  \bibfield  {author} {\bibinfo {author} {\bibfnamefont {T.}~\bibnamefont
  {Robson}}, \bibinfo {author} {\bibfnamefont {N.~J.}\ \bibnamefont {Cornish}},
  \ and\ \bibinfo {author} {\bibfnamefont {C.}~\bibnamefont {Liu}},\ }\href
  {\doibase 10.1088/1361-6382/ab1101} {\bibfield  {journal} {\bibinfo
  {journal} {Class. Quant. Grav.}\ }\textbf {\bibinfo {volume} {36}},\ \bibinfo
  {pages} {105011} (\bibinfo {year} {2019})},\ \Eprint
  {http://arxiv.org/abs/1803.01944} {arXiv:1803.01944 [astro-ph.HE]}
  \BibitemShut {NoStop}%
\bibitem [{\citenamefont {Yagi}\ and\ \citenamefont
  {Seto}(2011)}]{Yagi:2011wg}%
  \BibitemOpen
  \bibfield  {author} {\bibinfo {author} {\bibfnamefont {K.}~\bibnamefont
  {Yagi}}\ and\ \bibinfo {author} {\bibfnamefont {N.}~\bibnamefont {Seto}},\
  }\href {\doibase 10.1103/PhysRevD.83.044011} {\bibfield  {journal} {\bibinfo
  {journal} {Phys. Rev. D}\ }\textbf {\bibinfo {volume} {83}},\ \bibinfo
  {pages} {044011} (\bibinfo {year} {2011})},\ \bibinfo {note} {[Erratum:
  Phys.Rev.D 95, 109901 (2017)]},\ \Eprint {http://arxiv.org/abs/1101.3940}
  {arXiv:1101.3940 [astro-ph.CO]} \BibitemShut {NoStop}%
\bibitem [{\citenamefont {Kuroyanagi}\ \emph {et~al.}(2015)\citenamefont
  {Kuroyanagi}, \citenamefont {Nakayama},\ and\ \citenamefont
  {Yokoyama}}]{Kuroyanagi:2014qza}%
  \BibitemOpen
  \bibfield  {author} {\bibinfo {author} {\bibfnamefont {S.}~\bibnamefont
  {Kuroyanagi}}, \bibinfo {author} {\bibfnamefont {K.}~\bibnamefont
  {Nakayama}}, \ and\ \bibinfo {author} {\bibfnamefont {J.}~\bibnamefont
  {Yokoyama}},\ }\href {\doibase 10.1093/ptep/ptu176} {\bibfield  {journal}
  {\bibinfo  {journal} {PTEP}\ }\textbf {\bibinfo {volume} {2015}},\ \bibinfo
  {pages} {013E02} (\bibinfo {year} {2015})},\ \Eprint
  {http://arxiv.org/abs/1410.6618} {arXiv:1410.6618 [astro-ph.CO]} \BibitemShut
  {NoStop}%
\bibitem [{\citenamefont {Croon}\ \emph {et~al.}(2020)\citenamefont {Croon},
  \citenamefont {McKeen}, \citenamefont {Raj},\ and\ \citenamefont
  {Wang}}]{Croon:2020ouk}%
  \BibitemOpen
  \bibfield  {author} {\bibinfo {author} {\bibfnamefont {D.}~\bibnamefont
  {Croon}}, \bibinfo {author} {\bibfnamefont {D.}~\bibnamefont {McKeen}},
  \bibinfo {author} {\bibfnamefont {N.}~\bibnamefont {Raj}}, \ and\ \bibinfo
  {author} {\bibfnamefont {Z.}~\bibnamefont {Wang}},\ }\href {\doibase
  10.1103/PhysRevD.102.083021} {\bibfield  {journal} {\bibinfo  {journal}
  {Phys. Rev. D}\ }\textbf {\bibinfo {volume} {102}},\ \bibinfo {pages}
  {083021} (\bibinfo {year} {2020})},\ \Eprint
  {http://arxiv.org/abs/2007.12697} {arXiv:2007.12697 [astro-ph.CO]}
  \BibitemShut {NoStop}%
\bibitem [{\citenamefont {Mr{\'o}z}\ \emph
  {et~al.}(2024{\natexlab{a}})\citenamefont {Mr{\'o}z} \emph
  {et~al.}}]{Mroz:2024wia}%
  \BibitemOpen
  \bibfield  {author} {\bibinfo {author} {\bibfnamefont {P.}~\bibnamefont
  {Mr{\'o}z}} \emph {et~al.},\ }\href {\doibase 10.3847/2041-8213/ad8e68}
  {\bibfield  {journal} {\bibinfo  {journal} {Astrophys. J. Lett.}\ }\textbf
  {\bibinfo {volume} {976}},\ \bibinfo {pages} {L19} (\bibinfo {year}
  {2024}{\natexlab{a}})},\ \Eprint {http://arxiv.org/abs/2410.06251}
  {arXiv:2410.06251 [astro-ph.CO]} \BibitemShut {NoStop}%
\bibitem [{\citenamefont {Mr{\'o}z}\ \emph
  {et~al.}(2024{\natexlab{b}})\citenamefont {Mr{\'o}z} \emph
  {et~al.}}]{Mroz:2024mse}%
  \BibitemOpen
  \bibfield  {author} {\bibinfo {author} {\bibfnamefont {P.}~\bibnamefont
  {Mr{\'o}z}} \emph {et~al.},\ }\href {\doibase 10.1038/s41586-024-07704-6}
  {\bibfield  {journal} {\bibinfo  {journal} {Nature}\ }\textbf {\bibinfo
  {volume} {632}},\ \bibinfo {pages} {749} (\bibinfo {year}
  {2024}{\natexlab{b}})},\ \Eprint {http://arxiv.org/abs/2403.02386}
  {arXiv:2403.02386 [astro-ph.GA]} \BibitemShut {NoStop}%
\bibitem [{\citenamefont {Serpico}\ \emph {et~al.}(2020)\citenamefont
  {Serpico}, \citenamefont {Poulin}, \citenamefont {Inman},\ and\ \citenamefont
  {Kohri}}]{Serpico:2020ehh}%
  \BibitemOpen
  \bibfield  {author} {\bibinfo {author} {\bibfnamefont {P.~D.}\ \bibnamefont
  {Serpico}}, \bibinfo {author} {\bibfnamefont {V.}~\bibnamefont {Poulin}},
  \bibinfo {author} {\bibfnamefont {D.}~\bibnamefont {Inman}}, \ and\ \bibinfo
  {author} {\bibfnamefont {K.}~\bibnamefont {Kohri}},\ }\href {\doibase
  10.1103/PhysRevResearch.2.023204} {\bibfield  {journal} {\bibinfo  {journal}
  {Phys. Rev. Res.}\ }\textbf {\bibinfo {volume} {2}},\ \bibinfo {pages}
  {023204} (\bibinfo {year} {2020})},\ \Eprint
  {http://arxiv.org/abs/2002.10771} {arXiv:2002.10771 [astro-ph.CO]}
  \BibitemShut {NoStop}%
\bibitem [{\citenamefont {Carr}\ \emph {et~al.}(2010)\citenamefont {Carr},
  \citenamefont {Kohri}, \citenamefont {Sendouda},\ and\ \citenamefont
  {Yokoyama}}]{Carr:2009jm}%
  \BibitemOpen
  \bibfield  {author} {\bibinfo {author} {\bibfnamefont {B.~J.}\ \bibnamefont
  {Carr}}, \bibinfo {author} {\bibfnamefont {K.}~\bibnamefont {Kohri}},
  \bibinfo {author} {\bibfnamefont {Y.}~\bibnamefont {Sendouda}}, \ and\
  \bibinfo {author} {\bibfnamefont {J.}~\bibnamefont {Yokoyama}},\ }\href
  {\doibase 10.1103/PhysRevD.81.104019} {\bibfield  {journal} {\bibinfo
  {journal} {Phys. Rev. D}\ }\textbf {\bibinfo {volume} {81}},\ \bibinfo
  {pages} {104019} (\bibinfo {year} {2010})},\ \Eprint
  {http://arxiv.org/abs/0912.5297} {arXiv:0912.5297 [astro-ph.CO]} \BibitemShut
  {NoStop}%
\bibitem [{\citenamefont {Coogan}\ \emph {et~al.}(2021)\citenamefont {Coogan},
  \citenamefont {Morrison},\ and\ \citenamefont {Profumo}}]{Coogan:2020tuf}%
  \BibitemOpen
  \bibfield  {author} {\bibinfo {author} {\bibfnamefont {A.}~\bibnamefont
  {Coogan}}, \bibinfo {author} {\bibfnamefont {L.}~\bibnamefont {Morrison}}, \
  and\ \bibinfo {author} {\bibfnamefont {S.}~\bibnamefont {Profumo}},\ }\href
  {\doibase 10.1103/PhysRevLett.126.171101} {\bibfield  {journal} {\bibinfo
  {journal} {Phys. Rev. Lett.}\ }\textbf {\bibinfo {volume} {126}},\ \bibinfo
  {pages} {171101} (\bibinfo {year} {2021})},\ \Eprint
  {http://arxiv.org/abs/2010.04797} {arXiv:2010.04797 [astro-ph.CO]}
  \BibitemShut {NoStop}%
\bibitem [{\citenamefont {Drlica-Wagner}\ \emph {et~al.}(2019)\citenamefont
  {Drlica-Wagner} \emph {et~al.}}]{LSSTDarkMatterGroup:2019mwo}%
  \BibitemOpen
  \bibfield  {author} {\bibinfo {author} {\bibfnamefont {A.}~\bibnamefont
  {Drlica-Wagner}} \emph {et~al.} (\bibinfo {collaboration} {LSST Dark Matter
  Group}),\ }\href@noop {} {\  (\bibinfo {year} {2019})},\ \Eprint
  {http://arxiv.org/abs/1902.01055} {arXiv:1902.01055 [astro-ph.CO]}
  \BibitemShut {NoStop}%
\bibitem [{\citenamefont {Schmitz}(2021)}]{Schmitz:2020syl}%
  \BibitemOpen
  \bibfield  {author} {\bibinfo {author} {\bibfnamefont {K.}~\bibnamefont
  {Schmitz}},\ }\href {\doibase 10.1007/JHEP01(2021)097} {\bibfield  {journal}
  {\bibinfo  {journal} {JHEP}\ }\textbf {\bibinfo {volume} {01}},\ \bibinfo
  {pages} {097} (\bibinfo {year} {2021})},\ \Eprint
  {http://arxiv.org/abs/2002.04615} {arXiv:2002.04615 [hep-ph]} \BibitemShut
  {NoStop}%
\bibitem [{\citenamefont {Kavanagh}(2019)}]{Kavanagh_PBHbounds_2019}%
  \BibitemOpen
  \bibfield  {author} {\bibinfo {author} {\bibfnamefont {B.~J.}\ \bibnamefont
  {Kavanagh}},\ }\href {\doibase 10.5281/zenodo.3538999} {\enquote {\bibinfo
  {title} {{bradkav/PBHbounds: Release version}},}\ } (\bibinfo {year}
  {2019})\BibitemShut {NoStop}%
\bibitem [{\citenamefont {Agazie}\ \emph {et~al.}(2023)\citenamefont {Agazie}
  \emph {et~al.}}]{NANOGrav:2023gor}%
  \BibitemOpen
  \bibfield  {author} {\bibinfo {author} {\bibfnamefont {G.}~\bibnamefont
  {Agazie}} \emph {et~al.} (\bibinfo {collaboration} {NANOGrav}),\ }\href
  {\doibase 10.3847/2041-8213/acdac6} {\bibfield  {journal} {\bibinfo
  {journal} {Astrophys. J. Lett.}\ }\textbf {\bibinfo {volume} {951}},\
  \bibinfo {pages} {L8} (\bibinfo {year} {2023})},\ \Eprint
  {http://arxiv.org/abs/2306.16213} {arXiv:2306.16213 [astro-ph.HE]}
  \BibitemShut {NoStop}%
\bibitem [{\citenamefont {Antoniadis}\ \emph {et~al.}(2023)\citenamefont
  {Antoniadis} \emph {et~al.}}]{Antoniadis:2023ott}%
  \BibitemOpen
  \bibfield  {author} {\bibinfo {author} {\bibfnamefont {J.}~\bibnamefont
  {Antoniadis}} \emph {et~al.} (\bibinfo {collaboration} {EPTA, InPTA:}),\
  }\href {\doibase 10.1051/0004-6361/202346844} {\bibfield  {journal} {\bibinfo
   {journal} {Astron. Astrophys.}\ }\textbf {\bibinfo {volume} {678}},\
  \bibinfo {pages} {A50} (\bibinfo {year} {2023})},\ \Eprint
  {http://arxiv.org/abs/2306.16214} {arXiv:2306.16214 [astro-ph.HE]}
  \BibitemShut {NoStop}%
\bibitem [{\citenamefont {Reardon}\ \emph {et~al.}(2023)\citenamefont {Reardon}
  \emph {et~al.}}]{Reardon:2023gzh}%
  \BibitemOpen
  \bibfield  {author} {\bibinfo {author} {\bibfnamefont {D.~J.}\ \bibnamefont
  {Reardon}} \emph {et~al.},\ }\href {\doibase 10.3847/2041-8213/acdd02}
  {\bibfield  {journal} {\bibinfo  {journal} {Astrophys. J. Lett.}\ }\textbf
  {\bibinfo {volume} {951}},\ \bibinfo {pages} {L6} (\bibinfo {year} {2023})},\
  \Eprint {http://arxiv.org/abs/2306.16215} {arXiv:2306.16215 [astro-ph.HE]}
  \BibitemShut {NoStop}%
\bibitem [{\citenamefont {Xu}\ \emph {et~al.}(2023)\citenamefont {Xu} \emph
  {et~al.}}]{Xu:2023wog}%
  \BibitemOpen
  \bibfield  {author} {\bibinfo {author} {\bibfnamefont {H.}~\bibnamefont {Xu}}
  \emph {et~al.},\ }\href {\doibase 10.1088/1674-4527/acdfa5} {\bibfield
  {journal} {\bibinfo  {journal} {Res. Astron. Astrophys.}\ }\textbf {\bibinfo
  {volume} {23}},\ \bibinfo {pages} {075024} (\bibinfo {year} {2023})},\
  \Eprint {http://arxiv.org/abs/2306.16216} {arXiv:2306.16216 [astro-ph.HE]}
  \BibitemShut {NoStop}%
\bibitem [{\citenamefont {Punturo}\ \emph {et~al.}(2010)\citenamefont {Punturo}
  \emph {et~al.}}]{Punturo:2010zz}%
  \BibitemOpen
  \bibfield  {author} {\bibinfo {author} {\bibfnamefont {M.}~\bibnamefont
  {Punturo}} \emph {et~al.},\ }\href {\doibase 10.1088/0264-9381/27/19/194002}
  {\bibfield  {journal} {\bibinfo  {journal} {Class. Quant. Grav.}\ }\textbf
  {\bibinfo {volume} {27}},\ \bibinfo {pages} {194002} (\bibinfo {year}
  {2010})}\BibitemShut {NoStop}%
\bibitem [{\citenamefont {Hild}\ \emph {et~al.}(2011)\citenamefont {Hild} \emph
  {et~al.}}]{Hild:2010id}%
  \BibitemOpen
  \bibfield  {author} {\bibinfo {author} {\bibfnamefont {S.}~\bibnamefont
  {Hild}} \emph {et~al.},\ }\href {\doibase 10.1088/0264-9381/28/9/094013}
  {\bibfield  {journal} {\bibinfo  {journal} {Class. Quant. Grav.}\ }\textbf
  {\bibinfo {volume} {28}},\ \bibinfo {pages} {094013} (\bibinfo {year}
  {2011})},\ \Eprint {http://arxiv.org/abs/1012.0908} {arXiv:1012.0908 [gr-qc]}
  \BibitemShut {NoStop}%
\bibitem [{\citenamefont {Maggiore}\ \emph {et~al.}(2020)\citenamefont
  {Maggiore} \emph {et~al.}}]{ET:2019dnz}%
  \BibitemOpen
  \bibfield  {author} {\bibinfo {author} {\bibfnamefont {M.}~\bibnamefont
  {Maggiore}} \emph {et~al.} (\bibinfo {collaboration} {ET}),\ }\href {\doibase
  10.1088/1475-7516/2020/03/050} {\bibfield  {journal} {\bibinfo  {journal}
  {JCAP}\ }\textbf {\bibinfo {volume} {03}},\ \bibinfo {pages} {050} (\bibinfo
  {year} {2020})},\ \Eprint {http://arxiv.org/abs/1912.02622} {arXiv:1912.02622
  [astro-ph.CO]} \BibitemShut {NoStop}%
\bibitem [{\citenamefont {Weltman}\ \emph {et~al.}(2020)\citenamefont {Weltman}
  \emph {et~al.}}]{Weltman:2018zrl}%
  \BibitemOpen
  \bibfield  {author} {\bibinfo {author} {\bibfnamefont {A.}~\bibnamefont
  {Weltman}} \emph {et~al.},\ }\href {\doibase 10.1017/pasa.2019.42} {\bibfield
   {journal} {\bibinfo  {journal} {Publ. Astron. Soc. Austral.}\ }\textbf
  {\bibinfo {volume} {37}},\ \bibinfo {pages} {e002} (\bibinfo {year}
  {2020})},\ \Eprint {http://arxiv.org/abs/1810.02680} {arXiv:1810.02680
  [astro-ph.CO]} \BibitemShut {NoStop}%
\bibitem [{\citenamefont {Amaro-Seoane}\ \emph {et~al.}(2017)\citenamefont
  {Amaro-Seoane} \emph {et~al.}}]{LISA:2017pwj}%
  \BibitemOpen
  \bibfield  {author} {\bibinfo {author} {\bibfnamefont {P.}~\bibnamefont
  {Amaro-Seoane}} \emph {et~al.} (\bibinfo {collaboration} {LISA}),\
  }\href@noop {} {\  (\bibinfo {year} {2017})},\ \Eprint
  {http://arxiv.org/abs/1702.00786} {arXiv:1702.00786 [astro-ph.IM]}
  \BibitemShut {NoStop}%
\bibitem [{\citenamefont {Kawamura}\ \emph {et~al.}(2011)\citenamefont
  {Kawamura} \emph {et~al.}}]{Kawamura:2011zz}%
  \BibitemOpen
  \bibfield  {author} {\bibinfo {author} {\bibfnamefont {S.}~\bibnamefont
  {Kawamura}} \emph {et~al.},\ }\href {\doibase 10.1088/0264-9381/28/9/094011}
  {\bibfield  {journal} {\bibinfo  {journal} {Class. Quant. Grav.}\ }\textbf
  {\bibinfo {volume} {28}},\ \bibinfo {pages} {094011} (\bibinfo {year}
  {2011})}\BibitemShut {NoStop}%
\bibitem [{\citenamefont {Sathyaprakash}\ \emph {et~al.}(2012)\citenamefont
  {Sathyaprakash} \emph {et~al.}}]{Sathyaprakash:2012jk}%
  \BibitemOpen
  \bibfield  {author} {\bibinfo {author} {\bibfnamefont {B.}~\bibnamefont
  {Sathyaprakash}} \emph {et~al.},\ }\href {\doibase
  10.1088/0264-9381/29/12/124013} {\bibfield  {journal} {\bibinfo  {journal}
  {Class. Quant. Grav.}\ }\textbf {\bibinfo {volume} {29}},\ \bibinfo {pages}
  {124013} (\bibinfo {year} {2012})},\ \bibinfo {note} {[Erratum:
  Class.Quant.Grav. 30, 079501 (2013)]},\ \Eprint
  {http://arxiv.org/abs/1206.0331} {arXiv:1206.0331 [gr-qc]} \BibitemShut
  {NoStop}%
\bibitem [{\citenamefont {Evans}\ \emph {et~al.}(2021)\citenamefont {Evans}
  \emph {et~al.}}]{Evans:2021gyd}%
  \BibitemOpen
  \bibfield  {author} {\bibinfo {author} {\bibfnamefont {M.}~\bibnamefont
  {Evans}} \emph {et~al.},\ }\href@noop {} {\  (\bibinfo {year} {2021})},\
  \Eprint {http://arxiv.org/abs/2109.09882} {arXiv:2109.09882 [astro-ph.IM]}
  \BibitemShut {NoStop}%
\bibitem [{\citenamefont {Liebling}\ and\ \citenamefont
  {Palenzuela}(2016)}]{Liebling:2016orx}%
  \BibitemOpen
  \bibfield  {author} {\bibinfo {author} {\bibfnamefont {S.~L.}\ \bibnamefont
  {Liebling}}\ and\ \bibinfo {author} {\bibfnamefont {C.}~\bibnamefont
  {Palenzuela}},\ }\href {\doibase 10.1103/PhysRevD.94.064046} {\bibfield
  {journal} {\bibinfo  {journal} {Phys. Rev. D}\ }\textbf {\bibinfo {volume}
  {94}},\ \bibinfo {pages} {064046} (\bibinfo {year} {2016})},\ \Eprint
  {http://arxiv.org/abs/1607.02140} {arXiv:1607.02140 [gr-qc]} \BibitemShut
  {NoStop}%
\bibitem [{\citenamefont {Vanvlasselaer}\ and\ \citenamefont
  {Yin}(2026)}]{Vanvlasselaer:2026fay}%
  \BibitemOpen
  \bibfield  {author} {\bibinfo {author} {\bibfnamefont {M.}~\bibnamefont
  {Vanvlasselaer}}\ and\ \bibinfo {author} {\bibfnamefont {W.}~\bibnamefont
  {Yin}},\ }\href@noop {} {\  (\bibinfo {year} {2026})},\ \Eprint
  {http://arxiv.org/abs/2604.20762} {arXiv:2604.20762 [hep-ph]} \BibitemShut
  {NoStop}%
\bibitem [{\citenamefont {Lopez-Eiguren}\ \emph {et~al.}(2017)\citenamefont
  {Lopez-Eiguren}, \citenamefont {Lizarraga}, \citenamefont {Hindmarsh},\ and\
  \citenamefont {Urrestilla}}]{Lopez-Eiguren:2017dmc}%
  \BibitemOpen
  \bibfield  {author} {\bibinfo {author} {\bibfnamefont {A.}~\bibnamefont
  {Lopez-Eiguren}}, \bibinfo {author} {\bibfnamefont {J.}~\bibnamefont
  {Lizarraga}}, \bibinfo {author} {\bibfnamefont {M.}~\bibnamefont
  {Hindmarsh}}, \ and\ \bibinfo {author} {\bibfnamefont {J.}~\bibnamefont
  {Urrestilla}},\ }\href {\doibase 10.1088/1475-7516/2017/07/026} {\bibfield
  {journal} {\bibinfo  {journal} {JCAP}\ }\textbf {\bibinfo {volume} {07}},\
  \bibinfo {pages} {026} (\bibinfo {year} {2017})},\ \Eprint
  {http://arxiv.org/abs/1705.04154} {arXiv:1705.04154 [astro-ph.CO]}
  \BibitemShut {NoStop}%
\bibitem [{\citenamefont {Gonzalez}\ \emph {et~al.}(2023)\citenamefont
  {Gonzalez}, \citenamefont {Kitajima}, \citenamefont {Takahashi},\ and\
  \citenamefont {Yin}}]{Gonzalez:2022mcx}%
  \BibitemOpen
  \bibfield  {author} {\bibinfo {author} {\bibfnamefont {D.}~\bibnamefont
  {Gonzalez}}, \bibinfo {author} {\bibfnamefont {N.}~\bibnamefont {Kitajima}},
  \bibinfo {author} {\bibfnamefont {F.}~\bibnamefont {Takahashi}}, \ and\
  \bibinfo {author} {\bibfnamefont {W.}~\bibnamefont {Yin}},\ }\href {\doibase
  10.1016/j.physletb.2023.137990} {\bibfield  {journal} {\bibinfo  {journal}
  {Phys. Lett. B}\ }\textbf {\bibinfo {volume} {843}},\ \bibinfo {pages}
  {137990} (\bibinfo {year} {2023})},\ \Eprint
  {http://arxiv.org/abs/2211.06849} {arXiv:2211.06849 [hep-ph]} \BibitemShut
  {NoStop}%
\bibitem [{\citenamefont {Kitajima}\ \emph {et~al.}(2025)\citenamefont
  {Kitajima}, \citenamefont {Lee}, \citenamefont {Takahashi},\ and\
  \citenamefont {Yin}}]{Kitajima:2023kzu}%
  \BibitemOpen
  \bibfield  {author} {\bibinfo {author} {\bibfnamefont {N.}~\bibnamefont
  {Kitajima}}, \bibinfo {author} {\bibfnamefont {J.}~\bibnamefont {Lee}},
  \bibinfo {author} {\bibfnamefont {F.}~\bibnamefont {Takahashi}}, \ and\
  \bibinfo {author} {\bibfnamefont {W.}~\bibnamefont {Yin}},\ }\href {\doibase
  10.1088/1475-7516/2025/07/053} {\bibfield  {journal} {\bibinfo  {journal}
  {JCAP}\ }\textbf {\bibinfo {volume} {07}},\ \bibinfo {pages} {053} (\bibinfo
  {year} {2025})},\ \Eprint {http://arxiv.org/abs/2311.14590} {arXiv:2311.14590
  [hep-ph]} \BibitemShut {NoStop}%
\bibitem [{\citenamefont {Gonzalez}()}]{2911336}%
  \BibitemOpen
  \bibfield  {author} {\bibinfo {author} {\bibfnamefont {D.}~\bibnamefont
  {Gonzalez}},\ }\emph {\bibinfo {title} {{Stability of domain walls with
  biased initial conditions and their signatures in the CMB cosmic
  birefringence}}},\ \href@noop {} {\bibinfo {type} {Other thesis}}\BibitemShut
  {NoStop}%
\bibitem [{\citenamefont {Moroi}\ and\ \citenamefont
  {Yin}(2021{\natexlab{a}})}]{Moroi:2020has}%
  \BibitemOpen
  \bibfield  {author} {\bibinfo {author} {\bibfnamefont {T.}~\bibnamefont
  {Moroi}}\ and\ \bibinfo {author} {\bibfnamefont {W.}~\bibnamefont {Yin}},\
  }\href {\doibase 10.1007/JHEP03(2021)301} {\bibfield  {journal} {\bibinfo
  {journal} {JHEP}\ }\textbf {\bibinfo {volume} {03}},\ \bibinfo {pages} {301}
  (\bibinfo {year} {2021}{\natexlab{a}})},\ \Eprint
  {http://arxiv.org/abs/2011.09475} {arXiv:2011.09475 [hep-ph]} \BibitemShut
  {NoStop}%
\bibitem [{\citenamefont {Moroi}\ and\ \citenamefont
  {Yin}(2021{\natexlab{b}})}]{Moroi:2020bkq}%
  \BibitemOpen
  \bibfield  {author} {\bibinfo {author} {\bibfnamefont {T.}~\bibnamefont
  {Moroi}}\ and\ \bibinfo {author} {\bibfnamefont {W.}~\bibnamefont {Yin}},\
  }\href {\doibase 10.1007/JHEP03(2021)296} {\bibfield  {journal} {\bibinfo
  {journal} {JHEP}\ }\textbf {\bibinfo {volume} {03}},\ \bibinfo {pages} {296}
  (\bibinfo {year} {2021}{\natexlab{b}})},\ \Eprint
  {http://arxiv.org/abs/2011.12285} {arXiv:2011.12285 [hep-ph]} \BibitemShut
  {NoStop}%
\bibitem [{\citenamefont {Kofman}\ \emph {et~al.}(1994)\citenamefont {Kofman},
  \citenamefont {Linde},\ and\ \citenamefont {Starobinsky}}]{Kofman:1994rk}%
  \BibitemOpen
  \bibfield  {author} {\bibinfo {author} {\bibfnamefont {L.}~\bibnamefont
  {Kofman}}, \bibinfo {author} {\bibfnamefont {A.~D.}\ \bibnamefont {Linde}}, \
  and\ \bibinfo {author} {\bibfnamefont {A.~A.}\ \bibnamefont {Starobinsky}},\
  }\href {\doibase 10.1103/PhysRevLett.73.3195} {\bibfield  {journal} {\bibinfo
   {journal} {Phys. Rev. Lett.}\ }\textbf {\bibinfo {volume} {73}},\ \bibinfo
  {pages} {3195} (\bibinfo {year} {1994})},\ \Eprint
  {http://arxiv.org/abs/hep-th/9405187} {arXiv:hep-th/9405187} \BibitemShut
  {NoStop}%
\bibitem [{\citenamefont {Kofman}\ \emph {et~al.}(1997)\citenamefont {Kofman},
  \citenamefont {Linde},\ and\ \citenamefont {Starobinsky}}]{Kofman:1997yn}%
  \BibitemOpen
  \bibfield  {author} {\bibinfo {author} {\bibfnamefont {L.}~\bibnamefont
  {Kofman}}, \bibinfo {author} {\bibfnamefont {A.~D.}\ \bibnamefont {Linde}}, \
  and\ \bibinfo {author} {\bibfnamefont {A.~A.}\ \bibnamefont {Starobinsky}},\
  }\href {\doibase 10.1103/PhysRevD.56.3258} {\bibfield  {journal} {\bibinfo
  {journal} {Phys. Rev. D}\ }\textbf {\bibinfo {volume} {56}},\ \bibinfo
  {pages} {3258} (\bibinfo {year} {1997})},\ \Eprint
  {http://arxiv.org/abs/hep-ph/9704452} {arXiv:hep-ph/9704452} \BibitemShut
  {NoStop}%
\bibitem [{\citenamefont {Amin}\ \emph {et~al.}(2019)\citenamefont {Amin},
  \citenamefont {Fan}, \citenamefont {Lozanov},\ and\ \citenamefont
  {Reece}}]{Amin:2018kkg}%
  \BibitemOpen
  \bibfield  {author} {\bibinfo {author} {\bibfnamefont {M.~A.}\ \bibnamefont
  {Amin}}, \bibinfo {author} {\bibfnamefont {J.}~\bibnamefont {Fan}}, \bibinfo
  {author} {\bibfnamefont {K.~D.}\ \bibnamefont {Lozanov}}, \ and\ \bibinfo
  {author} {\bibfnamefont {M.}~\bibnamefont {Reece}},\ }\href {\doibase
  10.1103/PhysRevD.99.035008} {\bibfield  {journal} {\bibinfo  {journal} {Phys.
  Rev. D}\ }\textbf {\bibinfo {volume} {99}},\ \bibinfo {pages} {035008}
  (\bibinfo {year} {2019})},\ \Eprint {http://arxiv.org/abs/1802.00444}
  {arXiv:1802.00444 [hep-ph]} \BibitemShut {NoStop}%
\bibitem [{\citenamefont {Peccei}\ and\ \citenamefont
  {Quinn}(1977)}]{Peccei:1977hh}%
  \BibitemOpen
  \bibfield  {author} {\bibinfo {author} {\bibfnamefont {R.~D.}\ \bibnamefont
  {Peccei}}\ and\ \bibinfo {author} {\bibfnamefont {H.~R.}\ \bibnamefont
  {Quinn}},\ }\href {\doibase 10.1103/PhysRevLett.38.1440} {\bibfield
  {journal} {\bibinfo  {journal} {Phys. Rev. Lett.}\ }\textbf {\bibinfo
  {volume} {38}},\ \bibinfo {pages} {1440} (\bibinfo {year}
  {1977})}\BibitemShut {NoStop}%
\bibitem [{\citenamefont {Weinberg}(1978)}]{Weinberg:1977ma}%
  \BibitemOpen
  \bibfield  {author} {\bibinfo {author} {\bibfnamefont {S.}~\bibnamefont
  {Weinberg}},\ }\href {\doibase 10.1103/PhysRevLett.40.223} {\bibfield
  {journal} {\bibinfo  {journal} {Phys. Rev. Lett.}\ }\textbf {\bibinfo
  {volume} {40}},\ \bibinfo {pages} {223} (\bibinfo {year} {1978})}\BibitemShut
  {NoStop}%
\bibitem [{\citenamefont {Wilczek}(1978)}]{Wilczek:1977pj}%
  \BibitemOpen
  \bibfield  {author} {\bibinfo {author} {\bibfnamefont {F.}~\bibnamefont
  {Wilczek}},\ }\href {\doibase 10.1103/PhysRevLett.40.279} {\bibfield
  {journal} {\bibinfo  {journal} {Phys. Rev. Lett.}\ }\textbf {\bibinfo
  {volume} {40}},\ \bibinfo {pages} {279} (\bibinfo {year} {1978})}\BibitemShut
  {NoStop}%
\bibitem [{\citenamefont {Harari}\ and\ \citenamefont
  {Lousto}(1990)}]{Harari:1990cz}%
  \BibitemOpen
  \bibfield  {author} {\bibinfo {author} {\bibfnamefont {D.}~\bibnamefont
  {Harari}}\ and\ \bibinfo {author} {\bibfnamefont {C.}~\bibnamefont
  {Lousto}},\ }\href {\doibase 10.1103/PhysRevD.42.2626} {\bibfield  {journal}
  {\bibinfo  {journal} {Phys. Rev. D}\ }\textbf {\bibinfo {volume} {42}},\
  \bibinfo {pages} {2626} (\bibinfo {year} {1990})}\BibitemShut {NoStop}%
\bibitem [{\citenamefont {Felder}\ \emph
  {et~al.}(2001{\natexlab{a}})\citenamefont {Felder}, \citenamefont
  {Garcia-Bellido}, \citenamefont {Greene}, \citenamefont {Kofman},
  \citenamefont {Linde},\ and\ \citenamefont {Tkachev}}]{Felder:2000hj}%
  \BibitemOpen
  \bibfield  {author} {\bibinfo {author} {\bibfnamefont {G.~N.}\ \bibnamefont
  {Felder}}, \bibinfo {author} {\bibfnamefont {J.}~\bibnamefont
  {Garcia-Bellido}}, \bibinfo {author} {\bibfnamefont {P.~B.}\ \bibnamefont
  {Greene}}, \bibinfo {author} {\bibfnamefont {L.}~\bibnamefont {Kofman}},
  \bibinfo {author} {\bibfnamefont {A.~D.}\ \bibnamefont {Linde}}, \ and\
  \bibinfo {author} {\bibfnamefont {I.}~\bibnamefont {Tkachev}},\ }\href
  {\doibase 10.1103/PhysRevLett.87.011601} {\bibfield  {journal} {\bibinfo
  {journal} {Phys. Rev. Lett.}\ }\textbf {\bibinfo {volume} {87}},\ \bibinfo
  {pages} {011601} (\bibinfo {year} {2001}{\natexlab{a}})},\ \Eprint
  {http://arxiv.org/abs/hep-ph/0012142} {arXiv:hep-ph/0012142} \BibitemShut
  {NoStop}%
\bibitem [{\citenamefont {Felder}\ \emph
  {et~al.}(2001{\natexlab{b}})\citenamefont {Felder}, \citenamefont {Kofman},\
  and\ \citenamefont {Linde}}]{Felder:2001kt}%
  \BibitemOpen
  \bibfield  {author} {\bibinfo {author} {\bibfnamefont {G.~N.}\ \bibnamefont
  {Felder}}, \bibinfo {author} {\bibfnamefont {L.}~\bibnamefont {Kofman}}, \
  and\ \bibinfo {author} {\bibfnamefont {A.~D.}\ \bibnamefont {Linde}},\ }\href
  {\doibase 10.1103/PhysRevD.64.123517} {\bibfield  {journal} {\bibinfo
  {journal} {Phys. Rev. D}\ }\textbf {\bibinfo {volume} {64}},\ \bibinfo
  {pages} {123517} (\bibinfo {year} {2001}{\natexlab{b}})},\ \Eprint
  {http://arxiv.org/abs/hep-th/0106179} {arXiv:hep-th/0106179} \BibitemShut
  {NoStop}%
\bibitem [{\citenamefont {Berg}\ and\ \citenamefont
  {Luscher}(1981)}]{Berg:1981er}%
  \BibitemOpen
  \bibfield  {author} {\bibinfo {author} {\bibfnamefont {B.}~\bibnamefont
  {Berg}}\ and\ \bibinfo {author} {\bibfnamefont {M.}~\bibnamefont {Luscher}},\
  }\href {\doibase 10.1016/0550-3213(81)90568-X} {\bibfield  {journal}
  {\bibinfo  {journal} {Nucl. Phys. B}\ }\textbf {\bibinfo {volume} {190}},\
  \bibinfo {pages} {412} (\bibinfo {year} {1981})}\BibitemShut {NoStop}%
\end{thebibliography}%
\end{document}